\documentclass{article}

\usepackage{hyperref}
\usepackage[a4paper, top=25mm, left=10mm, right=10mm, bottom=30mm, headsep=10mm, footskip=12mm]{geometry}
\usepackage{authblk}

\usepackage[nointlimits]{amsmath}
\usepackage{siunitx}

\usepackage{caption}
\usepackage{graphicx}
\usepackage{subfigure}
\usepackage{xcolor}
\usepackage{lscape}

\usepackage[ruled,vlined]{algorithm2e}
\usepackage{algorithmic}

\usepackage{natbib}

\definecolor{mycolor_blue}{RGB}{66,124,161}
\definecolor{mycolor_grey}{RGB}{198,198,198}

\usepackage{tikz}
\usepackage{pgfplots}

\usetikzlibrary{patterns}

\pgfplotsset{
	kurze Legende/.style={
		legend image code/.code={
			\draw[##1,mark repeat=2,line width=0.6pt]
			plot coordinates {
				(0cm,0cm)
				(0.3cm,0cm)
			};
		}
	}
}

\pgfplotsset{
	compat = newest,
	scale only axis, 
	max space between ticks = 50pt,
	ticklabel style = {font=\footnotesize},
	legend style =  {font=\footnotesize},
	grid = major,
	grid style = {dotted},
	legend columns=1, 
	xtick pos=left,
	ytick pos=left
}

\newcommand{\longPicture}{
	\pgfplotsset{  
		width=0.8\textwidth,
		height=0.2\textwidth,
		ylabel style={text width=0.2\textwidth,align=center}
	}
}

\newcommand{\smallPicture}{
	\pgfplotsset{  
		width=0.35\textwidth,
		height=0.35\textwidth,
		ylabel style={text width=0.2\textwidth,align=center}
	}
}

\newcommand{\tinyPicture}{
	\pgfplotsset{  
		width=0.3\textwidth,
		height=0.3\textwidth,
		ylabel style={text width=0.2\textwidth,align=center}
	}
}

\newcommand{\analytiSolutionPictures}{
	\pgfplotsset{  
		width=0.24\textwidth,
		height=0.25\textwidth,
		ylabel style={text width=0.2\textwidth,align=center}
	}
}

\pgfplotsset{select coords between index/.style 2 args={
		x filter/.code={
			\ifnum\coordindex<#1\fi
			\ifnum\coordindex>#2\fi
		}
}}

\definecolor{color1}{HTML}{0060AD} % blau
\definecolor{color2}{HTML}{FF4500} % rot
\definecolor{color3}{HTML}{FFA500} % gelb
\definecolor{color4}{HTML}{006400} % gruen
\definecolor{color5}{HTML}{9400D3} % lila
\definecolor{color6}{HTML}{800000} % bordeauxrot/braun
\definecolor{color7}{HTML}{000000} % schwarz
\definecolor{color8}{HTML}{0000FF} % blau heller
\definecolor{color9}{HTML}{FF0000} % rot heller
\definecolor{mycolor_blue}{RGB}{66,124,161}% niklas blue
\definecolor{mycolor_grey}{RGB}{198,198,198} % niklas grey

\tikzstyle{line1} = [color=color7,semithick] 
\tikzstyle{line2} = [color=color2,densely dotted,semithick]
\tikzstyle{line3} = [color=color1,densely dashed,semithick]
\tikzstyle{line4} = [color=color5,dash dot,semithick]
\tikzstyle{line5} = [color=color4,dash dot dot,semithick]
\tikzstyle{line6} = [color=color6,semithick]

\tikzstyle{mark1}  = [color=color7,mark=x,mark size=2pt,mark options=solid,semithick] 
\tikzstyle{mark2}  = [color=color2,mark=o,mark size=2pt,mark options=solid,semithick]
\tikzstyle{mark3}  = [color=color1,mark=*,mark size=2pt,mark options=solid,semithick]
\tikzstyle{mark4}  = [color=color5,mark=triangle,mark size=2pt,mark options=solid,semithick]
\tikzstyle{mark5}  = [color=color4,mark=square,mark size=2pt,mark options=solid,semithick]
\tikzstyle{mark8}  = [color=color7,mark=o,mark size=2pt,mark options=solid,semithick]
\tikzstyle{mark9}  = [color=color7,mark=*,mark size=2pt,mark options=solid,semithick]
\tikzstyle{mark10} = [color=color7,mark=triangle,mark size=2pt,mark options=solid,semithick]

\begin{document}

\title{Adjoint Complement to the Volume-of-Fluid Method for Immiscible Flows}

\author[1]{Niklas K\"uhl\thanks{niklas.kuehl@tuhh.de}}
\author[2]{J\"orn Kr\"oger}
\author[3]{Martin Siebenborn}
\author[4]{Michael Hinze}
\author[1]{Thomas Rung}

\affil[1]{Hamburg University of Technology, Institute for Fluid Dynamics and Ship Theory, Am Schwarzenberg-Campus 4, D-21075 Hamburg, Germany}
\affil[2]{Hamburg Ship Model Basin, Bramfelder Straße 164, D-22305 Hamburg, Germany}
\affil[3]{University of Hamburg, Department of Mathematics, Bundesstrasse 55, D-20146 Hamburg, Germany}
\affil[4]{University Koblenz-Landau, Department of Mathematics, Campus Koblenz, Universit\"atsstrasse 1, D-56070 Koblenz}

\maketitle

\begin{abstract}
The paper is concerned with an adjoint complement to the Volume-of-Fluid (VoF) method for immiscible two-phase flows, e.g. air and water, which is widely used in marine engineering due to its computational efficiency. The particular challenge of the primal and the corresponding adjoint VoF-approach refers to the sharp interface treatment featuring discontinuous physical properties.
Both the continuous adjoint two-phase system (integration-by-parts) and the corresponding dual compressive convection schemes (summation-by-parts) are derived
for two prominent compressive convection schemes, namely the High Resolution Interface Capturing Scheme (HRIC) and Compressive Interface Capturing Scheme for Arbitrary Meshes (CICSAM).
 The dual scheme rigorously mirrors the primal Normalized-Variable-Diagram (NVD) stencils. 
  Attention is restricted to steady state applications. Thus both the primal and the dual procedures are performed in pseudo time and 
 %Instead of switching from upwind (UD) to downwind (DD) biased interpolation along the interface in the pseudo-transient forward simulation,
the 
  %subsequent 
  backward integration of the dual approach is performed around the (pseudo-temporal) converged primal field.
  %and locally switches from DD to UD. 
  Therefore, the adjoint system experiences the same time step size restrictions as the primal system, %\trung{verstehe ich nicht ganz:} 
  is independent of the primal time horizon and forms a robust as well as an a priori stable adjoint solution process.
  %\nkuehl{in the absence of adjoint two-phase related coupling terms}. 

%Motivated by several source/sink terms inherent to the adjoint VoF system, 
The paper analyses the primal and adjoint equations for an engineering model problem. An analytical solution to the model problem is initially presented, which displays that the adjoint part does not offer a unique, non-trivial solution. As a remedy, an additional diffusive concentration term is introduced to the adjoint concentration equation. The imposed heuristic modification violates the dual consistency but strongly regularizes the solution of the adjoint system. The modification can be justified by reference to phase-separating diffuse-interface models and inheres a free mobility-parameter. Numerical results obtained from the modified approach are benchmarked against the analytical solution for the model problem. Supplementary, the influence of the modification on the sensitivities obtained from simulations for the two-dimensional flow around a submerged hydrofoil at Froude and Reynolds numbers of practical interest are discussed for a range of mobility-parameters. The final application refers to a shape-optimization of a generic 3D underwater vehicle and underlines a negligible influence of the free mobility parameter, even for an objective functional that directly depends on the manipulated (dual) field quantity.
\end{abstract}


\section{Introduction}
The overall goal of adjoint analysis in the Computational Fluid Dynamics (CFD) context is usually the efficient computation of derivative information of an integral objective functional with respect to a general control parameter or control function. In case of Optimal Shape Design (OSD), the latter is defined by parts of the bounded calculation area. Using a steepest descent approach, the required shape-sensitivities are often preferably obtained by adjoint methods due to the independence of the computational costs from the number of design variables. When attention is given to CAD-free shape optimization, the number of design variables is usually large and tends to infinity for physically consistent shape optimization \citep{peter2010numerical}. For this reason, the adjoint method has been used in many scientific \citep{errico1997adjoint} and industrial \citep{jameson1995optimum,papadimitriou2007continuous,giannakoglou2008adjoint,othmer2014adjoint,kavvadias2015proper,papoutsis2016continuous,kapellos2019unsteady} fields of fluid dynamic shape optimization since its first use by \citep{pironneau1974optimum} and  \citep{jameson1988aerodynamic}. 
Because an adjoint (mathematical) system by definition follows from a primal (physical) system, the development of adjoint simulation methods lags naturally behind that of classical CFD approaches. Additionally, the adjoint system is at least partially closed to the underlying physics due to e.g. possibly unintuitive boundary conditions. One of the challenges posed to the adjoint approach is associated to differentiability issues or singularities of the underlying primal-flow model. Examples refer to shocks \citep{ulbrich2003adjoint, beckers2019duality}, singular boundary conditions of turbulence properties \citep{zymaris2009continuous, zymaris2010adjoint, papoutsis2015continuous, papoutsis2016continuous, manservisi2016numerical, manservisi2016optimal} or discontinuities associated to immiscible two-phase flows \citep{kroger2018adjoint}, which is the scope of this paper. 

\subsection{Industrial Marine CFD-strategies}
Industrial free-surface flow simulations featuring viscous and turbulent effects are nowadays dominated by Finite Volume (FV) approximations of the Reynolds-Averaged (RANS) or filtered (LES) Navier-Stokes equations for multiple fluid phases \citep{lafaurie1994modelling,larsson2003numerical,queutey2007interface,rung2009challenges,sadat2013cfd,schellin2011wave,shen2015dynamic}. When attention is restricted to non-cavitating flows, such simulations are nowadays usually performed with Volume-of-Fluid (VoF) procedures proposed by \citep{hirt1981volume} for two immiscible, incompressible phases i.e. air and water. Virtually all such procedures apply a nonlinear compressive approximation for the convective mixture fraction transport to keep the discrete air/water-interface sharp \citep{muzaferija1998computation,muzaferija1999two,rusche2002computational,so2011anti,ubbink1999method,waclawczyk2008comparison,zhang2014refined}. Moreover they often employ a pressure-correction or pressure-projection scheme that is embedded in a segregated solution process \citep{ferziger2008numerische}. The inherent conservation of mass, the superior numerical efficiency and the flexibility to capture ruptured free surfaces explain the dominance of VoF-procedures in industrial marine CFD.

\subsection{Adjoint CFD-strategies}
Two options of deriving the discrete adjoint equations are conceivable, i.e. following the continuous or discrete adjoint approach. The former derives the adjoint PDE system in continuous space and subsequently discretizes them. The latter exclusively operates in discrete space and formulates the adjoint of the discrete (linearized) primal flow by transposing the discrete primal operators. A detailed discussion of the merits and drawbacks of the respective baseline approaches is beyond the scope of the present paper and the interested reader is referred to \citep{peter2010numerical} or \citep{griewank2008evaluating}. The continuous adjoint approach is preferred herein since it provides a deeper insight into physical and mathematical relationships of the underlying problem and its mathematical model \citep{giles2000introduction}. This is particularly relevant in this study, which subsequently utilizes elements of the discrete adjoint approach to discretize the adjoint equations along a route described in \citep{stuck2013adjoint}.
 
Although a few publications on hydrodynamic optimization can be found, a holistic view on adjoint shape optimisation in free surface-flows is scarce. Ragab  \citep{ragab2001adjoint, ragab2003shape} developed a design framework for the optimisation of surface ships and submarines operating near a free surface based on potential flow methods. The optimisation is performed for wave resistance objectives, but also for inverse approaches to reach a prescribed pressure distribution and wave pattern. S\"oding \citep{soding2001hull, soding2001practical, soding2001resistance} employed the adjoint complement of a potential flow solver to reduce the resistance of merchant vessels. Soto and L\"ohner \citep{soto2001cfd}  and Soto et al. \citep{soto2004adjoint} applied an adjoint Euler-flow solver to reduce the resistance of a container vessel using a frozen free-surface approach. An adjoint Euler-flow method that incorporates an interface-tracking approach was used by Martinelli et al. \citep{martinelli2007adjoint} to reduce the wave resistance of an academic Wigley hull. 

The challenges  associated with the concentration transport in VoF-schemes have motivated the restriction of the previous research to either simplified VoF-schemes \citep{springer2015adjoint}, level-set based strategies \citep{palacios2012shape, palacios2013design} or diffusive interface schemes \citep{garcke2019optimal, hinze2011nonlinear, kuhl2020cahn} in interface capturing approaches. The primal VoF-approach transports the discontinuous flow properties along with the free surface.  In previous work related to shocks, similar differentiability problems were treated with shift differentiability \citep{ulbrich2002sensitivity, ulbrich2003adjoint} or the application of an artificial viscosity, with the purpose of filtering out physical solutions \citep{giles2010convergence}. Synthetic viscosities have also been used for error estimation in transient shallow-water flows as reported by Beckers et al \citep{beckers2019duality}. Moreover, the above-mentioned compressive approximations of the primal convective kinematics are based on heuristic, nonlinear expressions, which are cumbersome to translate into an adjoint context. Palacios et al. \citep{palacios2012shape,palacios2013design} thus opted for an adjoint level-set Euler-flow solver that is applied to 2D- and 3D obstacles in free-surface channel-flows. Springer and Urban \citep{springer2015adjoint} developed an adjoint VoF Navier-Stokes solver to identify the floatation position of ships, but neglect some of the adjoint coupling terms and compromised on the duality of the approximation. Their main objective referred to the equilibrium of the trim moment at steady state.

The present study pursues the work of \citep{kroger2018adjoint}, who suggested a heuristic approach to fully coupled primal/adjoint two-phase flow procedures. This approach is dedicated to industrial ship hydrodynamic simulations and will be further scrutinized in this paper. We aim to demonstrate the necessity of the modification by means of a model problem with practical relevance, where non-trivial analytical  solutions are only available when using the modification. In addition, it can be shown that the impact on the final sensitivity is
%-- for the model problem as well as for fully turbulent two/three dimensional two phase flows -- 
negligible in the investigated academic and practical cases. 
The remainder of the paper is organised as follows: Section (\ref{sec:derivation}) is devoted to the primal VoF-system and the associated objective based adjoint equations followed by a detailed derivation of the adjoint representation of compressive interface capturing schemes as well as implementation details of the 
adjoint two-phase terms in Section (\ref{sec:numerical}). The coupled primal/adjoint-system is subsequently investigated for a plane Couette-flow in Section (\ref{sec:couette}). The adjoint system is applied to the two-dimensional flow around a frequently investigated towed hydrofoil as well as the three-dimensional flow around generic underwater vehicle (\ref{sec:application}) at Froude and Reynolds of practical interest. Section (\ref{sec:discussion}) provides conclusions and outlines future options. Within the publication, Einstein's summation convention is used for lower-case Latin subscripts. Vectors and tensors are defined with reference to Cartesian coordinates and dimensional variables are consistently marked with an asterisk.
%\trung{ist das klug? - oder besser anders herum?} \nkuehl{da kann man nat. anfangen haare zu spalten. meine meinung: wenn man dimensionslose groessen mit einem sternchen versieht, muesste man das eigentlich (konsistent) immer bei reynolds-zahl und froude-zahl etc mitziehen. darueber hinaus tut es in der herleitung/interpretation gut, genau zu wissen mit was fuer einer groesse man es gerade zu tun hat.. ABER: am ende ist ist und bleibt es auslegungssache!}

\section{Mathematical Model}
\label{sec:derivation}
The paper is confined to the flow of two immiscible fluid phases ($a,b$) featuring constant bulk phase properties, i.e. density $\rho_\mathrm{a}^*, \rho_\mathrm{b}^*$  and dynamic viscosity $\mu_\mathrm{a}^*, \mu_\mathrm{b}^*$. The fluid $a$ is referred to as foreground fluid and fluid $b$ as background fluid.
%, which can basically also be defined the other way around. 
Both fluids are assumed to share the kinematic field along the route of the VoF-approach. The governing equations refer to 
the momentum and continuity equation for the mixture as well as a transport
equation for the volume concentration of the foreground phase. 
The residual form of the governing primal Navier-Stokes equations, that need to be solved for the pressure $p^*$, the velocity $v_i^*$ and the volume concentration $c$ read
\begin{alignat}{2}
R_\mathrm{i}: &  \rho^* \frac{\partial v_\mathrm{i}^*}{\partial t^*} + \rho^* v_\mathrm{j}^* \frac{\partial v_\mathrm{i}^*}{\partial x_\mathrm{j}^*} + \frac{\partial }{\partial x_\mathrm{j}^*} \left[ p^* \delta_\mathrm{ij} - 2 \mu_\mathrm{e}^* S_\mathrm{ij}^* \right] - \rho^* g_\mathrm{i}^* &&= 0 \label{equ:prima_momen} , \\
Q:& -\frac{\partial v_\mathrm{i}^*}{\partial x_\mathrm{i}^*} &&= 0 \label{equ:prima_mass} ,  \\
C:& \frac{\partial c}{\partial t^*} + v_\mathrm{j}^* \frac{\partial c}{\partial x_\mathrm{j}^*} &&= 0. \label{equ:prima_conce}
\end{alignat}
The unit coordinates and the shear rate tensor are denoted by  $\delta_\mathrm{ij}$ and $S_\mathrm{ij}^*$. The boundary conditions are given in Tab. (\ref{tab:bound_condi}). 
The framework supports laminar and Reynolds-averaged (modelled) turbulent flows (RANS). In the latter case, $v_i$ and $p$ correspond to Reynolds-averaged properties and $p$ is additionally augmented by a turbulent kinetic energy ($k$) term, i.e. $2 \rho k/3$. Along with the Bousinesq hypothesis, the dynamic viscosity $\mu_\mathrm{e} = \mu+ \mu_\mathrm{t}$ of turbulent flows consists of a molecular and a turbulent contribution ($\mu_\mathrm{t}$), and the system is closed using a two-equation turbulence model to determine $\mu_\mathrm{t}$ and $k$. Details of the turbulence modelling practice are omitted to safe space and can be found in textbooks, e.g. \citep{wilcox1998turbulence}. 
The system of Eqns. (\ref{equ:prima_momen})-(\ref{equ:prima_conce}) is closed by equations of state (EoS) to determine the local mixture properties from a concentration-based interpolation. Typically a simple linear interpolation is employed, viz.
\begin{align}
%\rho^* = c \rho_\mathrm{a}^* + \left( 1 - c \right) \rho_\mathrm{b}^*, \label{equ:mater_prope}
\rho^* = \rho_\mathrm{a}^* +  \frac{\rho_\Delta^*}{c_\Delta} \left( c - c_\mathrm{a} \right) , \label{equ:mater_prope}
\qquad
%\mu^* = c \mu_\mathrm{a}^* + \left( 1 - c \right) \mu_\mathrm{b}^*.
\mu^* = \mu_\mathrm{a}^* +  \frac{\mu_\Delta^*}{c_\Delta} \left( c - c_\mathrm{a} \right)
\end{align}
where $\rho_\Delta^* = \rho_\mathrm{a}^* - \rho_\mathrm{b}^*$, $\mu_\Delta^* = \mu_\mathrm{a}^* - \mu_\mathrm{b}^*$ and $c_\Delta = c_\mathrm{a} - c_\mathrm{b}$. In this paper we employ $c_\mathrm{a} = 1$ and $c_\mathrm{b} = 0$, thus the volume concentration is solved for the foreground fluid a only.  Physically, the concentration value is restricted to $c = 0$ and $c = 1$. Intermediate values might occur and denote to the free surface. They are usually suppressed by a compressive convective approximation to keep the interface sharp.

The present research is concerned with steady state problems, i.e. the solution is advanced in pseudo time and converges to a steady state. An important example refers to the drag of a vessel in calm-water conditions. The suggested strategies could, however, also be applied to unsteady problems. 

\subsection{Adjoint Volume-of-Fluid Model}
The adjoint equations depend on the underlying integral objective functional, viz.
\begin{align}
J^* = \int_{\Omega_\mathrm{O}^*} j_\Omega^* \, \mathrm{d} \Omega^*  + \int_{\Gamma_\mathrm{O}^*} j_\Gamma^* \, \mathrm{d} \Gamma^* \label{equ:general_objective}
\end{align}
over certain parts of the domain ($\Omega_\mathrm{O}^* \subseteq \Omega^* $) or its boundary ($\Gamma_\mathrm{O}^* \subseteq \Gamma^*$ ). Both integrands in Eqn. (\ref{equ:general_objective}) can depend on the field quantities of the primal system  (\ref{equ:prima_momen})-(\ref{equ:prima_conce}). Two exemplary objectives used in this paper read
\begin{align}
j_\Omega =   \frac{1}{2} \left[ c - c_\mathrm{t} \right]^2 \qquad j_\Gamma^* =  \left[ p^* \delta_{\mathrm{ij}} - 2 \mu^* S_{\mathrm{ij}}^* \right] n_\mathrm{j} d_\mathrm{i} , \label{equ:special_objective}
\end{align}
where the first aims at minimizing the deviation of  the concentration from a target distribution $c_\mathrm{t}$  and the second represents the fluid flow induced 
%(surface) 
force projected in a spatial direction $d_\mathrm{i}$.

Generally, the adjoint solution points backward in time and is linearised around the current primal flow solution. However, although the present primal procedure is unsteady, we are seeking for the steady-state solution, force and wave pattern. Hence, robust convergence but not time accuracy is required from the intermediate primal and dual flow fields. Therefore, the adjoint solution can be computed from the converged steady primal flow field and the adjoint equations belonging to the general functional (\ref{equ:general_objective}) follow from a Lagrangian formalism
\begin{align}
L^* = J^* + \int_{t^*} \int_{\Omega^*} \left[ \hat{v_\mathrm{i}}^* R_\mathrm{i}^* + \hat{p}^* Q^* + \hat{c}^* C^*  \right] \ \mathrm{d} \Omega^* \ \mathrm{d} t^* \label{equ:optim}.
\end{align}
The Lagrange multipliers $\hat{p}^*$, $\hat{v}_\mathrm{i}^*$ and $\hat{c}^*$ refer to the adjoint pressure, adjoint velocity, and adjoint concentration. The units of adjoint pressure and adjoint concentration are equal $\left[ \hat{p}^* \right] = \left[ \hat{c}^* \right] = [J^*] \, 1  / \mathrm{m}^3$. The unit of the adjoint velocity refers to $\left[ \hat{v}_i^* \right] = [J^*] \, 1  / \mathrm{(N \, s)}$.
As outlined before, the current work focuses on quasi-steady primal flows. Hence, for the converged primal solution the unsteady term vanishes and the primal time history is physically meaningless. 
However, the time-dependent adjoint terms are initially retained during the derivation and optionally serve to stabilize the adjoint solution process in terms of a pseudo-time integration. Therefore, all adjoint time steps are linearised around the same converged primal flow field.
 
First order optimality conditions demand the derivatives of the objective to disappear in all relevant directions at the optimal point, which leads to the adjoint system of equations \citep{heners2017adjoint} and allows the definition of a sensitivity rule along the design surface. By way of example, we restrict ourselves to those demonstrative contributions to the augmented objective functional $L^*$ in Eqn. (\ref{equ:optim}), which do not occur in the classical steady, single-phase system. These terms read
\begin{align}
\tilde{L}^* = \int_ {t^*} \int_{\Omega^*} \hat{v_\mathrm{i}}^* \rho^* \frac{\partial v_\mathrm{i}^*}{\partial t^*} + \hat{c}^* \left[ \frac{\partial c}{\partial t^*} + v_\mathrm{j}^* \frac{\partial c}{\partial x_\mathrm{j}^*} \right] \mathrm{d} \Omega^* \ \mathrm{d} t^* \label{equ:free_surf_rele}
 \; . 
\end{align}
A more detailed derivation can -- for example -- be found in \citep{kroger2018adjoint}. The continuity equation remains unaltered and the changes are confined to a (pseudo-) transient term in the momentum equation and the concentration transport. The derivative of Eqn. (\ref{equ:free_surf_rele}) in velocity direction reads
\begin{align}
\delta_{v} \tilde{L}^* \cdot \delta v_\mathrm{i}^* &= 
\underbrace{\left[ \int_{\Omega^*} \rho^* \hat{v_\mathrm{i}}^* \delta v_\mathrm{i}^* \mathrm{d} \Omega^* \right]_{t_0^*}^{t_1^*}}_{=0} + 
\int_{t^*} \int_{\Omega^*} \delta v_\mathrm{i}^* \left[ - \rho^* \frac{\partial \hat{v_\mathrm{i}}^*}{\partial t^*} + \delta_\mathrm{ij} \hat{c}^* \frac{\partial c}{\partial x_\mathrm{j}^*} \right] \mathrm{d} \Omega^* \ \mathrm{d} t^*\label{equ:velo_der}.
\end{align}
Optionally, the spatial derivative of the concentration can be shifted to its adjoint counterpart in Eqn. (\ref{equ:velo_der}) via integration by parts.
The derivatives in the directions of the fluid properties are linked to the  derivatives in concentration direction via the EoS (\ref{equ:mater_prope}) and read
\begin{align}
\delta_{c} \rho^* \cdot \delta c &=  \delta c  \frac{\rho_\Delta^*}{c_\Delta} =  \delta c \, \rho_\Delta^*, \qquad
\delta_c \mu^* \cdot \delta c = \delta c  \frac{\mu_\Delta^*}{c_\Delta} = \delta c \, \mu_\Delta^*\, . \label{equ:delta_term}
\end{align}
where $c_\Delta$ is assigned to unity in the present study. The final derivative of $\tilde{L}^*$ in the direction of $c$ takes the following form
\begin{align}
\delta_c \tilde{L}^* \cdot \delta c &= \underbrace{\left[ \int_{\Omega^*} \delta c \, \hat{c}^* \mathrm{d} \Omega^* \right]_{t_0^*}^{t_1^*}}_{=0} + \int_{t^*} \int_{\Gamma^*} \delta c \left[ v_\mathrm{j}^* \hat{c}^* n_\mathrm{j} \right] \mathrm{d} \Gamma^* +  \int_{\Omega^*} 
 \delta c \bigg[ \rho_\Delta^* \hat{v_\mathrm{i}} \underbrace{ \frac{\partial v_\mathrm{i}^*}{\partial t^*} }_{= 0} - \frac{\partial \hat{c}^*}{\partial t^*} - v_\mathrm{j}^* \frac{\partial \hat{c}^*}{\partial x_\mathrm{j}^*} \bigg] \mathrm{d} \Omega^* \ \mathrm{d} t^* \nonumber \\
&=  \int_{t^*} \int_{\Gamma^*} \delta c \left[ v_\mathrm{j}^* \hat{c}^* n_\mathrm{j} \right] \mathrm{d} \Gamma^* +  \int_{\Omega^*} \delta c \bigg[ - \frac{\partial \hat{c}^*}{\partial t^*} - v_\mathrm{j}^* \frac{\partial \hat{c}^*}{\partial x_\mathrm{j}^*} \bigg] \mathrm{d} \Omega^* \ \mathrm{d} t^* \label{equ:conc_der} \ ,
\end{align}
where a  steady state primal flow is assumed. Eqns. (\ref{equ:velo_der}) and (\ref{equ:conc_der}) are supplemented by the standard steady single-phase terms that were not explicitly mentioned herein. Moreover, three more derivatives in the direction of the concentration occur which originate from the density inherent to the primal convection and buoyancy term of the momentum equation as well as the viscosity derivative of the primal shear stress.

The requirement for vanishing derivatives in all respective directions yields the final set of adjoint equations that have to be solved throughout the domain:
\begin{alignat}{2}
\hat{R_\mathrm{i}}^*:& - \rho^* \frac{\partial \hat{v}_\mathrm{i}^*}{\partial t^*} - \rho^* v_\mathrm{j}^* \frac{\partial \hat{v}_\mathrm{i}^*}{\partial x_\mathrm{j}^*} +  \rho^* \hat{v}_\mathrm{j}^* \frac{\partial v_\mathrm{j}^*}{\partial x_\mathrm{i}^*} - c \frac{\partial \hat{c}^*}{\partial x_\mathrm{i}^*} + \frac{\partial}{\partial x_\mathrm{j}^*} \left[ \hat{p}^* \delta_\mathrm{ij} - 2 \mu^* \hat{S}_\mathrm{ij}^* \right] &&= - \frac{\partial j_\mathrm{\Omega}^*}{\partial v_\mathrm{i}^*} \label{equ:adjoi_momen} , \\
\hat{Q}^*:& -\frac{\partial \hat{v_\mathrm{i}}}{\partial x_\mathrm{i}^*} &&= - \frac{\partial j_\mathrm{\Omega}^*}{\partial p^*} \label{equ:adjoi_mass} , \\
\hat{C}^*:& -\frac{\partial \hat{c}^*}{\partial t^*} - v_\mathrm{j}^* \frac{\partial \hat{c}^*}{\partial x_\mathrm{j}^*} + \rho_\Delta^* \hat{v}_\mathrm{i}^* v_\mathrm{j}^* \frac{\partial v_\mathrm{i}^*}{\partial x_\mathrm{j}^*} + 2 \mu_\Delta^* S_\mathrm{ij}^* \frac{\partial \hat{v}_\mathrm{i}^*}{\partial x_\mathrm{j}^*} - \rho_\Delta^* \hat{v}_\mathrm{i}^* g_\mathrm{i}^* &&= - \frac{\partial j_\mathrm{\Omega}^*}{\partial c}. \label{equ:adjoi_conce}
\end{alignat}
The adjoint equations are similar to the primal equations. However, additional advection and cross-coupling terms occur. Above all, the adjoint concentration equation contains significantly more terms that scale with the density or viscosity difference of the two fluids.

The associated boundary conditions result from the boundary integrals of partial integration, which likewise have to disappear for all variations (cf. Tab. \ref{tab:bound_condi}). Interestingly, the Dirichlet condition of adjoint concentration switches from the inlet in primal to the outlet in adjoint mode. With the aid of remaining optimality criteria and a correct approximation of the primal and dual equations, a sensitivity rule can be determined along the controlled design wall \citep{kroger2016numerical, stuck2012adjoint, kuhl2019decoupling} 
\begin{align}
\delta_u J^* = -\int_{\Gamma_\mathrm{O}^*} s^* \, \mathrm{d} \Gamma_\mathrm{O}^* \qquad \mathrm{with} \qquad s^* = \mu^* \frac{\partial v_\mathrm{i}^*}{\partial x_\mathrm{j}^*} \frac{\partial \hat{v_\mathrm{i}}}{\partial x_k^*} n_\mathrm{j} n_k \label{equ:shape_derivative}.
\end{align}

\begin{table}
\centering
\begin{tabular}{|c||c|c|c|c|c|c|}
\hline
boundary type & $v_\mathrm{i}^*$  & $p^*$ & $c$ & $\hat{v}_\mathrm{i}^*$  & $\hat{p}^*$ & $\hat{c}^*$\\
\hline
\hline
inlet & $v_\mathrm{i}^* = v_\mathrm{in}^*$ & $\frac{\partial p^*}{\partial x_\mathrm{n}^*} = 0$ & $c = c_\mathrm{in}$ & $\hat{v}_\mathrm{i}^* = 0$ & $\frac{\partial \hat{p}^*}{\partial x_\mathrm{n}^*} = 0$ & $\frac{\partial \hat{c}^*}{\partial x_\mathrm{n}^*} = 0$ \\
\hline
outlet & $\frac{\partial v_\mathrm{i}^*}{\partial x_\mathrm{n}^*} = 0$ & $p^* = p_\mathrm{out}^*$ & $\frac{\partial c}{\partial x_\mathrm{n}^*} = 0$  &\multicolumn{2}{c|}{$\hat{p}^* n_\mathrm{i} = \rho^* \hat{v_\mathrm{i}} v_\mathrm{j}^* n_\mathrm{j} + \mu^* \frac{\partial \hat{v}_\mathrm{i}^*}{\partial x_\mathrm{n}^*} + \hat{c}^* c n_\mathrm{i}$} & $\hat{c}^* = \frac{1}{v_\mathrm{n}^*} \hat{v}_\mathrm{i}^* \mu_\Delta^* \frac{\partial v_\mathrm{n}^*}{\partial x_\mathrm{i}^*}$ \\
\hline
symmetry & $v_\mathrm{n}^* = 0$ & $\frac{\partial p^*}{\partial x_\mathrm{n}^*} = 0$ & $\frac{\partial c}{\partial x_\mathrm{n}^*} = 0$  & $\hat{v}_\mathrm{i}^* = 0$ &  $\hat{p}_\mathrm{n}^* = 0$ & $\frac{\partial \hat{c}^*}{\partial x_\mathrm{n}^*} = 0$\\
\hline
wall $\Gamma^* \backslash \Gamma_O^*$ & $v_\mathrm{i}^* = 0$ & $\frac{\partial p^*}{\partial x_\mathrm{n}^*} = 0$ & $\frac{\partial c}{\partial x_\mathrm{n}^*} = 0$  & $\hat{v}_\mathrm{i}^* = 0$ &  $\frac{\partial \hat{p}^*}{\partial n} = 0$ & $\frac{\partial \hat{c}^*}{\partial x_\mathrm{n}} = 0$\\
\hline
wall $\Gamma^* \subset \Gamma_O^*$ & $v_\mathrm{i}^* = 0$ & $\frac{\partial p^*}{\partial x_\mathrm{n}^*} = 0$ & $\frac{\partial c}{\partial x_\mathrm{n}^*} = 0$  & $\hat{v}_\mathrm{i}^* = -\frac{\partial j_\mathrm{\Gamma}^*}{\partial p^*}$ &  $\frac{\partial \hat{p}^*}{\partial x_\mathrm{n}^*} = 0$ & $\frac{\partial \hat{c}^*}{\partial x_\mathrm{n}^*} = 0$ \\
\hline
\end{tabular}
\caption{Boundary conditions for the primal and dual equations.}
\label{tab:bound_condi}
\end{table}

\subsection{Non-Dimensional Equations}
For a more detailed analysis, the primal (\ref{equ:prima_momen})-(\ref{equ:prima_conce}) and adjoint (\ref{equ:adjoi_momen})-(\ref{equ:adjoi_conce}) equations are non-dimensionalized. The employed reference quantities as well as the resulting non-dimensional field quantities are given in Tab. (\ref{tab:refe_data}). The non-dimensional primal equations read
\begin{alignat}{2}
R_\mathrm{i}:&  \frac{\rho}{\mathrm{St}} \ \frac{\partial v_\mathrm{i}}{\partial t} + \rho v_\mathrm{j} \frac{\partial v_\mathrm{i}}{\partial x_\mathrm{j}} + \frac{\partial }{\partial x_\mathrm{j}} \left[ (\mathrm{Eu}) \, p \delta_\mathrm{ij} - \frac{2 \mu}{\mathrm{Re}}  S_\mathrm{ij} \right] - \frac{\rho}{\mathrm{Fr}^2}  g_\mathrm{i} &&= 0, \label{equ:prima_momen_dile} \\
Q:& -\frac{\partial v_\mathrm{i}}{\partial x_\mathrm{i}} &&= 0, \label{equ:prima_mass_dile} \\
C:&  \frac{1}{\mathrm{St}} \frac{\partial c}{\partial t} + v_\mathrm{j} \frac{\partial c}{\partial x_\mathrm{j}} &&= 0 \label{equ:prima_conce_dile}.
\end{alignat}
An exemplary relationship between a field quantity and reference value reads $v_\mathrm{i}^* = V^* v_\mathrm{i}$. The employed non-dimensional flow parameters are defined by
\begin{alignat}{2}
\mathrm{St} = \frac{\mathrm{T}^* \mathrm{V}^*}{\mathrm{L}^*}  \quad 
\mathrm{(Strouhal)} ,\qquad 
\mathrm{Re} = \frac{\rho_\mathrm{b}^* \mathrm{V}^* \mathrm{L}^*}{\mu_\mathrm{b}^*} \quad 
\mathrm{(Reynolds)} ,  \qquad
\mathrm{Eu} = \frac{ \mathrm{P}^*}{\rho_\mathrm{b}^* {\mathrm{V}^*}^2} \quad
\mathrm{(Euler)} ,  \qquad
\mathrm{Fr} = \frac{\mathrm{V}^*}{ \sqrt{ G^* \mathrm{L}^* }} \quad  \mathrm{(Froude)}.
\end{alignat}
The Reynolds- and Froude-number are always formed with the density $\rho_\mathrm{b}^*$ and the viscosity $\mu_\mathrm{b}^*$, unless it is stated otherwise. The momentum Eqn. (\ref{equ:prima_momen_dile}) inheres two source terms which are not related to the velocity and display a well-known pairing in hydrostatic conditions (Fr $ \to 0$), e.g. $p + \rho \, g_\mathrm{j} \, x_\mathrm{j}$.  
Hence typical choices for the reference pressure are $\mathrm{P}^*=\rho_b^* {\mathrm{V}^*}^2$ (i.e. $\mathrm{Eu}=1$; hydrodynamics) or $\mathrm{P}^*=\rho_b^* \mathrm{G}^* \mathrm{L}^*$ (i.e. $\mathrm{Eu}=\mathrm{Fr}^{-2}$; hydrostatics). The non-dimensional set of adjoint equations belonging to the general functional (\ref{equ:general_objective}) reads:
\begin{alignat}{2}
\hat{R_\mathrm{i}}:& - \frac{\rho}{\mathrm{St}}  \frac{\partial \hat{v}_\mathrm{i}}{\partial t} - \rho v_\mathrm{j} \frac{\partial \hat{v}_\mathrm{i}}{\partial x_\mathrm{j}} + \rho \hat{v}_\mathrm{j} \frac{\partial v_\mathrm{j}}{\partial x_\mathrm{i}} - \left(\frac{C^*}{V^* \hat{V}^*}\right) \, c \frac{\partial \hat{c}}{\partial x_\mathrm{i}} + \frac{\partial }{\partial x_\mathrm{j}} \left[ \left(\frac{C^*}{V^* \hat{V}^*} \, \right) \hat{p} \delta_\mathrm{ij} -  \frac{2 \mu  }{\mathrm{Re}} \hat{S}_\mathrm{ij} \right] &&= - \left( \frac{J^* \rho_\mathrm{b}^* \hat{V}^*}{L^*} \right) \frac{\partial j_\mathrm{\Omega}}{\partial v_\mathrm{i}} \label{equ:adjoi_momen_dile} \\
\hat{Q}:& - \frac{\partial \hat{v_\mathrm{i}}}{\partial x_\mathrm{i}} &&= -  \left( \frac{J^* L^*}{\hat{V}^* P^*} \right) \frac{\partial j_\mathrm{\Omega}}{\partial p} \label{equ:adjoi_mass_dile} \\
\hat{C}:& - \frac{1}{\mathrm{St}} \frac{\partial \hat{c}}{\partial t} - 
 v_\mathrm{j} \frac{\partial \hat{c}}{\partial x_\mathrm{j}} + \left( \frac{V^* \hat{V}^*}{C^*} \right)  \rho_\Delta \hat{v}_\mathrm{i} v_\mathrm{j} \frac{\partial v_\mathrm{i}}{\partial x_\mathrm{j}} + \left( \frac{V^* \hat{V}^*}{C^*} \right)  \, \frac{2 \mu_\Delta}{\mathrm{Re}}  S_\mathrm{ij} \frac{\partial \hat{v}_\mathrm{i}}{\partial x_\mathrm{j}} - \left( \frac{V^* \hat{V}^*}{C^*} \right) \, 
\frac{\rho_\Delta } {\mathrm{Fr}^2} 
 \hat{v}_\mathrm{i} g_\mathrm{i} &&= - \left( \frac{J^* L^*}{\rho_\mathrm{b}^* V^* C^*} \right) \frac{\partial j_\mathrm{\Omega}}{\partial c} \label{equ:adjoi_conce_dile}
\end{alignat}
One can observe that the adjoint system generally displays a stronger coupling and seems challenging to solve due to the locally volatile characteristics. Two source terms that do not depend on $\hat{v}_i$ occur in the adjoint momentum Eqn. (\ref{equ:adjoi_momen_dile}), where the sum  $\hat{p} + c \hat{c}$ forms an adjoint hydrostatic pairing. The sum could also be condensed into an effective adjoint pressure along the route of the frequently employed scrambling of pressure and turbulent kinetic energy for Boussinesq-viscosity turbulence models. However, this would require a vanishing concentration gradient $\partial c / \partial x_\mathrm{i} = 0$, which is in general not defined along the sharp free surface. Obviously, the adjoint concentration Eqn. (\ref{equ:adjoi_conce_dile}) has a strong similarity to the primal momentum equation, and contains Froude-, Reynolds-, and Strouhal-number terms. As opposed to the primal and adjoint momentum equations, the Reynolds number contribution is not linked to the transported property. Hence three terms acting as pure source terms appear in Eqn. (\ref{equ:adjoi_conce_dile}): An augmented convection term, a Reynolds- and a Froude-number term. However, the adjoint concentration equation does not feature a hydrostatic correspondence or paring of variables, since  the  velocity scaling differs for all three source terms.
The missing links to the transported property and pairing options are the origin of a non-unique (solution) nature of Eqn. (\ref{equ:adjoi_conce_dile}). The issue scales with the property differences between the two fluids. As outlined in Sec. (\ref{sec:couette}) this can be addressed by the introduction of a (self-adjoint) diffusion term which does not depend on a specific property difference and exposes the preferred negligible influence on the computed sensitivity.

\begin{table}
\centering
\begin{tabular}{|c||c|c|c|c|c|c|c|c|c|c|}
\hline
field quantity & 
$x_\mathrm{i}^*$  & $v_\mathrm{i}^*$ & $g_\mathrm{i}^*$ & $p^*$ & $t^*$ & $\rho^*$ & $\mu^*$ & $\hat{p}^*$ & $\hat{c}^*$ & $\hat{
v}_\mathrm{i}^*$ \\
\hline
reference value & $\mathrm{L}^*$ & $\mathrm{V}^*$ & $\mathrm{G}^*$ & $\mathrm{P}^*$ & $\mathrm{T}^*$ & $\rho_\mathrm{b}^*$ & $\mu_\mathrm{b}^*$ & \multicolumn{2}{c|}{$\rho_\mathrm{b}^* C^*$}  & $\hat{V}^*$ \\
\hline
\end{tabular}
\caption{Quantities for the non-dimensionalisation of the governing equations.}
\label{tab:refe_data}
\end{table}

\newpage
\section{Numerical Procedure}
\label{sec:numerical}
The numerical procedure is based upon the Finite-Volume procedure FreSCo+ \citep{rung2009challenges}. Analogue to the use of integration-by-parts in deriving the continuous adjoint equations, summation-by-parts is used to derive the building blocks of the discrete (dual) adjoint expressions. A detailed derivation of this hybrid adjoint approach can be found in  \citep{stuck2012adjoint, stuck2013adjoint, kroger2016numerical, kroger2018adjoint} for the single phase system. Implementation details about the multi-phase approximation are given in the upcoming section.  The segregated algorithm uses a cell-centered, collocated storage arrangement for all transport properties. The implicit numerical approximation is second order accurate and supports polyhedral cells. Both, the primal and adjoint pressure-velocity coupling is based on a SIMPLE method and possible parallelization is realized by means of a domain decomposition approach \citep{yakubov2013hybrid, yakubov2015experience}. The parallel approach supports local mesh refinements, over-set techniques \citep{volkner2017analysis}, or fluid-structure interactions to simulate rigid, mechanically coupled, floating bodies \citep{luo2017computation, luo2019numerical}. In terms of a CAD-free optimisation approach, the computational grid can be adjusted using a Laplace-Beltrami \citep{stuck2012adjoint,kroger2015cad} or Steklov-Poincar\'e \citep{schulz2016computational} type (surface metric) approach.

\subsection{Adjoint to Compressive Convection Schemes}
In the absence of geometrical reconstruction techniques, one crucial part within a VoF procedure refers to the compressive approximation of the convective mixture fraction transport to keep the discrete interface sharp. A classical starting point for the derivation of prominent interface capturing schemes is the Normalised-Variable-Diagram (NVD), first proposed by Leonard \citep{leonard1991ultimate}.
In line with Fig. \ref{fig:1d_flow_schematic}, the general challenge reads: How to interpolate the (on cell level) discrete non-dimensional field value (e.g. $c^{ \mathrm{U}} = 1$ and $c^{ \mathrm{D}} = 0$) on the face F in between, so that during the numerical simulation (a) the transported field quantity is bounded and (b) the interface remains sharp. While (a) results in the Convective Boundedness Criterion (CBC) (e.g. $c^{ \mathrm{U}} \ge c^{ \mathrm{F}} \ge c^{ \mathrm{D}}$), (b) is a matter of the  diffusive/compressive character of the underlying numerical approximation.
Thus, we introduce a normalization based on all relevant cell quantities around F, viz.
\begin{alignat}{2}
c_\mathrm{n} &= \frac{c^{} - c^{ \mathrm{UU}}}{c^{ \mathrm{D}} - c^{ \mathrm{UU}}} \qquad &&\rightarrow \qquad c_\mathrm{n}^\mathrm{D} = 1, \qquad c_\mathrm{n}^\mathrm{UU} = 0 \qquad \mathrm{and} \qquad c_\mathrm{n}^\mathrm{D} \ge c_\mathrm{n}^\mathrm{U} \ge c_\mathrm{n}^\mathrm{UU} \label{equ:primal_normalization} \\
\hat{c}_\mathrm{n} &= \frac{\hat{c}^{*} - \hat{c}^{* \mathrm{DD}}}{\hat{c}^{* \mathrm{U}} - \hat{c}^{* \mathrm{DD}}} \qquad &&\rightarrow \qquad \hat{c}_\mathrm{n}^\mathrm{U} = 1, \qquad \hat{c}_\mathrm{n}^\mathrm{DD} = 0 \qquad \mathrm{and} \qquad \hat{c}_\mathrm{n}^\mathrm{U} \ge \hat{c}_\mathrm{n}^\mathrm{D} \ge \hat{c}_\mathrm{n}^\mathrm{DD}\label{equ:adjoint_normalization}
\end{alignat}
where we anticipate a mirrored normalisation of the adjoint face value $\hat{c}_\mathrm{n}$ around the face F.
The general dependence of $c^{ \mathrm{F}} (c^{ \mathrm{UU}},c^{ \mathrm{U}},c^{ \mathrm{D}})$ [$\hat{c}^{ \mathrm{F}} (\hat{c}^{ \mathrm{DD}},\hat{c}^{ \mathrm{D}},\hat{c}^{ \mathrm{U}})$] reduce to $c_\mathrm{n}^\mathrm{F} (c_\mathrm{n}^\mathrm{U})$ [$\hat{c}_\mathrm{n}^\mathrm{F} (\hat{c}_\mathrm{n}^\mathrm{D})$] in the normalized setting for the primal [adjoint] variable. Explicit implementations inhere an additional dependence on the face Courant number $\mathrm{Co}^\mathrm{F}$ \citep{leonard1991ultimate}. The relation is depicted in Fig. \ref{fig:nvd_primal_dual} for  three bounded baseline schemes, i.e. Upwind (UD), Central (CD) and Downwind (DD) Differencing Scheme. The regime between CD and UD (DD) is prone to be diffusive (compressive).  Thus, modern interface capturing schemes try to stay as long as possible in the compressive regime and fall back to UD when the  discrete local setting is insufficient. Two prominent examples for such compressive approximations refer to (a) the High Resolution Interface Capturing Scheme (HRIC) of \citep{muzaferija1998computation} and (b) the Compressive Interface Capturing Scheme for Arbitrary Meshes (CICSAM) of \citep{ubbink1999method} which are investigated in the upcoming lines.
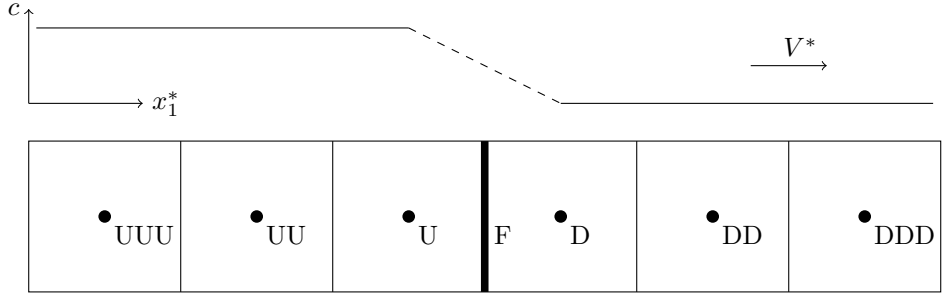
\begin{figure}
\centering
\begin{tikzpicture}
\draw (0,0) rectangle (12,2);
\draw (2,0) -- (2,2);
\draw (4,0) -- (4,2);
\draw[line width = 1mm] (6,0) -- (6,2);
\draw (8,0) -- (8,2);
\draw (10,0) -- (10,2);
\node[] at (1,1) [circle,fill,scale=0.5]{};
\node[anchor=north west] at (1,1) {UUU};
\node[] at (3,1) [circle,fill,scale=0.5]{};
\node[anchor=north west] at (3,1) {UU};
\node[] at (5,1) [circle,fill,scale=0.5]{};
\node[anchor=north west] at (5,1) {U};
\node[] at (7,1) [circle,fill,scale=0.5]{};
\node[anchor=north west] at (7,1) {D};
\node[] at (9,1) [circle,fill,scale=0.5]{};
\node[anchor=north west] at (9,1) {DD};
\node[] at (11,1) [circle,fill,scale=0.5]{};
\node[anchor=north west] at (11,1) {DDD};
\node[anchor=north west] at (6,1) {F};

\draw (0.1,3.5) -- (5,3.5);
\draw (7,2.5) -- (11.9,2.5);
\draw[dashed] (5,3.5) -- (7,2.5);
\draw[->] (9.5,3) -- (10.5,3) node[anchor=south east] {$V^*$};
\draw[thin,->] (0,2.5) -- (1.5,2.5) node[anchor=west] {$x_\mathrm{1}^*$};
\draw[thin,->] (0,2.5) -- (0,3.75) node[anchor=east] {$c$};
%\draw[thin,->] (0,2.5) -- (0,3.75) node[anchor=east] {$\phi^*$,$\hat{\phi}^*$};
\end{tikzpicture}
\caption{Schematic one dimensional view of the interpolation of primal field quantities $\phi^*$ on the face F from the adjacent cells located upstream (U) or downstream (D) of F under the flow field $V^*$.}
\label{fig:1d_flow_schematic}
\end{figure}
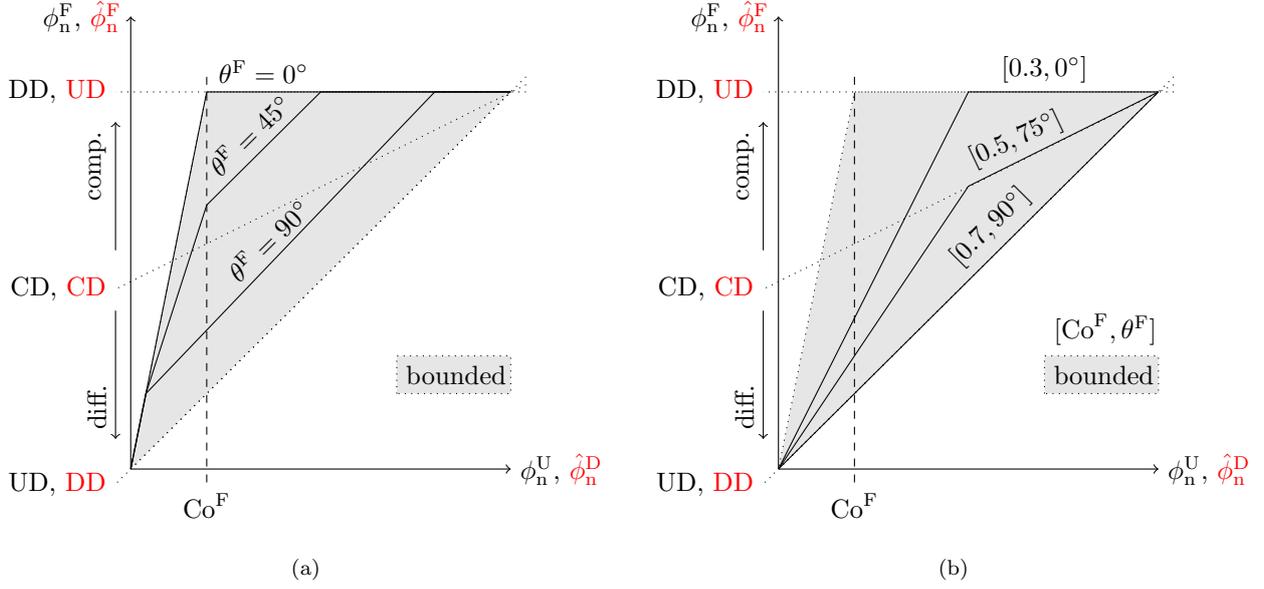
\begin{figure}
\centering
\subfigure[]{
\begin{tikzpicture}
\draw[thin,->] (0,0) -- (5,0) node[anchor=west] {\textcolor{black}{$\phi_\mathrm{n}^\mathrm{U}$}, \textcolor{red}{$\hat{\phi}_\mathrm{n}^\mathrm{D}$}};
\draw[thin,->] (0,0) -- (0,6) node[anchor=east] {\textcolor{black}{$\phi_\mathrm{n}^\mathrm{F}$}, \textcolor{red}{$\hat{\phi}_\mathrm{n}^\mathrm{F}$}};

\draw[dotted,fill=gray!20] (0,0) -- (5,5) -- (1,5) -- cycle;
\draw[dotted] (5.2,5.2) -- (-0.2,-0.2) node[anchor=east] {\textcolor{black}{UD}, \textcolor{red}{DD}};
\draw[dotted] (5.2,5) -- (-0.2,5) node[anchor=east] {\textcolor{black}{DD}, \textcolor{red}{UD}};
\draw[dotted] (5.2,5.095) -- (-0.2,2.385) node[anchor=east] {\textcolor{black}{CD}, \textcolor{red}{CD}};
\draw[dashed] (1.0,5.2) -- (1.0,-0.2) node[anchor=north] {$\mathrm{Co}^\mathrm{F}$};

%\draw[thin,->] (1.6666,3.3333) -- (1.6666-0.5,3.3333+0.8660) node[anchor=west] {\textcolor{black}{comp.}};
%\draw[thin,->] (1.6666,3.3333) -- (1.6666+0.5,3.3333-0.8660) node[anchor=east] {\textcolor{black}{diff.}};

\draw[] (0.0,0.0) -- (1.0,5.0) -- (5.0,5.0);
\node[anchor=south] at (1.75,5.0) {$\theta^\mathrm{F} = 0^\circ$};

\draw[] (0.0,0.0) -- (0.2,1.0) -- (1.0,3.5) -- (2.5,5.0) -- (5.0,5.0);
\node[anchor=south,rotate=45] at (1.75,4.25) {$\theta^\mathrm{F} = 45^\circ$};

\draw[] (0.0,0.0) -- (0.2,1.0) -- (4.0,5.0) -- (5.0,5.0);
\node[anchor=south,rotate=45] at (2.0,2.85) {$\theta^\mathrm{F} = 90^\circ$};

\draw[thin,->] (-0.2,2.1) -- (-0.2,0.4) node[rotate=90, anchor=south west] {\textcolor{black}{diff.}};
\draw[thin,->] (-0.2,2.9) -- (-0.2,4.6) node[rotate=90, anchor=south east] {\textcolor{black}{comp.}};

%\draw[dotted] (2.5,3.75) -- (2.5,0.00) node[anchor=north] {1/2};
%\draw[dotted] (2.5,3.75) -- (0.0,3.75) node[anchor=east] {3/4};

\draw[dotted,fill=gray!20] (3.5,1) -- (3.5,1.5) -- (5,1.5) -- (5,1) -- cycle node[anchor=north] {};
\node[anchor=west] at (3.5,1.25) {bounded};
%\node[anchor=west] at (3.5,1.85) {$[\mathrm{Co}^\mathrm{F},\theta]$};

\end{tikzpicture}
}
%\hspace{2cm}
\subfigure[]{
\begin{tikzpicture}
\draw[thin,->] (0,0) -- (5,0) node[anchor=west] {\textcolor{black}{$\phi_\mathrm{n}^\mathrm{U}$}, \textcolor{red}{$\hat{\phi}_\mathrm{n}^\mathrm{D}$}};
\draw[thin,->] (0,0) -- (0,6) node[anchor=east] {\textcolor{black}{$\phi_\mathrm{n}^\mathrm{F}$}, \textcolor{red}{$\hat{\phi}_\mathrm{n}^\mathrm{F}$}};

\draw[dotted,fill=gray!20] (0,0) -- (5,5) -- (1,5) -- cycle;
\draw[dotted] (5.2,5.2) -- (-0.2,-0.2) node[anchor=east] {\textcolor{black}{UD}, \textcolor{red}{DD}};
\draw[dotted] (5.2,5) -- (-0.2,5) node[anchor=east] {\textcolor{black}{DD}, \textcolor{red}{UD}};
\draw[dotted] (5.2,5.095) -- (-0.2,2.385) node[anchor=east] {\textcolor{black}{CD}, \textcolor{red}{CD}};
\draw[dashed] (1.0,5.2) -- (1.0,-0.2) node[anchor=north] {$\mathrm{Co}^\mathrm{F}$};

\draw[] (0.0,0.0) -- (2.5,5.0) -- (5.0,5.0);
\node[anchor=south] at (3.5,5.0) {$[0.3,0^\circ]$};

\draw[] (0.0,0.0) -- (2.5,3.75) -- (5.0,5.0);
\node[anchor=south,rotate=25] at (3.25,4.1) {$[0.5,75^\circ]$};

\draw[] (0.0,0.0) -- (2.5,2.5) -- (5.0,5.0);
\node[anchor=south,rotate=45] at (3.0,3.0) {$[0.7,90^\circ]$};

%\draw[thin,->] (1.6666,3.3333) -- (1.6666-0.5,3.3333+0.8660) node[anchor=west] {\textcolor{black}{comp.}};
%\draw[thin,->] (1.6666,3.3333) -- (1.6666+0.5,3.3333-0.8660) node[anchor=east] {\textcolor{black}{diff.}};

\draw[thin,->] (-0.2,2.1) -- (-0.2,0.4) node[rotate=90, anchor=south west] {\textcolor{black}{diff.}};
\draw[thin,->] (-0.2,2.9) -- (-0.2,4.6) node[rotate=90, anchor=south east] {\textcolor{black}{comp.}};

%\draw[dotted] (2.5,3.75) -- (2.5,0.00) node[anchor=north] {1/2};
%\draw[dotted] (2.5,3.75) -- (0.0,3.75) node[anchor=east] {3/4};

\draw[dotted,fill=gray!20] (3.5,1) -- (3.5,1.5) -- (5,1.5) -- (5,1) -- cycle node[anchor=north] {};
\node[anchor=west] at (3.5,1.25) {bounded};
\node[anchor=west] at (3.5,1.85) {$[\mathrm{Co}^\mathrm{F},\theta^\mathrm{F}]$};

\end{tikzpicture}
}
\caption{Sketch of the Normalized Variable Diagram (NVD) for two compressive convection schemes: a) Compressive Interface Capturing Scheme for Arbitrary Meshes (CICSAM) and b) High Resolution Interface Capturing Scheme (HRIC) as well as their adjoint analogue (red).}
\label{fig:nvd_primal_dual}
\end{figure}
Since both approaches are discrete by nature, the adjoint derivation and representation is also performed on discrete level. The discussion originates in the steady discrete analogue to the continuous Lagrangian (last term in Eqn. (\ref{equ:free_surf_rele})) for the transported property $c$, viz.
\begin{align}
\tilde{L}^* = \int_{\mathrm{\Omega^*}} \hat{c}^* v_\mathrm{k}^* \frac{\partial c}{\partial x_\mathrm{k}^*} \mathrm{d} \Omega^*
\qquad \overset{discretize}{\rightarrow}  \qquad 
\tilde{L}^* = \sum_\mathrm{P} \hat{c}^\mathrm{*P} \int_{\mathrm{\Omega^*(\mathrm{P})}} v_\mathrm{k}^* \frac{\partial c}{\partial x_\mathrm{k}^*} \mathrm{d} \Omega^* = \sum_\mathrm{P} \hat{c}^\mathrm{*P} \sum_\mathrm{F(P)} \left( \dot{V}^* c \right)^\mathrm{F}. \label{equ:discrete_adjoint}
\end{align}
The latter expression employs the continuity equation and a second order accurate midpoint rule to approximate integrals.
Contrary to the continuous relationship of Lagrangian multiplier and convective term, the dual variable is now multiplied with the discrete representation of the convective operator, expressed by discrete face-based volume fluxes $\dot{V}^* = v_\mathrm{k}^* n_\mathrm{k} \Delta \Gamma^*$ in a Finite-Volume framework. Thus, the investigated approaches differ in the approximation of $c^\mathrm{F}$ only.

\paragraph{HRIC}
The normalized face value $c_\mathrm{n}^\mathrm{F}$ 
%within the HRIC procedure 
is basically determined in three consecutive steps. In compact notation they read:
\begin{align}
c_\mathrm{n}^\mathrm{F} = 
\begin{cases}
c_\mathrm{n}^\mathrm{F,2} &: \mathrm{Co}^\mathrm{F} < \mathrm{Co}_\mathrm{l}^\mathrm{F} \\
c_\mathrm{n}^\mathrm{U} + \left(c_\mathrm{n}^\mathrm{F,2} - c_\mathrm{n}^\mathrm{U} \right) \frac{\mathrm{Co}_\mathrm{u}^\mathrm{F} -  \mathrm{Co}^\mathrm{F}}{\mathrm{Co}_\mathrm{u}^\mathrm{F} - \mathrm{Co}_\mathrm{l}^\mathrm{F}} &: \mathrm{Co}_\mathrm{l}^\mathrm{F} \le \mathrm{Co}^\mathrm{F} \le \mathrm{Co}_\mathrm{u}^\mathrm{F} \\
c_\mathrm{n}^\mathrm{U} &: \mathrm{Co}^\mathrm{F} > \mathrm{Co}_\mathrm{u}^\mathrm{F}
\end{cases},
\ \ \
c_\mathrm{n}^\mathrm{F,2} = \gamma^\mathrm{F} c_\mathrm{n}^\mathrm{F,3} + \left(1 - \gamma^\mathrm{F} \right) c_\mathrm{n}^\mathrm{U},
\ \ \
c_\mathrm{n}^\mathrm{F,3} = 
\begin{cases}
2 c_\mathrm{n}^\mathrm{U} &: c_\mathrm{n}^\mathrm{U} \in \left[0,0.5 \right] \\
1 &: c_\mathrm{n}^\mathrm{U} \in \left[0.5,1 \right] \\
c_\mathrm{n}^\mathrm{U} &: c_\mathrm{n}^\mathrm{U} \not\in \left[0,1 \right]
\end{cases} \label{equ:hric_principal}
\end{align}
where three additional quantities enter the approximation: Two ensure a stable coefficient matrix while preserving a positive main diagonal via the definition of an upper (lower) Courant number $\mathrm{Co}_\mathrm{u}^\mathrm{F}$ ($\mathrm{Co}_\mathrm{l}^\mathrm{F}$). Frequently employed values are $\mathrm{Co}_\mathrm{l}^\mathrm{F} = 0.3$ and $\mathrm{Co}_\mathrm{u}^\mathrm{F} = 0.7$. The third guarantees the flow to free surface alignment $\gamma^\mathrm{F} = \sqrt{|\mathrm{cos}(\theta^\mathrm{F})|}$ where $\theta^\mathrm{F}$ represents the angle between the free surface normal and the flow direction. Mind that that also slightly different HRIC formulations exist. e.g. \citep{park2009volume} and we choose this one exemplary only.

The continuous adjoint system is derived for quasi steady situations over an integration within a pseudo-time. Thus, after the primal integration until convergence, the discrete Courant number distribution as well as concentration distribution is known. Based on Eqn. (\ref{equ:hric_principal}) we can clearly decide how the value of each face in Eqn. (\ref{equ:discrete_adjoint}) is approximated. Algorithmically we need to decide between three situation: 1) pure UD, 2) pure DD as well as 3) an interpolation between both based on $\mathrm{Co}^\mathrm{F}$.

Therefore, the derivation of the adjoint HRIC scheme needs to be done for all scenarios in the last case distinction in Eqn. (\ref{equ:hric_principal}), viz. $c_\mathrm{n}^\mathrm{F,3} = 2 c_\mathrm{n}^\mathrm{U}$, $c_\mathrm{n}^\mathrm{F,3} = 1$ as well as $c_\mathrm{n}^\mathrm{F,3} = c_\mathrm{n}^\mathrm{U}$.
Exemplary, we derive the adjoint to $c_\mathrm{n}^\mathrm{F,3} = 2 c_\mathrm{n}^\mathrm{U}$. Both other scenarios follow the same approach. Based on Eqn. (\ref{equ:primal_normalization}), the expression can be cast to $ c^\mathrm{F,3} = 2 c^\mathrm{U} - c^\mathrm{UU}$ and the Lagrangian Eqn. (\ref{equ:discrete_adjoint}) for the control volumes UU, U, D and DD (cf. Fig. \ref{fig:nvd_primal_dual}) reads:
\begin{alignat}{10}
\tilde{L}^* = ... 
&+ \hat{c}^\mathrm{*UU} &&\dot{V}^* \big[\big(2 c^{UU} &&- c^\mathrm{UUU} &&\big) &&-  \big( 2c^{UUU} &&- c^\mathrm{UUUU} &&\big) \big] \nonumber \\
&+ \hat{c}^\mathrm{*U}  &&\dot{V}^* \big[\big(2 c^{U}  &&- c^\mathrm{UU}  &&\big) &&-  \big( 2c^{UU}  &&- c^\mathrm{UUU}  &&\big) \big] \\
&+ \hat{c}^\mathrm{*D}  &&\dot{V}^* \big[\big(2 c^{D}  &&- c^\mathrm{U}   &&\big) &&-  \big( 2c^{U}   &&- c^\mathrm{UU}   &&\big) \big] \nonumber \\
&+ \hat{c}^\mathrm{*DD} &&\dot{V}^* \big[\big(2 c^{DD} &&- c^\mathrm{D}   &&\big) &&-  \big( 2c^{D}   &&- c^\mathrm{U}    &&\big) \big] + ... . \nonumber
\end{alignat}
A variation of $\tilde{L}^*$ and a subsequent isolation of all variation terms resembles the summation by parts characteristics. Similar to the continuous derivation, first order optimality conditions finally yield the adjoint face value, viz.
\begin{alignat}{10}
%\delta_\mathrm{c} \tilde{L}^* \cdot \delta c^{(\cdot)} = ... 
%&+ \hat{c}^\mathrm{*UU} &&\dot{V}^* \big[\big(2  \delta c^{UU} &&-  \delta c^\mathrm{UUU} &&\big) &&- \big( 2 \delta c^{UUU} &&-  \delta c^\mathrm{UUUU} &&\big) \big] \nonumber \\
%&+ \hat{c}^\mathrm{*U}  &&\dot{V}^* \big[\big(2  \delta c^{U}  &&-  \delta c^\mathrm{UU}  &&\big) &&- \big( 2 \delta c^{UU} &&-  \delta c^\mathrm{UUU} &&\big) \big] \nonumber \\
%&+ \hat{c}^\mathrm{*D}  &&\dot{V}^* \big[\left(2  \delta c^{D}  &&-  \delta c^\mathrm{U}   &&\big) &&- \big( 2 \delta c^{U} &&-  \delta c^\mathrm{UU} &&\big) \big] \nonumber \\
%&+ \hat{c}^\mathrm{*DD} &&\dot{V}^* \big[\big(2  \delta c^{DD} &&-  \delta c^\mathrm{D}   &&\big) &&- \big( 2 \delta c^{D} &&-  \delta c^\mathrm{U} &&\big) \big] + ...  \nonumber \\
%\Leftrightarrow \qquad
\delta_\mathrm{c} \tilde{L}^* \cdot \delta c^{(\cdot)} = ... 
&- \delta c^{UU} &&\dot{V}^* \big[\big( 2 \hat{c}^{*U}   &&-  \hat{c}^\mathrm{*D}    &&\big) &&- \big( 2 \hat{c}^{*UU} &&-  \hat{c}^\mathrm{*U}   &&\big) \big] \nonumber \\
&- \delta c^{U}  &&\dot{V}^* \big[\big( 2 \hat{c}^{*D}   &&-  \hat{c}^\mathrm{*DD}   &&\big) &&- \big( 2 \hat{c}^{*U}  &&-  \hat{c}^\mathrm{*D}   &&\big) \big] \label{equ:isolated_hric} \\
&- \delta c^{D}  &&\dot{V}^* \big[\big( 2 \hat{c}^{*DD}  &&-  \hat{c}^\mathrm{*DDD}  &&\big) &&- \big( 2 \hat{c}^{*D}  &&-  \hat{c}^\mathrm{*DD}  &&\big) \big] \nonumber \\
&- \delta c^{DD} &&\dot{V}^* \big[\big( 2 \hat{c}^{*DDD} &&-  \hat{c}^\mathrm{*DDDD} &&\big) &&- \big( 2 \hat{c}^{*DD} &&-  \hat{c}^\mathrm{*DDD} &&\big) \big] + ... \qquad \overset{!}{=} 0 \qquad \forall \, \delta c^{(\cdot)} \nonumber
\end{alignat}
Apparently, the adjoint face value interpolation results in mirroring the primal stencil \citep{stuck2013adjoint}. For the unknown face value within the adjoint HRIC scheme in Fig. \ref{fig:nvd_primal_dual} (first inner bracket in Eqn. (\ref{equ:isolated_hric})), we end up with $\hat{c}^\mathrm{*F} = 2 \hat{c}^\mathrm{*D} - \hat{c}^\mathrm{*DD}$ or $\hat{c}_\mathrm{n}^\mathrm{F} = 2 \hat{c}_\mathrm{n}^\mathrm{D}$ after applying the adjoint normalization from Eqn. (\ref{equ:adjoint_normalization}). 
Finally, the adjoint HRIC scheme to Eqn. (\ref{equ:hric_principal}) reads:
\begin{align}
\hat{c}_\mathrm{n}^\mathrm{F} = 
\begin{cases}
\hat{c}_\mathrm{n}^\mathrm{F,2} &: \mathrm{Co}^\mathrm{F} < \mathrm{Co}_\mathrm{l}^\mathrm{F} \\
\hat{c}_\mathrm{n}^\mathrm{D} + \left(\hat{c}_\mathrm{n}^\mathrm{F,2} - \hat{c}_\mathrm{n}^\mathrm{D} \right) \frac{\mathrm{Co}_\mathrm{u}^\mathrm{F} -  \mathrm{Co}^\mathrm{F}}{\mathrm{Co}_\mathrm{u}^\mathrm{F} - \mathrm{Co}_\mathrm{l}^\mathrm{F}} &: \mathrm{Co}_\mathrm{l}^\mathrm{F} \le \mathrm{Co}^\mathrm{F} \le \mathrm{Co}_\mathrm{u}^\mathrm{F} \\
\hat{c}_\mathrm{n}^\mathrm{D} &: \mathrm{Co}^\mathrm{F} > \mathrm{Co}_\mathrm{u}^\mathrm{F}
\end{cases},
\ \ \
\hat{c}_\mathrm{n}^\mathrm{F,2} = \gamma^\mathrm{F} \hat{c}_\mathrm{n}^\mathrm{F,3} + \left(1 - \gamma^\mathrm{F} \right) \hat{c}_\mathrm{n}^\mathrm{D},
\ \ \
\hat{c}_\mathrm{n}^\mathrm{F,3} = 
\begin{cases}
2 \hat{c}_\mathrm{n}^\mathrm{D} &: \hat{c}_\mathrm{n}^\mathrm{D} \in \left[0,0.5 \right] \\
1 &: \hat{c}_\mathrm{n}^\mathrm{D} \in \left[0.5,1 \right] \\
\hat{c}_\mathrm{n}^\mathrm{D} &: \hat{c}_\mathrm{n}^\mathrm{D} \not\in \left[0,1 \right]
\end{cases} \label{equ:adjoint_hric_principal}
\end{align}
where all upwind approximations are switched to downwind and vice versa.

\paragraph{CICSAM}
The normalized face value $c_\mathrm{n}^\mathrm{F}$ within the CICSAM procedure is determined via a blending between the Hyper-C (HC) und the ULTIMATE-QUICKEST (UQ) scheme. The latter is inspired by a QUICK (Quadratic Interpolation of Convective Kinematics) approach for the face value. A compact notation reads:
\begin{align}
c_\mathrm{n}^\mathrm{F} &= \gamma^\mathrm{F} c_\mathrm{n}^\mathrm{F, HC} + (1 - \gamma^\mathrm{F})c_\mathrm{n}^\mathrm{F, UQ} \qquad \mathrm{with} \label{equ:cicsam_principal} \\
c_\mathrm{n}^\mathrm{F, HC} &= \begin{cases}
\mathrm{min} \left(1, \frac{c_\mathrm{n}^\mathrm{U}}{\mathrm{Co}^\mathrm{F}} \right) &: 0 \le c_\mathrm{n}^\mathrm{U} \le 1 \nonumber \\
c_\mathrm{n}^\mathrm{U} &: c_\mathrm{n}^\mathrm{U} \not\in \left[0,1 \right]
\end{cases}
\ \  \mathrm{and} \ \
c_\mathrm{n}^\mathrm{F, UQ} = \begin{cases}
\mathrm{min} \left(c_\mathrm{n}^\mathrm{F, HC}, \frac{8 \mathrm{Co}^\mathrm{F} c_\mathrm{n}^\mathrm{U} + \left(1 - \mathrm{Co}^\mathrm{F}\right) \left(6 c_\mathrm{n}^\mathrm{U} + 3 \right)}{8} \right) &: 0 \le c_\mathrm{n}^\mathrm{U} \le 1\\
c_\mathrm{n}^\mathrm{U} &: c_\mathrm{n}^\mathrm{U} \not\in \left[0,1 \right]
\end{cases}. 
\end{align}
Again, a blending factor is introduced that accounts for the free surface to flow alignment, viz. $\gamma^\mathrm{F} = \mathrm{min} ((1 + \mathrm{cos}(2\, \theta^\mathrm{F}))/2,1)$ where $\theta^\mathrm{F}$ represents the relation between interface normal and the vector connecting adjacent cells \citep{ubbink1999method}. Again, we would like to point out that this is only one of many formulations.
Whether HC or UQ is used, CICSAM falls back to pure UD for inadequate situations, which in turn switches to DD in adjoint mode.
If HC is active, its first case switches between DD ($c_\mathrm{n}^\mathrm{F, HC} = 1$) and UD ($c_\mathrm{n}^\mathrm{F, HC} = c_\mathrm{n}^\mathrm{U} / \mathrm{Co}^\mathrm{F}$). Thus, the adjoint to HC reads ($\hat{c}_\mathrm{n}^\mathrm{F, HC} = 1$) or ($\hat{c}_\mathrm{n}^\mathrm{F, HC} = \hat{c}_\mathrm{n}^\mathrm{D} / \mathrm{Co}^\mathrm{F}$).

The delicate term is the first QUICK like case within the UQ scenario. A denormalization offers $c^\mathrm{F} =  [\mathrm{Co}^\mathrm{F} + 3] \, c^\mathrm{U}/4 + [\mathrm{Co}^\mathrm{F} - 1] \, c^\mathrm{UU}/8  + [1 -\mathrm{Co}^\mathrm{F}] \, 3 c^\mathrm{D}/8$ or $c^\mathrm{F} = 3/8 \, c^\mathrm{D} + 3/4 \, c^\mathrm{U} - 1/8 \, c^\mathrm{UU}$ for $\mathrm{Co}^\mathrm{F} \rightarrow 0$ and $c^\mathrm{F} = c^\mathrm{U} $ for $\mathrm{Co}^\mathrm{F} \rightarrow 1$. Again, the discrete Lagrangian (\ref{equ:discrete_adjoint}) is constructed for the control volumes UU, U, D and DD (cf. Fig. \ref{fig:nvd_primal_dual}):
\begin{alignat}{10}
\tilde{L}^* = ... 
&+ \hat{c}^\mathrm{*UU} &&\dot{V}^* \big[ \big(r \, c^\mathrm{UU} &&+ s \, c^\mathrm{UUU}  &&+ t \, c^\mathrm{U}&&\big) && -  \big( r \, c^\mathrm{UUU} &&+ s \, c^\mathrm{UUUU}  &&+ t \,  c^\mathrm{UU} &&\big) \big] \nonumber \\
&+ \hat{c}^\mathrm{*U} &&\dot{V}^* \big[ \big(r \, c^\mathrm{U} &&+ s \, c^\mathrm{UU}  &&+ t \, c^\mathrm{D}&&\big) && -  \big( r \, c^\mathrm{UU} &&+ s \, c^\mathrm{UUU}  &&+ t \, c^\mathrm{U} &&\big) \big] \\
&+ \hat{c}^\mathrm{*D} &&\dot{V}^* \big[ \big(r \, c^\mathrm{D} &&+ s \, c^\mathrm{U}  &&+ t \, c^\mathrm{DD}&&\big) && -  \big( r \, c^\mathrm{U} &&+ s \, c^\mathrm{UU}  &&+ t \, c^\mathrm{D} &&\big) \big] \nonumber  \\
&+ \hat{c}^\mathrm{*DD} &&\dot{V}^* \big[ \big(r \, c^\mathrm{DD} &&+ s \, c^\mathrm{D}  &&+ t \, c^\mathrm{DDD}&&\big) && -  \big( r \, c^\mathrm{D} &&+ s \, c^\mathrm{U}  &&+ t \, c^\mathrm{DD} &&\big) \big] + ... .  \nonumber 
\end{alignat}
where the coefficients $r = [\mathrm{Co}^\mathrm{F} + 3]/4$, $s = [\mathrm{Co}^\mathrm{F} - 1]/8$ and $t = 3[1 -\mathrm{Co}^\mathrm{F}]/8$ are defined to shorten the notation. 
%For the sake of simplicity we assume a constant face Courant number. 
In line with the adjoint HRIC derivation, a variation of $\tilde{L}^*$ and a subsequent isolation of all variations yields:
\begin{alignat}{10}
%\delta_\mathrm{c^*} \tilde{L}^* \cdot \delta c^{(\cdot)} = ... 
%&+ \hat{c}^\mathrm{*UU} &&\dot{V}^* \big[ \big(r \, \delta c^\mathrm{UU} &&+ s \, \delta c^\mathrm{UUU}  &&+ t \, \delta c^\mathrm{U}&&\big) && -  \big( r \, \delta c^\mathrm{UUU} &&+ s \, \delta c^\mathrm{UUUU}  &&+ t \,  \delta c^\mathrm{UU} &&\big) \big] \\
%&+ \hat{c}^\mathrm{*U} &&\dot{V}^* \big[ \big(r \, \delta c^\mathrm{U} &&+ s \, \delta c^\mathrm{UU}  &&+ t \, \delta c^\mathrm{D}&&\big) && -  \big( r \, \delta c^\mathrm{UU} &&+ s \, \delta c^\mathrm{UUU}  &&+ t \, \delta c^\mathrm{U} &&\big) \big] \\
%&+ \hat{c}^\mathrm{*D} &&\dot{V}^* \big[ \big(r \, \delta c^\mathrm{D} &&+ s \, \delta c^\mathrm{U}  &&+ t \, \delta c^\mathrm{DD}&&\big) && -  \big( r \, \delta c^\mathrm{U} &&+ s \, \delta c^\mathrm{UU}  &&+ t \, \delta c^\mathrm{D} &&\big) \big] \\
%&+ \hat{c}^\mathrm{*DD} &&\dot{V}^* \big[ \big(r \, \delta c^\mathrm{DD} &&+ s \, \delta c^\mathrm{D}  &&+ t \, \delta c^\mathrm{DDD}&&\big) && -  \big( r \, \delta c^\mathrm{D} &&+ s \, \delta c^\mathrm{U}  &&+ t \, \delta c^\mathrm{DD} &&\big) \big] + ... . \\
%\Leftrightarrow \qquad
\delta_\mathrm{c^*} \tilde{L}^* \cdot \delta c^{(\cdot)} = ... 
&- \delta c^\mathrm{UU} &&\dot{V}^* \big[ \big(r \, \hat{c}^\mathrm{*U} &&+ s \, \hat{c}^\mathrm{*D}  &&+ t \, \hat{c}^\mathrm{*UU}&&\big) && -  \big( r \, \hat{c}^\mathrm{*UU} &&+ s \, \hat{c}^\mathrm{*U}  &&+ t \,  \hat{c}^\mathrm{*UUU} &&\big) \big]  \nonumber \\
&- \delta c^\mathrm{U} &&\dot{V}^* \big[ \big(r \, \hat{c}^\mathrm{*D} &&+ s \, \hat{c}^\mathrm{*DD}  &&+ t \, \hat{c}^\mathrm{*U}&&\big) && -  \big( r \, \hat{c}^\mathrm{*U} &&+ s \, \hat{c}^\mathrm{*D}  &&+ t \,  \hat{c}^\mathrm{*UU} &&\big) \big] \label{equ:isolated_cicsam}  \\
&- \delta c^\mathrm{D} &&\dot{V}^* \big[ \big(r \, \hat{c}^\mathrm{*DD} &&+ s \, \hat{c}^\mathrm{*DDD}  &&+ t \, \hat{c}^\mathrm{*D}&&\big) && -  \big( r \, \hat{c}^\mathrm{*D} &&+ s \, \hat{c}^\mathrm{*DD}  &&+ t \,  \hat{c}^\mathrm{*U} &&\big) \big]  \nonumber \\
&- \delta c^\mathrm{DD} &&\dot{V}^* \big[ \big(r \, \hat{c}^\mathrm{*DDD} &&+ s \, \hat{c}^\mathrm{*DDDD}  &&+ t \, \hat{c}^\mathrm{*DD}&&\big) && -  \big( r \, \hat{c}^\mathrm{*DD} &&+ s \, \hat{c}^\mathrm{*DDD}  &&+ t \,  \hat{c}^\mathrm{*D} &&\big) \big] + ... \qquad \overset{!}{=} 0 \qquad \forall \, \delta c^{(\cdot)}. \nonumber 
\end{alignat}
For the unknown face value in Fig. \ref{fig:nvd_primal_dual} within the adjoint CICSAM scheme (first inner bracket in Eqn. (\ref{equ:isolated_cicsam})), we end up with $\hat{c}^\mathrm{*F} = r \, \hat{c}^\mathrm{*D} + s \, \hat{c}^\mathrm{*DD}  + t \, \hat{c}^\mathrm{*U}$ or $\hat{c}_\mathrm{n}^\mathrm{F} = 8 \mathrm{Co}^\mathrm{F} \hat{c}_\mathrm{n}^\mathrm{D} + (1 - \mathrm{Co}^\mathrm{F}) (6 \hat{c}_\mathrm{n}^\mathrm{D} + 3 )$ after applying the adjoint normalization from Eqn. (\ref{equ:adjoint_normalization}) where we used $r+s+t = 1$.
The adjoint CICSAM scheme to Eqn. (\ref{equ:cicsam_principal}) reads:
\begin{align}
\hat{c}_\mathrm{n}^\mathrm{F} &= \gamma^\mathrm{F} \hat{c}_\mathrm{n}^\mathrm{F, HC} + (1 - \gamma^\mathrm{F})\hat{c}_\mathrm{n}^\mathrm{F, UQ} \qquad \mathrm{with} \\
\hat{c}_\mathrm{n}^\mathrm{F, HC} &= \begin{cases}
\mathrm{min} \left(1, \frac{\hat{c}_\mathrm{n}^\mathrm{D}}{\mathrm{Co}^\mathrm{F}} \right) &: 0 \le \hat{c}_\mathrm{n}^\mathrm{D} \le 1 \nonumber \\
\hat{c}_\mathrm{n}^\mathrm{D} &: \hat{c}_\mathrm{n}^\mathrm{D} \not\in \left[0,1 \right]
\end{cases}
\ \  \mathrm{and} \ \
\hat{c}_\mathrm{n}^\mathrm{F, UQ} = \begin{cases}
\mathrm{min} \left(\hat{c}_\mathrm{n}^\mathrm{F, HC}, \frac{8 \mathrm{Co}^\mathrm{F} \hat{c}_\mathrm{n}^\mathrm{D} + \left(1 - \mathrm{Co}^\mathrm{F}\right) \left(6 \hat{c}_\mathrm{n}^\mathrm{D} + 3 \right)}{8} \right) &: 0 \le \hat{c}_\mathrm{n}^\mathrm{D} \le 1\\
\hat{c}_\mathrm{n}^\mathrm{D} &: \hat{c}_\mathrm{n}^\mathrm{D} \not\in \left[0,1 \right]
\end{cases}.
\end{align}
Again, downwind and upwind are rigorously exchanged.

\paragraph{A Note on Additional Variational Contributions} Eqn. (\ref{equ:hric_principal}) as well as (\ref{equ:cicsam_principal}) inhere several implicit dependencies on the flow field, e.g. via the local Courant number or the free surface orientation. These NVD dependencies are kept frozen during the derivation of the adjoint HRIC / CICSAM scheme. A strategy also used for the adjoint to limited convection schemes, e.g. Total Variation Diminishing (TVD) schemes \citep{stuck2013adjoint}.

\subsubsection{Approximation of Adjoint Two-Phase Terms}
Terms within the adjoint system Eqns. (\ref{equ:adjoi_momen_dile}) - (\ref{equ:adjoi_conce_dile}) originating from the variable fluid properties are approximated in line with the concept of hybrid adjoint, e.g. swapping derivation and discretization (cf. \citep{stuck2013adjoint}). Hence, analogue to Eqn. (\ref{equ:discrete_adjoint}) the Lagrangian is investigated on discrete level.
The discussion refers to the symbolic finite-volume approximation of a variable $\hat{\phi}^*$ located in the center of a control volume P with size $\Delta \Omega^\mathrm{*P}$. Relations with adjacent control volumes NB yield one line within the discrete system, i.e. $A^\mathrm{*P} \hat{\phi}^\mathrm{*P} - \Sigma_\mathrm{NB} A^\mathrm{*NB} \hat{\phi}^\mathrm{*NB} = S^*_\mathrm{\hat{\phi}^*} \Delta \Omega^\mathrm{*P}$.
For the three source/sink terms on the left-hand side of the adjoint concentration Equ. \ref{equ:adjoi_conce_dile} we end up with:
\begin{alignat}{10}
\tilde{L}^* 
&= \int_{\mathrm{\Omega^*}} \hat{v}_\mathrm{i}^* \rho^* g_\mathrm{i}^* \mathrm{d} \Omega^*
\quad &&\overset{discretize}{\rightarrow}  \quad 
\tilde{L}^* 
= \sum_\mathrm{P} \left[ \hat{v}_\mathrm{i}^\mathrm{*} \rho^* g_\mathrm{i}^* \Delta \Omega^* \right]^\mathrm{P}
\quad &&\overset{derive}{\rightarrow} \quad 
\delta_\mathrm{c}\tilde{L}^* \cdot   \delta c = \sum_\mathrm{P} \delta c^\mathrm{P} \big[ \underbrace{ \rho_\mathrm{\Delta}^* \hat{v}_\mathrm{i}^\mathrm{*} g_\mathrm{i}^*}_{ S_\mathrm{\hat{c}^*}^*} \Delta \Omega^* \big]^\mathrm{P} \nonumber \\
\tilde{L}^* 
&= \int_{\mathrm{\Omega^*}} \hat{v}_\mathrm{i}^* \rho^* v_\mathrm{k}^* \frac{\partial v_\mathrm{i}^*}{\partial x_\mathrm{k}^*} \mathrm{d} \Omega^*
\quad &&\overset{discretize}{\rightarrow}  \quad 
\tilde{L}^* 
= \sum_\mathrm{P} \hat{v}_\mathrm{i}^\mathrm{*P}  \bigg[ \rho^*  v_\mathrm{k}^* \frac{\partial v_\mathrm{i}^*}{\partial x_\mathrm{k}^*} \Delta \Omega^* \bigg]^\mathrm{P}
\quad &&\overset{derive}{\rightarrow} \quad 
\delta_\mathrm{c} \tilde{L}^*  \cdot \delta c =  \sum_\mathrm{P} \delta c^\mathrm{P} \bigg[ \underbrace{ \rho_\mathrm{\Delta}^* \hat{v}_\mathrm{i}^\mathrm{*} v_\mathrm{k}^* \frac{\partial v_\mathrm{i}^*}{\partial x_\mathrm{k}	^*}}_{ S^*_\mathrm{\hat{c}^*}} \Delta \Omega^* \bigg]^\mathrm{P} \nonumber  \\
\tilde{L}^* 
&= \int_{\mathrm{\Omega^*}} \hat{v}_\mathrm{i}^* \frac{\partial 2 \mu^* S_\mathrm{ik}^*}{\partial x_\mathrm{k}^*} \mathrm{d} \Omega^*
\quad &&\overset{discretize}{\rightarrow}  \quad 
\tilde{L}^* 
= \sum_\mathrm{P} \hat{v}_\mathrm{i}^\mathrm{*P} \sum_\mathrm{F(P)} \big[ 2 \mu^* S_\mathrm{ik}^* \Delta \Gamma_\mathrm{k}^* \big]^\mathrm{F}
\quad &&\overset{derive}{\rightarrow} \quad 
\delta_\mathrm{c} \tilde{L}^*  \cdot \delta c =- \sum_\mathrm{P} \delta c^\mathrm{P} \underbrace{\sum_\mathrm{F(P)}2 \mu_\Delta^*  \bigg[  S_\mathrm{ik}^\mathrm{*} \frac{\partial  \hat{v}_\mathrm{i}^*}{\partial x_\mathrm{k}^*} \bigg]^\mathrm{F}}_{S^*_\mathrm{\hat{c}^*} \Delta \Omega^*} . \nonumber
\end{alignat}
The product between primal and adjoint concentration equation introduces the multi-phase information to the adjoint velocity field, which in turn enters the sensitivity derivative Eqn. (\ref{equ:shape_derivative}). A robust and consistent approximation of this term is crucial, since a neglect resembles a frozen VoF approach.
The interpolation of fluid data from the concentration distribution does not follow from a physical conservation equation but rather represents a basic modeling assumption. Thus, it is possible to shift the active region of the adjoint concentration within the adjoint momentum equation via a formal re-definition of $c_\mathrm{a} \rightarrow c_\mathrm{a} - \alpha$ and $c_\mathrm{b} \rightarrow c_\mathrm{b} + \alpha$. Reasonable values refer to $\alpha=0$ ($\alpha=1$) to activate the term in the foreground (background) phase a (b). To obtain a consistent dual formulation, the three source/sink terms on the right side of the adjoint concentration equation experience a sign change also, since $c_\Delta = c_\mathrm{a} - c_\mathrm{b} = 1 \rightarrow -1 = c_\mathrm{a} - c_\mathrm{b} - 2 \alpha = c_\Delta$ (cf. Eqn. (\ref{equ:delta_term})).
In line with the hybrid adjoint strategy we end up with the following approximation
\begin{align}
\tilde{L}^* 
&= \int_{\mathrm{\Omega^*}} \hat{c}^* v_\mathrm{k}^* \frac{\partial c}{\partial x_\mathrm{k}^*} \mathrm{d} \Omega^*
\quad &&\overset{discr.}{\rightarrow}  \quad 
\tilde{L}^* 
= \sum_\mathrm{P} \hat{c}^\mathrm{*P}  \bigg[v_\mathrm{k}^* \frac{\partial c}{\partial x_\mathrm{k}^*} \Delta \Omega^* \bigg]^\mathrm{P}
\quad &&\overset{deri.}{\rightarrow} \quad 
\delta_\mathrm{v_\mathrm{i}^*} \tilde{L}^*  \cdot \delta v_\mathrm{i}^* =  -\sum_\mathrm{P} \delta v_\mathrm{i}^\mathrm{*P} \bigg[ \underbrace{ \left(c - \alpha \right) \frac{\partial \hat{c}^\mathrm{*} }{\partial x_\mathrm{i}^*}}_{ S^*_\mathrm{\hat{v}_\mathrm{i}^*}} \Delta \Omega^* \bigg]^\mathrm{P} \nonumber. \label{equ:shift_active_region}
\end{align}

\paragraph{Interpretation of Primal vs. Dual Time Horizon} Originating from integration by parts, the adjoint time integration is inverted, e.g. declining with respect to the physical time. The latter is the reason why the complete primal solution has to be stored during the (forward) integration to be available to the adjoint solver for its subsequent (backward) integration. Several memory-reduced methods such as one-shot (piggy-bag) methods \citep{gherman2005preconditioning, kuruvila1994airfoil, ozkaya2009single} for pseudo time stepping simulations or check-pointing approaches \citep{giering1998recipes, griewank2000algorithm, hinze2005revolve, hinze2006optimal} were suggested. They reduce the memory requirements at the expense of enhanced (partially rerunning) computing efforts \citep{kapellos2019unsteady} 
%\nkuehl{War das so? nachlesen! Eventuell wurde nur im Frequenzbereich gerechnet?} \trung{Nein im Zeitbereich!!}. 
The formulation of the present adjoint problem is pseudo-time dependent. 
Although the primal flow problem is formerly unsteady, we are seeking for the steady-state wave field at $t \rightarrow \infty$ ($\Delta t \rightarrow \infty$). Thus, time accuracy is not required from the intermediate temporary solutions of the primal/dual flow fields which in turn greatly simplifies the approach: Also necessary on the primal side, time stepping can be interpreted as a relaxation during the numerical integration and is solely used to ensure/improve the stability of the segregated algorithm, as discussed in the upcoming section. Thus, formally no consistency demands arise w.r.t. (with respect to) discretization and evaluation of the primal/dual unsteady terms, e.g. different time step stencils and sizes are valid.
Algorithmically, however, the adjoint solver should know the primal Courant number distribution in order to consistently invert the primary NVD board, e.g. switch from UD to DD and vice versa.

\paragraph{Stability and Time-Step Requirements} During the approximation of the primal VoF equation, its main diagonal coefficient of the system matrix is affected if the approximation of the convective term switches locally from UD to DD. Using a first order time implicit Euler discretization of $D c \, / \, Dt = 0$ yields the following main diagonal entry for upwind and downwind interpolation of $c^\mathrm{F}$, viz.
\begin{align}
\left[\frac{c^\mathrm{P,t} - c^\mathrm{P,t-1}}{\Delta t^*} \right] \Delta \Omega^* + \sum_{\mathrm{F(P)}} \dot{V}^\mathrm{*F,t} c^\mathrm{F,t} = 0 
\qquad \rightarrow \qquad
A^\mathrm{*P,t}= \frac{\Delta \Omega^*}{\Delta t^*} + \sum_{\mathrm{F(P})} 
\begin{cases} 
\mathrm{max}\left( \dot{V}^\mathrm{*F,t},0 \right) &:\mathrm{UD}\\
\mathrm{min}\left( \dot{V}^\mathrm{*F,t},0 \right) &:\mathrm{DD}
\end{cases}.
\end{align}
In order to preserve diagonal dominance and avoid a zero crossing introduced by DD, the time step $\Delta t$ must be chosen carefully to comply with $\Delta t \, \dot{V}^\mathrm{*F} / \Delta \Omega^\mathrm{P} = \mathrm{Co} < 1/2$, or even lower in 3D flows (e.g. $\mathrm{Co} \leq 0.3$). Since this time step dependency originates from the concentration equation only, sub-cycling strategies can be employed to increase the efficiency of the temporal integration (cf. \citep{ubbink1999method,manzke2013sub}). Basically, sub-cycling retards the simulation within admissible Courant number bounds and subdivides the temporal integration based on an invalid (too large) time step into several valid (smaller) sub time steps as exemplary sketched in Alg. (\ref{alg:sub_cycling}). Special attention should be paid on (un)synchronization as well as in-between gradient computation and we refer to Manzke \citep{manzke2018development} for a detailed discussion.
In adjoint mode, the compressive convection schemes switch from DD to UD along the critical inter-facial region. A closer look at the main diagonal coefficient of $-D \hat{c}^* \, / \, Dt = 0$ reveals a similar observation compared to the primal discretization
\begin{align}
-\left[\frac{\hat{c}^\mathrm{*P,t+1} - \hat{c}^\mathrm{*P,t}}{\Delta t^*} \right] \Delta \Omega^* - \sum_{\mathrm{F(P)}} \dot{V}^\mathrm{*F,t} \hat{c}^\mathrm{*F,t} = 0 
\qquad \rightarrow \qquad
A^\mathrm{*P,t}= \frac{\Delta \Omega^*}{\Delta t^*} - \sum_{\mathrm{F(P)}} 
\begin{cases} 
\mathrm{min}\left( \dot{V}^\mathrm{*F,t},0 \right) &:\mathrm{DD}\\
\mathrm{max}\left( \dot{V}^\mathrm{*F,t},0 \right) &:\mathrm{UD}
\end{cases} \label{equ:adjoint_concentration_matrix}.
\end{align}
Mind that the right-hand side of Eqn. (\ref{equ:adjoi_conce_dile}) is not zero and inheres several source/sink terms which are independent of $\hat c^*$. In contrast to downwinding during the primal integration, now an upwind interpolation of the adjoint face value $\hat{c}^\mathrm{*F}$ counteracts a positive main diagonal. Thus, the adjoint (backward) integration facilitates a sub-cycling procedure for the dual concentration equation also, as schematically coded in Alg. (\ref{alg:sub_cycling}).
Basically, the primal/dual time step sizes as well as the number sub-cycles do not have to match. Nevertheless, coincidence of time-step size and sub-cycles yield an adjoint simulation that is always UD-stable at cells and faces that are primal DD-stable.

\begin{algorithm}
define $\Delta t$, $\mathrm{Co}_\mathrm{tar}$ and $N$ \\
\While{nTimeSteps $\leq$ maxTimeSteps}{
   \While{ (nOuterIter $\leq$ maxOuterIter) or (residual $\leq$ maxResidual)}{
      evaluate state equation \\
      solve linearized momentum equations \\
      solve pressure correction equation \\
      solve turbulence equations \\
      \fcolorbox{white}{mycolor_grey!100}{\parbox{0.3\textwidth}{$\Delta t = \Delta t / N$ \\ 
                                                                 \For{sub-cycle=1,N}{ solve concentration equation}
                                                                 $\Delta t = \Delta t \, N$}}
   }
   %$\mathrm{Co}_\mathrm{max} = \mathrm{max}\left[ \mathrm{Co} \, \left(c \left(c - 1\right) \right) \right]$ \\
   $\Delta t = \gamma \, \mathrm{Co}_\mathrm{tar}/ \mathrm{Co}_\mathrm{max}  + (1-\gamma) \Delta t$
}
\While{nAdTimeSteps $\leq$ maxAdTimeSteps}{
   \While{ (nAdOuterIter $\leq$ maxAdOuterIter) or (adResidual $\leq$ maxAdResidual)}{
      solve adjoint momentum equations \\
      solve adjoint pressure correction equation \\
      \fcolorbox{white}{mycolor_grey!100}{\parbox{0.35\textwidth}{$\Delta t = \Delta t / N$ \\ 
                                                                 \For{adSub-cycle=1,N}{ solve adjoint concentration equation}
                                                                 $\Delta t = \Delta t \, N$}}
   }
}
compute shape derivative Eqn. (\ref{equ:shape_derivative})
\caption{(Pseudo) temporal integration of the primal and adjoint two-phase system based on a sub-cycling approach for the computation of the shape derivative Eqn. (\ref{equ:shape_derivative}). The adaptive time step estimation employs a relaxation which is typically assigned to $\gamma = 0.3$. }
\label{alg:sub_cycling}
\end{algorithm}

\newpage
\section{Plane Couette Flow}
\label{sec:couette}
In this section, the two-phase model is scrutinized for a plane Couette flow. The wall-bounded, homogeneous shear flow is an important paradigm of fluids engineering. The access to available analytical solutions for the primal (\ref{equ:prima_momen_dile})-(\ref{equ:prima_conce_dile}) as well as the dual (\ref{equ:adjoi_momen_dile})-(\ref{equ:adjoi_conce_dile}) problem makes this case particularly interesting. 
Moreover, the case illustrates two important aspects: (a) the non-unique solution behaviour of the adjoint concentration equation and (b) a remedy with negligible impact on the computed sensitivities.  

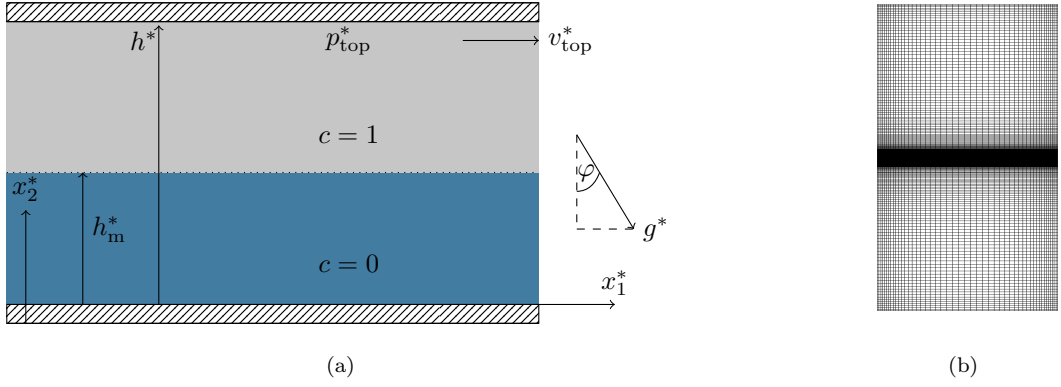
\begin{figure}
	\centering
	\subfigure[]{
		\begin{tikzpicture}
\filldraw[pattern=north east lines, pattern color=black] (0,0) rectangle (7,0.25);
\filldraw[fill=mycolor_blue!100, draw=none] (0,0.25) rectangle (7,2);
\filldraw[fill=mycolor_grey!100, draw=none] (0,2.0) rectangle (7,4);
\draw[thin] (0,0) -- (7,0);
\draw[thin] (0,0.25) -- (7,0.25);
\filldraw[pattern=north east lines, pattern color=black] (0,4) rectangle (7,4.25);
\draw[thin] (0,4) -- (7,4);
\draw[thin] (0,4.25) -- (7,4.25);
\draw[thin,->] (6,3.75) -- (7,3.75) node[anchor=west] {$v_\mathrm{top}^*$};
\draw[dotted] (0,2.0) -- (7,2.0);
\draw (4.5,0.8) node[] {$c = 0$};
\draw (4.5,2.5) node[] {$c = 1$};
\draw[thin,->] (0,0.25) -- (8,0.25) node[anchor=south] {$x_\mathrm{1}^*$};
\draw[thin,->] (0.25,0) -- (0.25,1.5) node[anchor=south] {$x_\mathrm{2}^*$};

%\draw[thin,->] (7.5,2.5) -- (7.5,1.25) node[anchor=west] {$g$};
\draw[thin,->] (7.5,2.5) -- (8.25,1.25) node[anchor=west] {$g^*$};
\draw[dashed] (7.5,2.5) -- (7.5,1.25);
\draw[dashed] (7.5,1.25) -- (8.25,1.25);
\draw [thin] (7.5,1.75) to [out=0,in=250] (7.8,2.0);
\draw (7.625,2.0) node[] {$\varphi$};

\draw[thin,->] (1.,0.25) -- (1.,2.0);
\draw (1.0,1.25) node[anchor=west] {$h_\mathrm{m}^*$};
\draw[thin,->] (2,0.25) -- (2,3.95);
\draw (1.5,3.75) node[anchor=west] {$h^*$};
\draw (4.5,3.75) node[] {$p_\mathrm{top}^*$};
\end{tikzpicture}
	}
	\hspace{2cm}
	\subfigure[]{
		\includegraphics[scale=0.595]{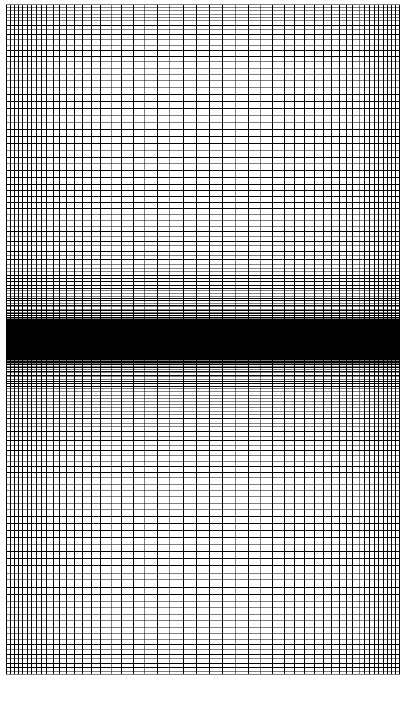}
	}
	\caption{Plane Couette flow case ($\mathrm{Re}_\mathrm{h} = 100$, $\mathrm{Fn}=5$): (a) Sketch of the investigated Couette flow where both fluids are immiscible and share a sharp interface that introduces discontinuous fluid properties. (b) Computational grid for the verification study in Sec. (\ref{sec:veri_stud}).}
	\label{fig:couet_flow}
\end{figure}

Fig. \ref{fig:couet_flow} illustrates the two-phase flow example and the employed computational grid. The laminar flow is considered to be steady and in a fully developed state. The interface normal between the to immiscible fluids is directed orthogonal to the wall boundaries and a body-force with variable angle of attack acts on the flow ($g_1^* = \mathrm{sin} \left( \varphi \right) g^*$, $g_2^* =- \mathrm{cos} \left( \varphi \right) g^*$). In accordance with the VoF-approach we assume that the fluid properties change abruptly across the interface from the foreground to the background fluid. A no-slip condition is imposed on the upper and lower wall. The mixture fraction displays a jump at the interface and the non-dimensional primal momentum Eqn. (\ref{equ:prima_momen_dile}) shrinks as follows:
\begin{alignat}{2}
R_1 &: - \frac{1}{\mathrm{Re}} \frac{\partial }{\partial x_2} \left[ \mu \frac{\partial v_1}{\partial {x_2}} \right] - \frac{1}{\mathrm{Fr}^2} \rho g_1 &&= 0 \, ,  \\
R_2 &: \mathrm{Eu} \frac{\partial p}{\partial x_2} - \frac{1}{\mathrm{Fr}^2} \rho g_2 &&= 0
\end{alignat}
Using $L^*=h^*$, $V^*= v_\mathrm{top}^*$, $P^* = \rho_\mathrm{b}^*V^{*2}$ as well as $G^*=g^*$. The analytical solution to the primal flow is  integrated to 
\begin{align}
 0 \leq x_2 <  \frac{1}{2}:& \begin{cases} 
c = 0 \\
v_1 =  \frac{1}{\mu_\mathrm{b}} \left(  \frac{1}{2} T_\mathrm{b} {x_2}^2 + C_\mathrm{1b} x_2 + C_\mathrm{2b} \right)  \\
p = K_\mathrm{b} x_2 + C_\mathrm{3b}
\end{cases} \label{equ:primal_result_bottom} \\
  \frac{1}{2} < x_2 \leq 1:& \begin{cases} 
c = 1 \\
v_1 =  \frac{1}{\mu_\mathrm{a}} \left( \frac{1}{2} T_\mathrm{a} {x_2}^2 + C_\mathrm{1a} x_2 + C_\mathrm{2a} \right) \\
p = K_\mathrm{a} x_2 + C_\mathrm{3a} \, , 
\end{cases} \label{equ:primal_result_top} 
\end{align}
with the following integration constants
\begin{alignat}{3}
C_\mathrm{1a} &=\frac{\mu_\mathrm{a} \left[ \frac{1}{4} T_\mathrm{b} - \frac{1}{2} T_\mathrm{a} + 2 \right] - \frac{3}{4} T_\mathrm{a}}{1 + \mu_\mathrm{a}}, \hspace{1cm}  && C_\mathrm{2a} = \mu_a - \frac{1}{2} T_\mathrm{a} - C_\mathrm{1a}, \hspace{1cm} && C_\mathrm{3a} = - K_\mathrm{a} \\
C_\mathrm{1b} &= \frac{1}{2}  \left( T_\mathrm{a} - T_\mathrm{b} \right) + C_\mathrm{1a}, && C_\mathrm{2b} = 0 && C_\mathrm{3b} = - \frac{1}{2} \left(K_\mathrm{a} + K_\mathrm{b} \right)
\end{alignat}
and abbreviations
\begin{alignat}{2}
T_\mathrm{a} &= -\frac{\mathrm{Re}}{\mathrm{Fr}^2} \rho_\mathrm{a} g_1, \hspace{1cm} && K_\mathrm{a} = \frac{1}{\mathrm{Fr}^2 \mathrm{Eu}} \rho_\mathrm{a} g_2 \\
T_\mathrm{b} &= -\frac{\mathrm{Re}}{\mathrm{Fr}^2} \rho_\mathrm{b}  g_1, \hspace{1cm} && K_\mathrm{b} = \frac{1}{\mathrm{Fr}^2 \mathrm{Eu}} \rho_\mathrm{b} g_2.
\end{alignat}
%The solutions for the pressure and the primary velocity are $C^0$ continuous. 
%
The adjoint Eqn. (\ref{equ:adjoi_momen_dile}) and (\ref{equ:adjoi_conce_dile}) belonging to a force functional Eqn. (\ref{equ:general_objective}) can also be simplified under the assumptions made for the primal flow and Eqn. (\ref{equ:primal_result_bottom})-(\ref{equ:primal_result_top}), viz. 
\begin{alignat}{2}
\hat{R}_1 &: \frac{\partial }{\partial x_2} \left[ \mu \frac{\partial \hat{v}_1}{\partial {x_2}} \right] &&= 0  \, ,\\
\hat{R}_2 &: \left( \frac{C}{V} \right) \frac{\partial \hat{p}}{\partial x_2} - \left( \frac{C}{V} \right) c \frac{\partial \hat{c}}{\partial x_2} 
+  \rho \hat{v}_1 \label{equ:simpl_secon_adjoin_momen}
 \frac{\partial v_1}{\partial x_2} &&= 0 \, ,  \\
\hat{C} &: \frac{\mu_\Delta}{\mathrm{Re}} \frac{\partial v_1}{\partial x_2} \frac{\partial \hat{v}_1}{\partial x_2} - \frac{\rho_\Delta}{\mathrm{Fr}^2}  \hat{v}_1 g_1 &&= 0 \, .
\end{alignat}
where the product between primal and adjoint concentration is active in the upper fluid, $\alpha = 0$ (see Eqn. (\ref{equ:shift_active_region})).
As opposed to the primal system, additional coupling terms occur in the transverse momentum ($\hat{R_2}$)  and the concentration balance ($\hat{C}$). Similar to the primal problem, the adjoint velocity and pressure could be determined from the two momentum equations provided that the adjoint concentration is known. However, the adjoint concentration equation does not support the determination of $\hat{c}$, but restricts the adjoint velocity further towards a constant. Moreover the product between the normal gradients of the adjoint and the primal velocities refers to the sensitivity which curiously vanishes according to Eqn. (\ref{equ:adjoi_conce_dile}).
Mind, that the adjoint equations and their solution depend on the underlying objective functional. However, a general objective Eqn. (\ref{equ:general_objective}) does not inhere the adjoint concentration and would not meaningful improve the right side towards an accessible solution. Using a force objective, the boundary values of the adjoint velocity are assigned to unity in the negative direction of the minimized force along the objective boundaries and zero in all other cases. This would indeed allow for a solution of the adjoint velocity and formally also $\hat{p}$, but does still not provide a solution for the adjoint concentration, which in turn might impair the solution for the pressure. The problem is admittedly motivated by the chosen uni-directional shear flow example, which however is fairly relevant in practical situations.  

\subsection{Heuristic modification}
The non-unique (solution) nature of the problem is related to the grossly simplified inter-phase physics of the VoF-procedure. Related problems are inhibited by subtracting an additional adjoint diffusion term $\mathcal{D}_{\hat{c}}$ on the left-hand side of the adjoint concentration Eqn. (\ref{equ:adjoi_conce}), viz.
\begin{align}
\mathcal{D}_{\hat{c}} =  \nu_{\hat{c}} \frac{\partial^2 \hat{c}}{\partial x_k^{2}}  \, ,
\end{align}
where $\nu_{\hat c} = \nu_{\hat c}^*  \, \rho_b^*/\mu_b^*$. The heuristic approach needs further justification and involves an artificial kinematic viscosity $ \nu_{\hat{c}}^* $ of dimension $[m^2/s]$ which remains a free parameter. The suggested modification essentially borrows an element of diffuse-interface models, e.g.  the Cahn-Hilliard (CH) model as described by \citep{kuhl2020cahn}. The CH-model describes the desired phase separation by the use of a nonlinear viscosity. As opposed to the VoF-scheme, the admissible values are not restricted to $c=0$ and $c=1$ and can vary between $c\in[-\epsilon,1+\epsilon]$ where $\epsilon$ depends on the underlying chemical potential \citep{garcke2019optimal, hinze2011nonlinear}. Accordingly, the immiscibility condition $Dc/Dt =0$ experiences a non-zero right-hand side. In the limit of a vanishing interface thickness, the CH-equation of the primal mixture fraction reads
\begin{align}
 \frac{D c}{D t^*} -  \ \frac{\partial}{\partial x_k^*} \left( \nu_c^* \; \frac{\partial c }{\partial x_k^*} \right) 
=0  \, , \quad {\rm with} \; \, \nu_c^*=  4M^* \,  (6c^{2}-6c+1) \; . 
\label{CH-Term}
\end{align}
Here $M^*$ is a spatially constant mobility factor of dimension $[m^2/s]$, which is often assigned to an empirically chosen or numerically measured value (cf. \citep{kuhl2020cahn}). 
The gradient diffusion displays a nonlinear normalised diffusivity $\nu_c^*/M^* = 2(6c^{2}-6c+1)$, which is part of a potential and exposed to sign changes. It is zero at $c=0.5 \pm \sqrt{1/12}$,  positive in outer regime and negative in between the zeros. The sign change promotes the desired phase separation. This avoids the need for compressive, downwind-biased  approximations of convective kinematics to separate the two fluids, which is inherent to VoF. Mind that downwinding also refers to the introduction of negative diffusion. The related CH-solution for the mixture fraction should closely resemble the VoF-solution, which is strictly speaking confined to zero or unity values.

Corresponding to the diffusion term of the  primal CH-type model Eqn. (\ref{CH-Term}), an additional adjoint term $ \nu_{\hat{c}}^* \; \partial^2 \hat{c}^*/ \partial x_k^{*2}$ occurs, that employs an adjoint diffusivity, i.e.  $ \nu_{\hat c}^*= 4M^*(6c^{2}-6c+1)$. Although the sign of $ \nu_{\hat c}^*$ formally changes, it is inherently positive in the region of small or large mixture fraction values, i.e. for the two valid VoF-states. Moreover the magnitude is afflicted with the empiricism inherent to the (positive) mobility value $M^*$. This motivates the introduction of $\mathcal{D}_{\hat{c}} $ to the adjoint concentration equation. In the limit of immiscible fluids the linearised  diffusivity $\nu_{\hat c}^*$ obtained from a frozen concentration field ansatz of a CH-model would be positive and identical for the two limit states of $c$.
The modified adjoint concentration equation for the simplified Couette flow takes the following dimensionless residual form
\begin{align}
\hat{C}:\frac{\mu_\Delta}{\mathrm{Re}} \frac{\partial v_1}{\partial x_2} \frac{\partial \hat{v}_1}{\partial x_2} - \frac{\rho_\Delta}{\mathrm{Fr}^2}  \hat{v}_1 g_1 -  \left( \frac{\mathrm{C}^*}{\mathrm{V}^*} \right)  \frac{\nu_{\hat c}}{\mathrm{Re}} \frac{\partial^2 \hat{c}}{\partial {x_2}^2} = 0 \label{equ:couet_diffu}.
\end{align}
The introduction of the additional diffusion term supports an analytical description of the adjoint concentration field: Due to the viscosity jump along the phase boundary, four integration constants have to be determined. Since the synthetic viscosity is constant, it is possible to achieve a smooth solution in the entire field by means of zero and first order coupling conditions along the phase transition regime. The third unknown is computed based on a Neumann condition on the upper channel side. However, the break of dual consistency in the domain has to be continued also at the boundary: The introduction of the diffusive term in Eqn. (\ref{equ:couet_diffu}) requires at least one Dirichlet boundary condition to determine all integration constants. This is contrary to the demand of adjoint analysis, which requires the invariable use of Neumann boundary conditions at walls (see Tab. (\ref{tab:bound_condi})). For this reason, a Dirichlet value is prescribed on the bottom of the channel for the adjoint concentration, which does not change the qualitative curve, but forces a quantitative fixation. Mind that this is also the case for the primal and adjoint pressure. It is justifiable since the adjoint concentration enters the adjoint momentum balance based on its gradient.

The solution of the adjusted adjoint system can be constructed as follows: The force functional (\ref{equ:general_objective}) enters the adjoint system via the optimisation direction that is defined in main flow direction $d_\mathrm{i} = \delta_\mathrm{i1}$. Thus, the boundary condition for the adjoint velocity reads $ \hat{v}_\mathrm{bottom,1} = -1 $ on the lower and $\hat{v}_\mathrm{top,1} = 0 $ on the upper channel side. Likewise, adjoint pressure  and adjoint concentration are prescribed on the lower channel wall, whereas Neumann conditions are imposed on top of the channel.  An integration provides the following analytical solution of the entire system:
\begin{align}
 0 \leq x_2 <  \frac{1}{2}:& \begin{cases} 
\hat{c} =  L_\mathrm{b} \left[ \frac{\mu_\Delta}{\mu_\mathrm{b}} \left( \frac{1}{6} T_\mathrm{b} {x_2}^3 + \frac{1}{2} C_\mathrm{1b} {x_2}^2 \right) + \frac{ \rho_\Delta}{\rho_\mathrm{b}} T_\mathrm{b}  \left( \frac{1}{6} {x_2}^3 - \frac{1}{2 P_\mathrm{b}} {x_2}^2 \right) \right] + C_\mathrm{4b} x_2 + C_\mathrm{5b} \\
\hat{v}_1 =  P_\mathrm{b} x_2 - 1 \\
\hat{p} = -\frac{V}{C} \left[ \frac{1}{3} P_\mathrm{b} T_\mathrm{b} {x_2}^3 + \frac{1}{2} \left( P_\mathrm{b} C_\mathrm{1b} - T_\mathrm{b} \right) {x_2}^2 - C_\mathrm{1b} x_2 \right] + C_\mathrm{6b}
\end{cases} \label{equ:adjoint_result_bottom} \\
 \frac{1}{2} < x_2 \leq 1:& \begin{cases} 
\hat{c} =  L_\mathrm{a} \left[ \frac{\mu_\Delta}{\mu_\mathrm{a}} \left( \frac{1}{6} T_\mathrm{a} {x_2}^3 + \frac{1}{2} C_\mathrm{1a} {x_2}^2 \right)  +  \frac{ \rho_\Delta}{\rho_\mathrm{b}} T_\mathrm{b} \left( \frac{1}{6} {x_2}^3 - \frac{1}{2} {x_2}^2 \right) \right] + C_\mathrm{4a} x_2 + C_\mathrm{5a} \\
\hat{v}_1 = P_\mathrm{a} \left(  x_2 - 1 \right)   \hspace{45.5mm} \\
\hat{p} =  \hat{c} -\frac{V}{C} \frac{\rho_\mathrm{a}}{\mu_\mathrm{a}} P_\mathrm{a} \left[ \frac{1}{3} T_\mathrm{a} {x_2}^3 + \frac{1}{2} \left(  C_\mathrm{1a} - T_\mathrm{a} \right) {x_2}^2 - C_\mathrm{1a} x_2 \right] + C_\mathrm{6a}
\end{cases}  \label{equ:adjoint_result_top} 
\end{align}
with the following integration constants: 
%\trung{wozu benoetigen wir hier die Null-Konstanten??}
\begin{align}
C_\mathrm{4a} &= -L_\mathrm{a} \left[ \frac{\mu_\Delta}{\mu_\mathrm{a}} \left( \frac{1}{2} T_\mathrm{a} + C_\mathrm{1a}  \right) - \frac{1}{2} T_\mathrm{b} \rho_\Delta  \right]\, ,  \hspace{2cm} C_\mathrm{5b} = 0\, ,  \hspace{2cm} C_\mathrm{6b} = 0\, , \\
C_\mathrm{4b} &= -L_\mathrm{a} \left[ \frac{\mu_\Delta}{\mu_\mathrm{a}} \left( \frac{3}{8} T_\mathrm{a} + \frac{1}{2} C_\mathrm{1a} \right) - \frac{1}{8} T_\mathrm{b}  \rho_\Delta  \right] - L_\mathrm{b} \left[\frac{\mu_\Delta}{\mu_\mathrm{b}} \left( \frac{1}{8} T_\mathrm{b} + \frac{1}{2} C_\mathrm{1b} \right) + T_\mathrm{b}  \rho_\Delta  \left( \frac{1}{8} - \frac{1}{2 P_\mathrm{b}} \right) \right]\, , \\
C_\mathrm{5a} &= L_\mathrm{a} \left[ \frac{\mu_\Delta}{\mu_\mathrm{a}}  \left( \frac{1}{24} T_\mathrm{a} + \frac{1}{8} C_\mathrm{1a} \right) - \frac{1}{12} T_\mathrm{b} \rho_\Delta  \right]  - L_\mathrm{b} \left[ \frac{\mu_\Delta}{\mu_\mathrm{b}} \left( \frac{1}{24} T_\mathrm{b} + \frac{1}{8} C_\mathrm{1b} \right) - T_\mathrm{b}  \rho_\Delta^*  \left( \frac{1}{24} - \frac{1}{8 P_\mathrm{b}} \right) \right]\, , \\
C_\mathrm{6a} &= -\frac{\mathrm{V}^*}{\mathrm{C}^*} \left[ \frac{1}{24} P_\mathrm{b} T_\mathrm{b} + \frac{1}{8} \left( P_\mathrm{b} C_\mathrm{1b} - T_\mathrm{b} \right) - \frac{1}{2} C_\mathrm{1a} - \frac{\rho_\mathrm{a}}{\mu_\mathrm{a}} P_\mathrm{a} \left[ \frac{1}{24} T_\mathrm{a} + \frac{1}{8} \left( C_\mathrm{1a} - T_\mathrm{a} \right) - \frac{1}{2} C_\mathrm{1a} \right] \right] - \hat{c}\bigg|_{x_2 = \frac{1}{2}}
\end{align}
as well as the upcoming abbreviations:
\begin{alignat}{2}
P_\mathrm{a} &= \frac{2 \mu_\mathrm{b}}{\mu_\mathrm{a} + \mu_\mathrm{b}}, \hspace{1cm} &&L_\mathrm{a} = \frac{1}{\nu_{\hat{c}}} \frac{\mathrm{V}^*}{\mathrm{C}^*} P_\mathrm{a} \\
P_\mathrm{b} &= \frac{2 \mu_\mathrm{a}}{\mu_\mathrm{a} + \mu_\mathrm{b}}, \hspace{1cm} &&L_\mathrm{b} =  \frac{1}{\nu_{\hat{c}}} \frac{\mathrm{V}^*}{\mathrm{C}^*} P_\mathrm{b}.
\end{alignat}
Interestingly, the solution of the adjoint velocity and therefore also the sensitivity does not depend on $L_\mathrm{a}$ or $L_\mathrm{b}$ and is by far less tedious compared to the solution of adjoint concentration and adjoint pressure. Thus the mobility parameter that scales the heuristic diffusivity $\nu_{\hat c}$ only governs the adjoint hydrostatic field. In the limit of vanishing diffusivity $\nu_{\hat c}$ we would impose a very large sink term while approaching to the wall, which would challenge the numerical solution of the adjoint pressure and the adjoint concentration in the upper fluid regime, where adjoint pressure sees the adjoint concentration abruptly. Mind that we can shift the active region of $\hat{c}$ in accordance to Eqn. (\ref{equ:shift_active_region}). The disappearance of the constants $T_\mathrm{a}$ and $T_\mathrm{b}$ for $g_1 \rightarrow 0$ (or $\hat{v}_\mathrm{i} g_\mathrm{i}  \rightarrow  0$), which strongly simplifies the solutions of adjoint concentration and adjoint pressure due to the cancellation of all cubic terms in the analytic solutions, supports the regularization characteristics of the synthetic viscosity.

From now on, the complete modified adjoint concentration equation takes the following non dimensional form
\begin{align}
\hat{C}: \hspace{1cm}  - \frac{1}{\mathrm{St}} \frac{\partial \hat{c}}{\partial t} - v_\mathrm{j} \frac{\partial \hat{c}}{\partial x_\mathrm{j}} 
+\left( \frac{\mathrm{V}^* \mathrm{\hat{V}}^*}{\mathrm{C}^*} \right)  \rho_\Delta \hat{v}_\mathrm{i} v_\mathrm{j} \frac{\partial v_\mathrm{i}}{\partial x_\mathrm{j}} 
+ \left( \frac{\mathrm{V}^*\mathrm{\hat{V}}^*}{\mathrm{C}^*} \right) \frac{2 {\mu_\Delta}}{\mathrm{Re}}  S_\mathrm{ij} \frac{\partial \hat{v}_\mathrm{i}}{\partial x_\mathrm{j}} 
-  \left( \frac{\mathrm{V}^*\mathrm{\hat{V}}^*}{\mathrm{C}^*} \right) \frac{\rho_\Delta }{\mathrm{Fr}^2} \hat{v}_\mathrm{i} g_\mathrm{i} 
- \frac{\nu_{\hat{c}}}{Re}  \frac{\partial^2 \hat{c}}{\partial {x_\mathrm{j}}^2} = 0 . \label{equ:fina_adjoi_conce_dile}
\end{align}
The additional last term is implicitly treated within the segregated FV-framework, resulting in a positive impact on the main diagonal of the system matrix and supports the balance for vanishing Reynolds/Froude numbers or under extreme strain rates when different fluid properties occur. The associated break of dual consistency provides a strongly regularized adjoint system.

\subsection{Verification}
\label{sec:veri_stud}
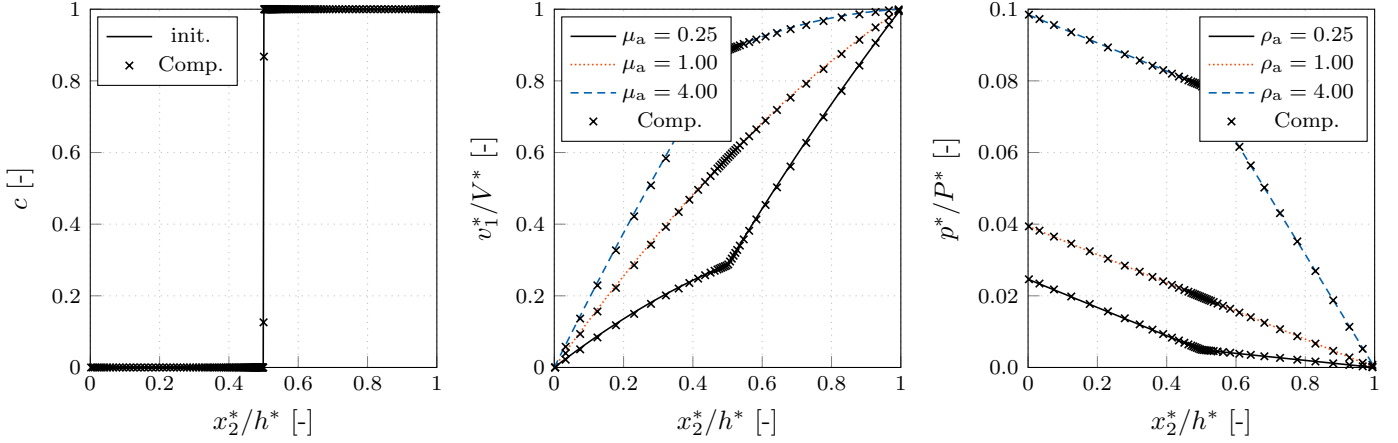
\begin{figure}
\centering
\analytiSolutionPictures
\begin{tikzpicture}
\begin{axis}[
 xlabel style={text width=0.25\textwidth,align=center},
 ylabel style={text width=0.25\textwidth,align=center},
 xlabel={$x_\mathrm{2}^* / h^*$ [-]},
 ylabel={$c$ [-]},
 xmin=0,xmax=1,
 legend style={at={(0.02,0.98)},anchor=north west},
 ymin=0.0,ymax=1.0
]
\addplot [line1] table[x expr={\thisrowno{0}},y expr={\thisrowno{1}}, col sep=space] {data/Couette_Results_Varphi_10_matRatio_025_concentration_analytic.dat};
\addplot [mark1, only marks] table[x expr={\thisrowno{0}},y expr={\thisrowno{1}}]{data/Couette_Results_Varphi_10_matRatio_025_concentration_FreSCo.dat};
 
\addlegendentry{init.};
\addlegendentry{Comp.}; 
 
\end{axis}
\end{tikzpicture}
\begin{tikzpicture}
\begin{axis}[
 ylabel style={text width=0.25\textwidth,align=center},
 xlabel={$x_\mathrm{2}^* / h^*$ [-]},
 ylabel={$v_\mathrm{1}^* / V^*$ [-]},
 xmin=0,xmax=1,
 legend style={at={(0.02,0.98)},anchor=north west},
 ymin=0.0,ymax=1.0
]
\addplot [line1] table[x expr={\thisrowno{0}},y expr={\thisrowno{1}}]{data/Couette_Results_Varphi_10_matRatio_025_velocity_analytic.dat};
\addplot [line2] table[x expr={\thisrowno{0}},y expr={\thisrowno{1}}]{data/Couette_Results_Varphi_10_matRatio_1_velocity_analytic.dat};
\addplot [line3] table[x expr={\thisrowno{0}},y expr={\thisrowno{1}}]{data/Couette_Results_Varphi_10_matRatio_4_velocity_analytic.dat};
\addplot [mark1, only marks,mark repeat=5] table[x expr={\thisrowno{0}},y expr={\thisrowno{1}}]{data/Couette_Results_Varphi_10_matRatio_025_velocity_FreSCo.dat};
\addplot [mark1, only marks,mark repeat=5] table[x expr={\thisrowno{0}},y expr={\thisrowno{1}}]{data/Couette_Results_Varphi_10_matRatio_1_velocity_FreSCo.dat};
\addplot [mark1, only marks,mark repeat=5] table[x expr={\thisrowno{0}},y expr={\thisrowno{1}}]{data/Couette_Results_Varphi_10_matRatio_4_velocity_FreSCo.dat};
 
\addlegendentry{$\mu_\mathrm{a} = 0.25$};
\addlegendentry{$\mu_\mathrm{a} = 1.00$};
\addlegendentry{$\mu_\mathrm{a} = 4.00$};
\addlegendentry{Comp.};

\end{axis}
\end{tikzpicture}
\begin{tikzpicture}
\begin{axis}[
 ylabel style={text width=0.25\textwidth,align=center},
 xlabel={$x_\mathrm{2}^* / h^*$ [-]},
 ylabel={$p^* / P^*$ [-]},
 xmin=0,xmax=1,
 ymin=0.0,ymax=0.1,
 ytick={0.0,0.02,0.04,0.06,0.08, 0.1},
 yticklabels={0.0,0.02,0.04,0.06,0.08, 0.1},
]
\addplot [line1] table[x expr={\thisrowno{0}},y expr={\thisrowno{1}}]{data/Couette_Results_Varphi_10_matRatio_025_pressure_analytic.dat};
\addplot [line2] table[x expr={\thisrowno{0}},y expr={\thisrowno{1}}]{data/Couette_Results_Varphi_10_matRatio_1_pressure_analytic.dat};
\addplot [line3] table[x expr={\thisrowno{0}},y expr={\thisrowno{1}}]{data/Couette_Results_Varphi_10_matRatio_4_pressure_analytic.dat};
\addplot [mark1, only marks,mark repeat=5] table[x expr={\thisrowno{0}},y expr={\thisrowno{1}}]{data/Couette_Results_Varphi_10_matRatio_025_pressure_FreSCo.dat};
\addplot [mark1, only marks,mark repeat=5] table[x expr={\thisrowno{0}},y expr={\thisrowno{1}}]{data/Couette_Results_Varphi_10_matRatio_1_pressure_FreSCo.dat};
\addplot [mark1, only marks,mark repeat=5] table[x expr={\thisrowno{0}},y expr={\thisrowno{1}}]{data/Couette_Results_Varphi_10_matRatio_4_pressure_FreSCo.dat};
 
\addlegendentry{$\rho_\mathrm{a} = 0.25$};
\addlegendentry{$\rho_\mathrm{a} = 1.00$};
\addlegendentry{$\rho_\mathrm{a} = 4.00$};
\addlegendentry{Comp.};

\end{axis}
\end{tikzpicture}
\caption{Plane Couette flow case ($\mathrm{Re}_\mathrm{h} = 100$, $\mathrm{Fn}=5$): Primal results of the concentration (left), velocity (centre) and pressure (right) for the planar Couette flow for Froude and Reynolds number of $ \mathrm{Re}_\mathrm{h}= 100$ and $\mathrm{Fr} = 5$ as well as density (viscosity) ratios of $\rho_\mathrm{a} = \rho_\mathrm{a}^* / \rho_\mathrm{b}^* = 1 / 4$ ($\mu_\mathrm{a} = \mu_\mathrm{a}^* / \mu_\mathrm{b}^* = 1 / 4$), $\rho_\mathrm{a} = 1$ ($\mu_\mathrm{a} = 1$) and $\rho_\mathrm{a} = 4$ ($\mu_\mathrm{a} = 4$) under a gravity angle of $\varphi = 10^\circ$.}
\label{fig:primal_results}
\end{figure}

In this section the implementation of the primal and adjoint system is verified. The planar two-phase Couette flow illustrated by Fig. \ref{fig:couet_flow} is solved with periodic boundary conditions for Reynolds and Froude numbers of $\mathrm{Re} = 100$ and $\mathrm{Fr} = 5$. The dimensionless analysis employs the following characteristic quantities: $\mathrm{L}^* = h^*$, $\mathrm{V}^* = \mathrm{C}^* = v_\mathrm{top}^*$, $\mathrm{P} = \rho_b^* \mathrm{V}^{*2}$ ($\mathrm{Eu} = 1$) and $\mathrm{G}^* = g^*$. Background fluid properties $\rho_b^*$ and $\mu_b^*$ are again used as reference data. The corresponding numerical grid is depicted on the right side of Fig. \ref{fig:couet_flow} and the required non-dimensional density (viscosity) ratios are assigned to exemplary values of $\rho_\mathrm{a} = \rho_\mathrm{a}^* / \rho_\mathrm{b}^* = 1 / 4$ ($\mu_\mathrm{a} = \mu_\mathrm{a}^* / \mu_\mathrm{b}^* = 1 / 4$), $\rho_\mathrm{a} = 1$ ($\mu_\mathrm{a} = 1$) as well as $\rho_\mathrm{a} = 4$ ($\mu_\mathrm{a} = 4$) and therefore range from a lighter and less viscous to a heavier and more viscous upper fluid. Convective primal momentum fluxes are discretized using a first-order upwind differencing schemes whereas diffusion employs central differences. The approximation of convective concentration transport is realized with the HRIC approach. A comparison of the analytical primal solutions (\ref{equ:primal_result_bottom}, \ref{equ:primal_result_top}) with the numerical results for a gravity angle of $\varphi = 10^\circ$ is displayed in Fig. \ref{fig:primal_results}.
The numerical implementation resolves the ideally sharp interface between both fluid phases within 2-4 cells, which is a common observation for VoF-procedures that employ compressive approximations for the convective term.

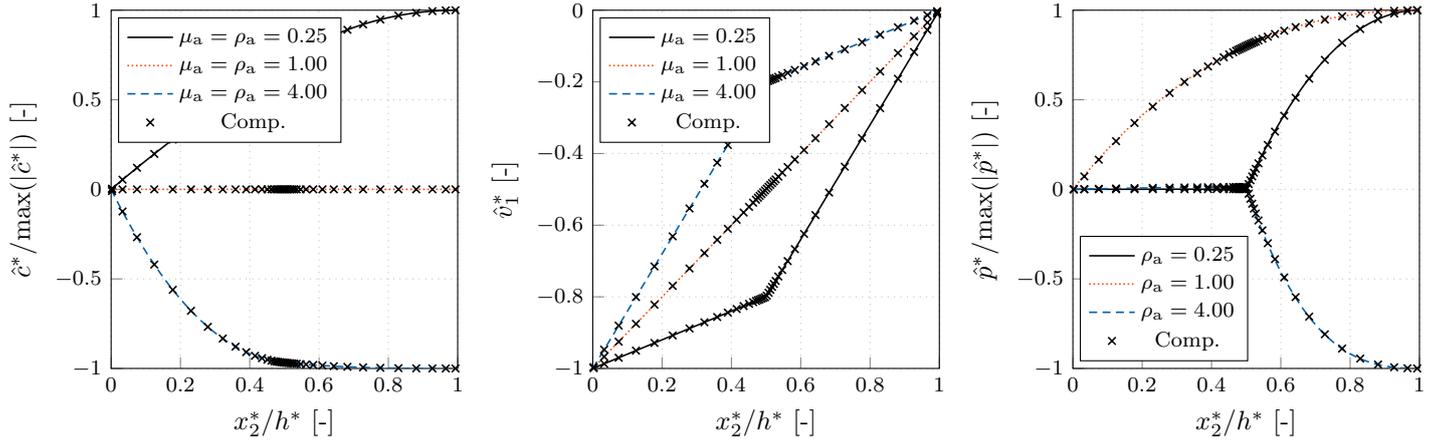
\begin{figure}
\centering
\analytiSolutionPictures
\begin{tikzpicture}
\begin{axis}[
 xlabel style={text width=0.25\textwidth,align=center},
 ylabel style={text width=0.25\textwidth,align=center},
 xlabel={$x_\mathrm{2}^* / h^*$ [-]},
 ylabel={$\hat{c}^* / \mathrm{max}(|\hat{c}^*|)$ [-]},
 xmin=0,xmax=1,
 legend style={at={(0.02,0.98)},anchor=north west},
 ymin=-1.0,ymax=1.0
]
\addplot [line1] table[x expr={\thisrowno{0}},y expr={\thisrowno{1}}]{data/Couette_Results_Varphi_10_matRatio_025_adconcentration_analytic_nuAdC_1E-05_NORM.dat};
\addplot [line2] table[x expr={\thisrowno{0}},y expr={\thisrowno{1}}]{data/Couette_Results_Varphi_10_matRatio_1_adconcentration_analytic_nuAdC_1E-05_NORM.dat};
\addplot [line3] table[x expr={\thisrowno{0}},y expr={\thisrowno{1}}]{data/Couette_Results_Varphi_10_matRatio_4_adconcentration_analytic_nuAdC_1E-05_NORM.dat};
\addplot [mark1, only marks, mark repeat=5] table[x expr={\thisrowno{0}},y expr={\thisrowno{1}}]{data/Couette_Results_Varphi_10_matRatio_025_adconcentration_FreSCo_nuAdC_1E-05_NORM.dat};
\addplot [mark1, only marks, mark repeat=5] table[x expr={\thisrowno{0}},y expr={\thisrowno{1}}]{data/Couette_Results_Varphi_10_matRatio_1_adconcentration_FreSCo_nuAdC_1E-05_NORM.dat};
\addplot [mark1, only marks, mark repeat=5] table[x expr={\thisrowno{0}},y expr={\thisrowno{1}}]{data/Couette_Results_Varphi_10_matRatio_4_adconcentration_FreSCo_nuAdC_1E-05_NORM.dat};
 
\addlegendentry{$\mu_\mathrm{a} = \rho_\mathrm{a} = 0.25$};
\addlegendentry{$\mu_\mathrm{a} = \rho_\mathrm{a} = 1.00$};
\addlegendentry{$\mu_\mathrm{a} = \rho_\mathrm{a} = 4.00$}; 
\addlegendentry{Comp.};
 
\end{axis}
\end{tikzpicture}
\begin{tikzpicture}
\begin{axis}[
 ylabel style={text width=0.25\textwidth,align=center},
 xlabel={$x_\mathrm{2}^* / h^*$ [-]},
 ylabel={$\hat{v}_\mathrm{1}^*$ [-]},
 xmin=0,xmax=1,
 legend style={at={(0.02,0.98)},anchor=north west},
 ymin=-1.0,ymax=0.0
]
\addplot [line1] table[x expr={\thisrowno{0}},y expr={\thisrowno{1}}]{data/Couette_Results_Varphi_10_matRatio_025_advelocity_analytic_nuAdC_1E-05.dat};
\addplot [line2] table[x expr={\thisrowno{0}},y expr={\thisrowno{1}}]{data/Couette_Results_Varphi_10_matRatio_1_advelocity_analytic_nuAdC_1E-05.dat};
\addplot [line3] table[x expr={\thisrowno{0}},y expr={\thisrowno{1}}]{data/Couette_Results_Varphi_10_matRatio_4_advelocity_analytic_nuAdC_1E-05.dat};
\addplot [mark1, only marks,mark repeat=5] table[x expr={\thisrowno{0}},y expr={\thisrowno{1}}]{data/Couette_Results_Varphi_10_matRatio_025_advelocity_FreSCo_nuAdC_1E-05.dat};
\addplot [mark1, only marks,mark repeat=5] table[x expr={\thisrowno{0}},y expr={\thisrowno{1}}]{data/Couette_Results_Varphi_10_matRatio_1_advelocity_FreSCo_nuAdC_1E-05.dat};
\addplot [mark1, only marks,mark repeat=5] table[x expr={\thisrowno{0}},y expr={\thisrowno{1}}]{data/Couette_Results_Varphi_10_matRatio_4_advelocity_FreSCo_nuAdC_1E-05.dat};
 
\addlegendentry{$\mu_\mathrm{a} = 0.25$};
\addlegendentry{$\mu_\mathrm{a} = 1.00$};
\addlegendentry{$\mu_\mathrm{a} = 4.00$};
\addlegendentry{Comp.};

\end{axis}
\end{tikzpicture}
\begin{tikzpicture}
\begin{axis}[
 ylabel style={text width=0.25\textwidth,align=center},
 xlabel={$x_\mathrm{2}^* / h^*$ [-]},
 ylabel={$\hat{p}^* / \mathrm{max}(|\hat{p}^*|)$ [-]},
 legend style={at={(0.02,0.02)},anchor=south west},
 xmin=0,xmax=1,
 ymin=-1.0,ymax=1.0
]
\addplot [line1] table[x expr={\thisrowno{0}},y expr={\thisrowno{1}}]{data/Couette_Results_Varphi_10_matRatio_025_adpressure_analytic_nuAdC_1E-05_NORM.dat};
\addplot [line2] table[x expr={\thisrowno{0}},y expr={\thisrowno{1}}]{data/Couette_Results_Varphi_10_matRatio_1_adpressure_analytic_nuAdC_1E-05_NORM.dat};
\addplot [line3] table[x expr={\thisrowno{0}},y expr={\thisrowno{1}}]{data/Couette_Results_Varphi_10_matRatio_4_adpressure_analytic_nuAdC_1E-05_NORM.dat};
\addplot [mark1, only marks,mark repeat=5] table[x expr={\thisrowno{0}},y expr={\thisrowno{1}}]{data/Couette_Results_Varphi_10_matRatio_025_adpressure_FreSCo_nuAdC_1E-05_NORM.dat};
\addplot [mark1, only marks,mark repeat=5] table[x expr={\thisrowno{0}},y expr={\thisrowno{1}}]{data/Couette_Results_Varphi_10_matRatio_1_adpressure_FreSCo_nuAdC_1E-05_NORM.dat};
\addplot [mark1, only marks,mark repeat=5] table[x expr={\thisrowno{0}},y expr={\thisrowno{1}}]{data/Couette_Results_Varphi_10_matRatio_4_adpressure_FreSCo_nuAdC_1E-05_NORM.dat};
 
\addlegendentry{$\rho_\mathrm{a} = 0.25$};
\addlegendentry{$\rho_\mathrm{a} = 1.00$};
\addlegendentry{$\rho_\mathrm{a} = 4.00$};
\addlegendentry{Comp.};

\end{axis}
\end{tikzpicture}
\caption{Plane Couette flow case ($\mathrm{Re}_\mathrm{h} = 100$, $\mathrm{Fn}=5$): Adjoint results of the planar Couette flow for a synthetic viscosity of $ \nu_{\hat{c}} = 10^{-3}$ for Froude and Reynolds number of $ \mathrm{Re}_\mathrm{h}= 100$ and $\mathrm{Fr} = 5$ as well as density (viscosity) ratios of $\rho_\mathrm{a} = \rho_\mathrm{a}^* / \rho_\mathrm{b}^* = 1 / 4$ ($\mu_\mathrm{a} = \mu_\mathrm{a}^* / \mu_\mathrm{b}^* = 1 / 4$), $\rho_\mathrm{a} = 1$ ($\mu_\mathrm{a} = 1$) and $\rho_\mathrm{a} = 4$ ($\mu_\mathrm{a} = 4$) under a gravity angle of $\varphi = 10^\circ$.}
\label{fig:adjoint_results}
\end{figure}

\begin{figure}
\centering
\subfigure[]{
\centering
\tinyPicture
\begin{tikzpicture}
\begin{axis}[
 xlabel style={text width=0.25\textwidth,align=center},
 ylabel style={text width=0.25\textwidth,align=center},
 xlabel={$x_\mathrm{2}^* / h^*$ [-]},
 ylabel={$\hat{c}^* / (\rho_\mathrm{b}^* C)$ [-]},
 xmin=0,xmax=1,
 legend style={at={(0.02,0.98)},anchor=north west},
 ymode=log
]
\addplot [line1] table[x expr={\thisrowno{0}},y expr={\thisrowno{1}}]{data/Couette_Results_Varphi_10_matRatio_025_adconcentration_analytic_nuAdC_1E-05.dat};
%\addplot [line2] table[x expr={\thisrowno{0}},y expr={\thisrowno{1}}]{data/Couette_Results_Varphi_10_matRatio_025_adconcentration_analytic_nuAdC_1E-04.dat};
\addplot [line2] table[x expr={\thisrowno{0}},y expr={\thisrowno{1}}]{data/Couette_Results_Varphi_10_matRatio_025_adconcentration_analytic_nuAdC_1E-03.dat};
%\addplot [line4] table[x expr={\thisrowno{0}},y expr={\thisrowno{1}}]{data/Couette_Results_Varphi_10_matRatio_025_adconcentration_analytic_nuAdC_1E-02.dat};
\addplot [line3] table[x expr={\thisrowno{0}},y expr={\thisrowno{1}}]{data/Couette_Results_Varphi_10_matRatio_025_adconcentration_analytic_nuAdC_1E-01.dat};
\addplot [mark1, only marks, mark repeat=5] table[x expr={\thisrowno{0}},y expr={\thisrowno{1}}]{data/Couette_Results_Varphi_10_matRatio_025_adconcentration_FreSCo_nuAdC_1E-05.dat};
%\addplot [mark1, only marks, mark repeat=5] table[x expr={\thisrowno{0}},y expr={\thisrowno{1}}]{data/Couette_Results_Varphi_10_matRatio_025_adconcentration_FreSCo_nuAdC_1E-04.dat};
\addplot [mark1, only marks, mark repeat=5] table[x expr={\thisrowno{0}},y expr={\thisrowno{1}}]{data/Couette_Results_Varphi_10_matRatio_025_adconcentration_FreSCo_nuAdC_1E-03.dat};
%\addplot [mark1, only marks, mark repeat=5] table[x expr={\thisrowno{0}},y expr={\thisrowno{1}}]{data/Couette_Results_Varphi_10_matRatio_025_adconcentration_FreSCo_nuAdC_1E-02.dat};
\addplot [mark1, only marks, mark repeat=5] table[x expr={\thisrowno{0}},y expr={\thisrowno{1}}]{data/Couette_Results_Varphi_10_matRatio_025_adconcentration_FreSCo_nuAdC_1E-01.dat};
 
%\addlegendentry{$\mu_\mathrm{a} = \rho_\mathrm{a} = 0.25$};
%\addlegendentry{$\mu_\mathrm{a} = \rho_\mathrm{a} = 1.00$};
%\addlegendentry{$\mu_\mathrm{a} = \rho_\mathrm{a} = 4.00$}; 
%\addlegendentry{Comp.};
 
\end{axis}
\end{tikzpicture}
}
\hspace{1cm}
\subfigure[]{
\centering
\tinyPicture
\begin{tikzpicture}
\begin{axis}[
 ylabel style={text width=0.25\textwidth,align=center},
 xlabel={$x_\mathrm{2}^* / h^*$ [-]},
 ylabel={$\hat{p}^* / (\rho_\mathrm{b}^* C)$ [-]},
 xmin=0,xmax=1,
 legend style={at={(0.02,0.98)},anchor=north west},
 ymode=log
]
\addplot [line1] table[x expr={\thisrowno{0}},y expr={\thisrowno{1}}]{data/Couette_Results_Varphi_10_matRatio_025_adpressure_analytic_nuAdC_1E-05.dat};
%\addplot [line1] table[x expr={\thisrowno{0}},y expr={\thisrowno{1}}]{data/Couette_Results_Varphi_10_matRatio_025_adpressure_analytic_nuAdC_1E-04.dat};
\addplot [line2] table[x expr={\thisrowno{0}},y expr={\thisrowno{1}}]{data/Couette_Results_Varphi_10_matRatio_025_adpressure_analytic_nuAdC_1E-03.dat};
%\addplot [line1] table[x expr={\thisrowno{0}},y expr={\thisrowno{1}}]{data/Couette_Results_Varphi_10_matRatio_025_adpressure_analytic_nuAdC_1E-02.dat};
\addplot [line3] table[x expr={\thisrowno{0}},y expr={\thisrowno{1}}]{data/Couette_Results_Varphi_10_matRatio_025_adpressure_analytic_nuAdC_1E-01.dat};
\addplot [mark1, only marks,mark repeat=5] table[x expr={\thisrowno{0}},y expr={\thisrowno{1}}]{data/Couette_Results_Varphi_10_matRatio_025_adpressure_FreSCo_nuAdC_1E-05.dat};
%\addplot [mark1, only marks,mark repeat=5] table[x expr={\thisrowno{0}},y expr={\thisrowno{1}}]{data/Couette_Results_Varphi_10_matRatio_025_adpressure_FreSCo_nuAdC_1E-04.dat};
\addplot [mark1, only marks,mark repeat=5] table[x expr={\thisrowno{0}},y expr={\thisrowno{1}}]{data/Couette_Results_Varphi_10_matRatio_025_adpressure_FreSCo_nuAdC_1E-03.dat};
%\addplot [mark1, only marks,mark repeat=5] table[x expr={\thisrowno{0}},y expr={\thisrowno{1}}]{data/Couette_Results_Varphi_10_matRatio_025_adpressure_FreSCo_nuAdC_1E-02.dat};
\addplot [mark1, only marks,mark repeat=5] table[x expr={\thisrowno{0}},y expr={\thisrowno{1}}]{data/Couette_Results_Varphi_10_matRatio_025_adpressure_FreSCo_nuAdC_1E-01.dat};
 
\addlegendentry{$\nu_\mathrm{\hat{c}} = 10^{-3}$};
\addlegendentry{$\nu_\mathrm{\hat{c}} = 10^{-1}$};
\addlegendentry{$\nu_\mathrm{\hat{c}} = 1$};
\addlegendentry{Comp.};

\end{axis}
\end{tikzpicture}
}
\caption{Plane Couette flow case ($\mathrm{Re}_\mathrm{h} = 100$, $\mathrm{Fn}=5$): Numerical and analytical results of a) adjoint concentration and b) adjoint pressure of the planar Couette flow for different (synthetic) viscosity magnitudes that indicates the regularization characteristics of the heuristic modification.}
\label{fig:different_nuadC}
\end{figure}
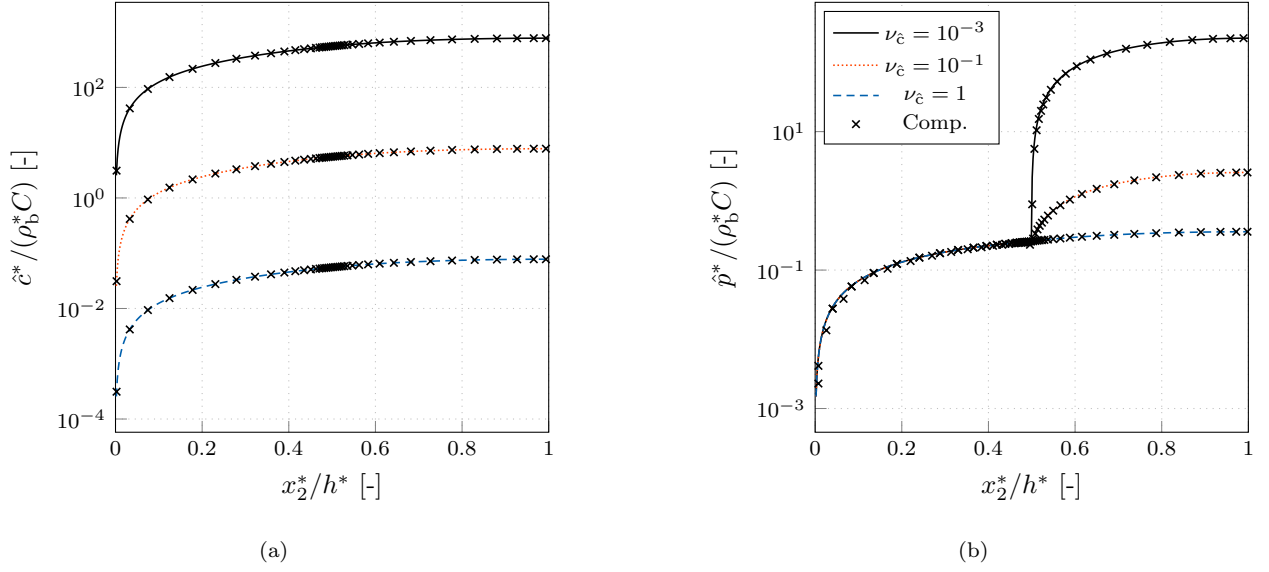

In adjoint mode, the solution is computed for an exemplary heuristic viscosity of $ \nu_{\hat{c}} = 10^{-3}$. Again, periodic boundary conditions are used in longitudinal direction. In contrast to the primal problem, adjoint field quantities are specified on the lower channel side, where the adjoint velocity is assigned to $ \hat{v}_\mathrm{bottom,1} = -\delta_\mathrm{1,1} = -1$. Dirichlet values for adjoint pressure and concentration are set to zero at the bottom wall. The adjoint momentum equation utilizes a first-order downwind differencing scheme  and the adjoint concentration employs the adjoint HRIC scheme (\ref{equ:adjoint_hric_principal}) to approximate convective fluxes. Due to its self adjoint nature, diffusive fluxes are discretized with central differences again. Fig. \ref{fig:adjoint_results} verifies the implementation of the adjoint system against the analytical solutions  (\ref{equ:adjoint_result_bottom}, \ref{equ:adjoint_result_top}). 
The resulting adjoint concentration is smooth in the entire domain, whereas the adjoint pressure buckles along the free surface region. This can be attributed to the abrupt introduction of the adjoint concentration to the upper fluid domain 
%following Eqn. (\ref{equ:adjoint_result_top})  
in accordance with $\alpha = 0$ 
%\trung{wo ist alpha in Glg. 51?}.
However, introducing a (heuristic) viscosity attenuates the adjoint concentration magnitude, thus a larger value of $\nu_{\hat{c}}$ smoothens the solution of the adjoint pressure. In Fig. \ref{fig:different_nuadC}, the synthetic viscosity is increased step by step to underline its regularizing characteristics with respect the adjoint pressure. This is especially relevant for real-world flows, in which adjoint pressure-velocity coupling is a critical aspect of the numerical stability. This is the motivation for the following section.

\newpage
\section{Validation}
\begin{figure}
\centering
\subfigure[]{
\centering
\includegraphics[scale=1]{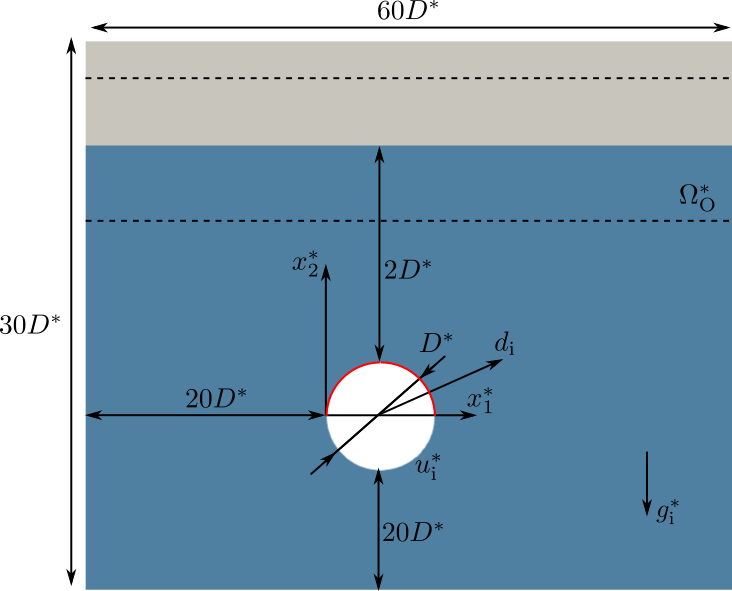}
}
\hspace{1cm}
\subfigure[]{
\centering
\includegraphics[scale=1]{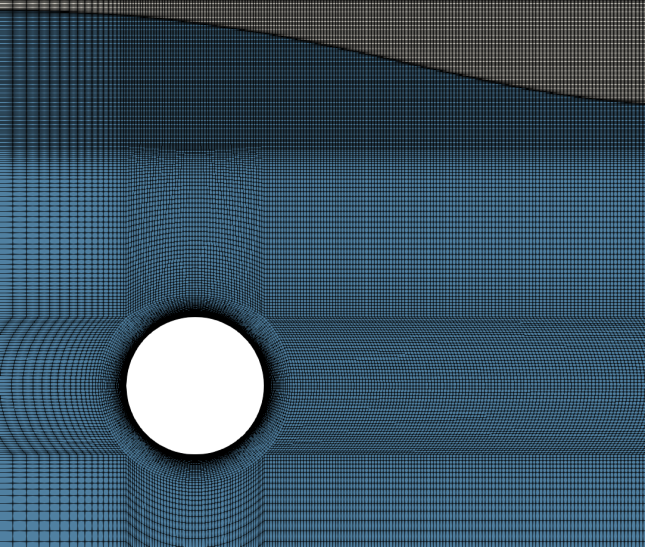}
}
\caption{Submerged cylinder case ($\mathrm{Re}_\mathrm{D} = 20$, $\mathrm{Fn}=0.75$): (a) Schematic drawing of the initial configuration around the controlled cylinder shape $u_\mathrm{i}^*$ (red) and (b) structured numerical grid around the cylinder and the free surface.}
\label{fig:cylinder_fn_075}
\end{figure}

Following a  successfully verification of the implementation in the previous chapter, this section is devoted to a validation the adjoint VoF approach. For this reason a cylindrical shape is placed twice its diameter $D^*$ below the initial free surface, cf. Fig. \ref{fig:cylinder_fn_075} (a). The study is performed for a laminar flow at $\mathrm{Re}_\mathrm{D^*} = v_\mathrm{1}^* D^*/\nu_\mathrm{b}^* = \SI{20}{}$ and $\mathrm{Fn} = v_\mathrm{1}^*/\sqrt{G^* 2 D^*} = \SI{0.75}{}$, based on the gravitational acceleration $G^*$, the inflow velocity $v_\mathrm{1}^*$ and the kinematic viscosity of the water $\nu_\mathrm{b}^*$. The two-dimensional domain has a length and height of $60 \, D^*$ and $30 \, D^*$, where the inlet and bottom boundaries are located 20 diameters away from the origin. A dimensionless wave length of $\lambda = \lambda^*/ D^* = 2 \, \pi \, \mathrm{Fn}^2 = 3.534$ is expected. To ensure the independence of the objective functional with respect to spatial discretization, a grid study was first conducted. The utilized structured numerical grid is displayed in Fig. \ref{fig:cylinder_fn_075} (b) and consists of approximately $\SI{215000}{}$ control volumes and the controlled cylinder shape is
discretized with 500 surface patches along the circumference. At the inlet, velocity and concentration values are prescribed, slip walls are used along the top and bottom boundaries and a hydrostatic pressure boundary is employed along the outlet. The convective term for momentum is approximated using the QUICK scheme. 
A compressive HRIC scheme was used to approximate the convective fluxes of the concentration equation.
The wall normal distance of the first grid layer reads $y^+ \approx \SI{0.01}{}$ and the free surface refinement employs approximately  $\delta x_\mathrm{1}^* / \lambda^* = 1/100 = \delta x_\mathrm{2}^* / \lambda^*$ cells in the longitudinal as well as in the normal direction. According to Alg. (\ref{alg:sub_cycling}) the integration in pseudo time applies an adaptive time step size based on $\mathrm{Co}_\mathrm{tar} = 0.2$ which is embedded in five sub-cycles.

The aim of the investigation is to compare the continuously adjoint sensitivity from Eqn. (\ref{equ:shape_derivative}) against locally evaluated, second order finite difference formulae, viz. $\delta_{\mathrm{u}^* = \tilde{\mathrm{u}}^*} J^* = [J(\tilde{u}_\mathrm{i}^* + \epsilon^* n_\mathrm{i}) - J(\tilde{u}_\mathrm{i}^* - \epsilon^* n_\mathrm{i})] / (2 \, \epsilon^*)$ at several positions $\tilde{u}_\mathrm{i}^*$ of the control $u_\mathrm{i}^*$, cf. Fig. \ref{fig:cylinder_fn_075} (a). Only the upper half of the cylinder is investigated for different magnitudes of the perturbation $\epsilon^* / D^* \in [10^{-4},10^{-5},10^{-6} ]$. The local boundary perturbations are transported based on a Laplacian mesh deformation approach into the (normal) interior domain as well as based on a Gaussian filter with a width of 4 times the discrete surface element width into the (tangential) neighbourhood, \citep{kroger2015cad, kuhl2019decoupling}. 

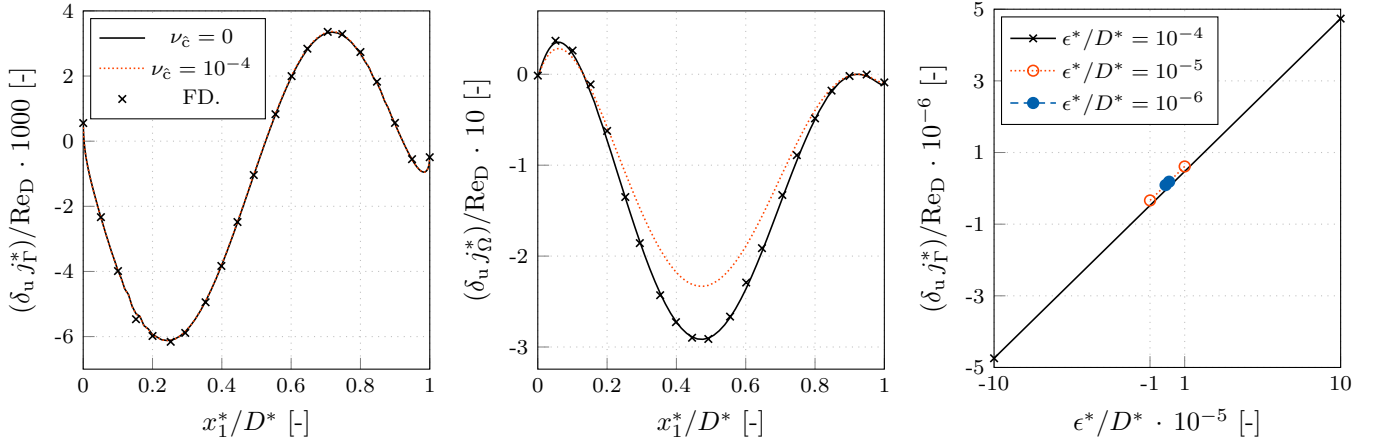
\begin{figure}
\centering
\analytiSolutionPictures
\begin{tikzpicture}
\begin{axis}[
 xlabel style={text width=0.25\textwidth,align=center},
 ylabel style={text width=0.25\textwidth,align=center},
 xlabel={$x_\mathrm{1}^* / D^*$ [-]},
 ylabel={$(\delta_\mathrm{u} \, j_\mathrm{\Gamma}^*) /\mathrm{Re}_\mathrm{D} \cdot 1000$ [-]},
 xmin=0,xmax=1,
 legend style={at={(0.02,0.98)},anchor=north west},
 ymin=-7000,ymax=4000,
 ytick={-8000,-6000,-4000,-2000,0,2000,4000},
 yticklabels={-8,-6,-4,-2,0,2,4},
]
\addplot [line1] table[x expr={\thisrowno{0}},y expr={\thisrowno{1}}]{data/Cylinder_drift_sensitivity.dat};
%\addplot [line2] table[x expr={\thisrowno{0}},y expr={\thisrowno{2}}]{data/Cylinder_drift_sensitivity.dat};
%\addplot [line3] table[x expr={\thisrowno{0}},y expr={\thisrowno{3}}]{data/Cylinder_drift_sensitivity.dat};
%\addplot [line4] table[x expr={\thisrowno{0}},y expr={\thisrowno{4}}]{data/Cylinder_drift_sensitivity.dat};
%\addplot [line5] table[x expr={\thisrowno{0}},y expr={\thisrowno{5}}]{data/Cylinder_drift_sensitivity.dat};
\addplot [line2] table[x expr={\thisrowno{0}},y expr={\thisrowno{6}}]{data/Cylinder_drift_sensitivity.dat};
\addplot [mark1, only marks] table[x expr={\thisrowno{0}},y expr={\thisrowno{2}}]{data/Cylinder_drift_FD.dat};
 
\addlegendentry{$\nu_\mathrm{\hat{c}} = 0$};
\addlegendentry{$\nu_\mathrm{\hat{c}} = 10^{-4}$};
\addlegendentry{FD.}; 
 
\end{axis}
\end{tikzpicture}
\begin{tikzpicture}
\begin{axis}[
 xlabel style={text width=0.25\textwidth,align=center},
 ylabel style={text width=0.25\textwidth,align=center},
 xlabel={$x_\mathrm{1}^* / D^*$ [-]},
 ylabel={$(\delta_\mathrm{u} \, j_\mathrm{\Omega}^*) /\mathrm{Re}_\mathrm{D} \cdot 10$ [-]},
 xmin=0,xmax=1,
 legend style={at={(0.02,0.98)},anchor=north west},
 ytick={-0.3,-0.2,-0.1,0,0.1},
 yticklabels={-3,-2,-1,0,1},
]
\addplot [line1] table[x expr={\thisrowno{0}},y expr={\thisrowno{1}}]{data/Cylinder_inver_sensitivity.dat};
%\addplot [line2] table[x expr={\thisrowno{0}},y expr={\thisrowno{2}}]{data/Cylinder_inver_sensitivity.dat};
%\addplot [line3] table[x expr={\thisrowno{0}},y expr={\thisrowno{3}}]{data/Cylinder_inver_sensitivity.dat};
%\addplot [line4] table[x expr={\thisrowno{0}},y expr={\thisrowno{4}}]{data/Cylinder_inver_sensitivity.dat};
%\addplot [line5] table[x expr={\thisrowno{0}},y expr={\thisrowno{5}}]{data/Cylinder_inver_sensitivity.dat};
\addplot [line2] table[x expr={\thisrowno{0}},y expr={\thisrowno{6}}]{data/Cylinder_inver_sensitivity.dat};
\addplot [mark1, only marks] table[x expr={\thisrowno{0}},y expr={\thisrowno{4}}]{data/Cylinder_inver_FD.dat};
 
%\addlegendentry{$\nu_\mathrm{\hat{c}} = 0$};
%\addlegendentry{$\nu_\mathrm{\hat{c}} = 0 $};
%\addlegendentry{FD.}; 
 
\end{axis}
\end{tikzpicture}
\begin{tikzpicture}
\begin{axis}[
 xlabel style={text width=0.25\textwidth,align=center},
 ylabel style={text width=0.25\textwidth,align=center},
 xlabel={$\epsilon^* / D^* \cdot 10^{-5}$ [-]},
 ylabel={$(\delta_\mathrm{u} \, j_\mathrm{\Gamma}^*) /\mathrm{Re}_\mathrm{D} \cdot 10^{-6}$ [-]},
 xmin=-1E-04,xmax=1E-04,
 legend style={at={(0.02,0.98)},anchor=north west},
 ymin=-5E-06,ymax=5E-06,
 xtick={-1E-04,-1E-05,1E-05,1E-04},
 xticklabels={-10,-1,1,10},
 scaled x ticks = false,
 ytick={-0.000005,-0.000003,-0.000001,0.000001,0.000003,0.000005},
 yticklabels={-5,-3,-1,1,3,5},
 scaled y ticks = false
]
%
%\addplot [line1,mark1] table[x expr={\thisrowno{0}},y expr={\thisrowno{1}}]{data/Cylinder_drift_Linear_Regime_4.dat};
\addplot [line1,mark1] table[x expr={\thisrowno{0}},y expr={\thisrowno{1}}]{data/Cylinder_drift_Linear_Regime_3.dat};
\addplot [line2,mark2] table[x expr={\thisrowno{0}},y expr={\thisrowno{1}}]{data/Cylinder_drift_Linear_Regime_2.dat};
\addplot [line3,mark3] table[x expr={\thisrowno{0}},y expr={\thisrowno{1}}]{data/Cylinder_drift_Linear_Regime_1.dat};

\addlegendentry{$\epsilon^* / D^* = 10^{-4}$};
\addlegendentry{$\epsilon^* / D^* = 10^{-5}$};
\addlegendentry{$\epsilon^* / D^* = 10^{-6}$};

\end{axis}
\end{tikzpicture}
\caption{Submerged cylinder case (Fn=0.75): Continuous as well as discrete finite-difference (FD) based sensitivity derivative along the upper cylinder side for left) a drift functional ($d_\mathrm{i} = [\sqrt(2), \sqrt{2}]^\mathrm{T} / 2$), middle) the target concentration objective ($\Omega_\mathrm{O}^* = [-5D^*,D^*] \times [25 D^*,5 D^*]$) as well as right) three exemplary finite (force functional) system answers at $x_\mathrm{1}^* / D^* = 1/4$.}
\label{fig:fd_study}
\end{figure}

Fig. \ref{fig:fd_study} depicts the results of one adjoint sensitivity evaluation for two exemplary objective: A boundary based force (left) objective into the direction $d_\mathrm{i} = [\sqrt(2), \sqrt{2}]^\mathrm{T} / 2$ as well as a volumetric target concentration (center) objective with a habitat along $\Omega_\mathrm{O}^* = [-5D^*,D^*] \times [25 D^*,5 D^*]$. For each objective, two adjoint system are constructed, employing either no ($\nu_\mathrm{\hat{c}} = 0$) or a moderate non-zero ($\nu_\mathrm{\hat{c}} = 10^{-4}$) synthetic viscosity.  
% resulting in four (backward) simulations.
%
Additionally, the results for several FD-studies are marked at 21 discrete positions obtained from 42 additional (forward) simulations using the intermediate perturbation size $\epsilon^* / D^* =10^{-5}$.  The consistent adjoint sensitivities agree almost perfectly with the discrete FD results. However, a break of dual consistency due to a  non-zero synthetic viscosity $ \nu_{\hat c}$  causes the sensitivity results to deviate for the concentration-based objective, especially in regions of high sensitivity. Since the volume functional depends exclusively on the primal concentration distribution, a manipulation of the dual concentration field is much more critical compared to the boundary based force objective. This is an important finding for practical marine engineering applications,  which often refer to the minimization of boundary based force objectives.
%Additionally, potential roots of the shape derivative are partly ignored.
%
Mind that the linearity of the FD-analysis has been verified using a the sequence of three perturbation magnitudes for both functionals. An exemplary documentation of the systems linear answer is displayed in the right graph of Figure \ref{fig:fd_study}, which refers to 
%for the three considered perturbations
an exemplary surface position $x_\mathrm{1}^* / D^* = 1/4$ for the force functional.

\newpage
\section{Application}
\label{sec:application}
\subsection{Submerged Hydrofoil}
\begin{figure}
\centering
\subfigure[]{
\centering
\includegraphics[scale=1]{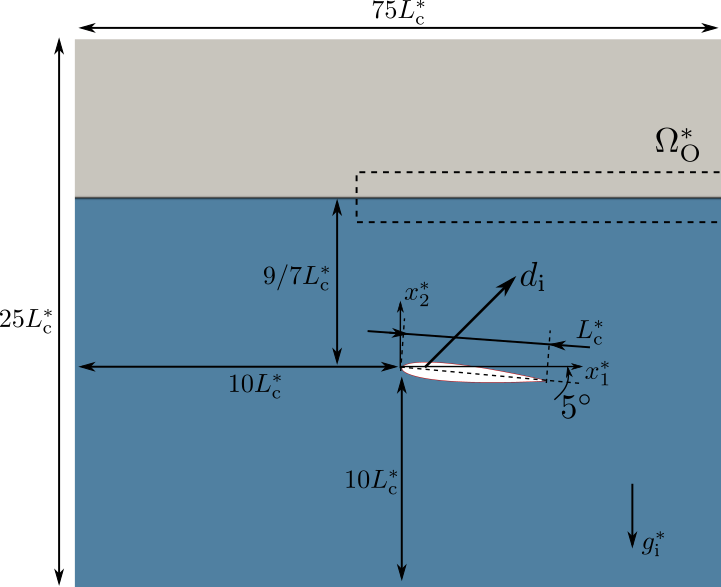}
}
\hspace{1cm}
\subfigure[]{
\centering
\includegraphics[scale=1]{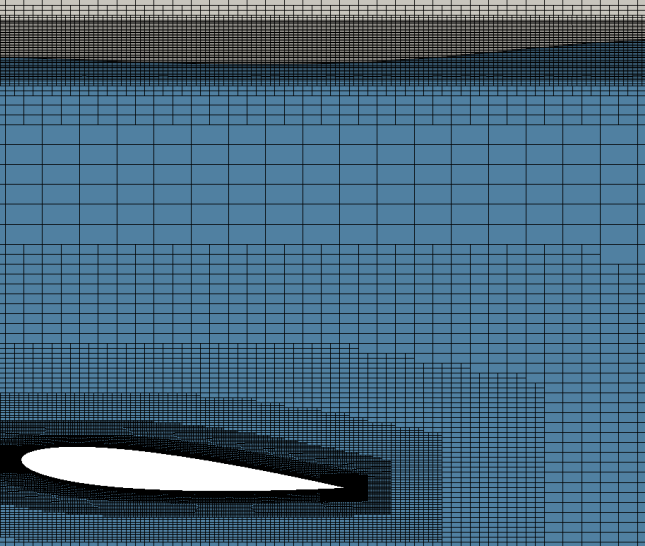}
}
\caption{Submerged hydrofoil case ($\mathrm{Re}_\mathrm{D} = \SI{144 855}{}$, $\mathrm{Fn}=0.567$): (a) Schematic drawing of the initial configuration and (b) unstructured numerical grid around the foil and the free surface.}
\label{fig:dunca_foil}
\end{figure}

This section examines the introduced adjoint two-phase system on a submerged NACA0012 hydrofoil at  $\SI{5}{\degree}$ incidence and aims at a deeper insight on the influence of the modified adjoint concentration equation, both on local and integral level. 
Figure \ref{fig:dunca_foil} (a) provides a sketch of the experiment reported by Duncan \citep{duncan1981experimental,duncan1983breaking}. The chord length to submergence ratio at the leading edge of the foil reads $L_\mathrm{c}^*/L^*= 7/9$. The study is performed for a turbulent flow at $\mathrm{Re} = v_\mathrm{1}^* L_\mathrm{c}^*/\nu_\mathrm{b}^* = \SI{144 855}{}$ and $\mathrm{Fn} = v_\mathrm{1}^*/\sqrt{G^* L^*} = \SI{0.567}{}$, based on the gravitational acceleration $G^*$, the inflow velocity $v_\mathrm{1}^*$ and the kinematic viscosity of the water $\nu_\mathrm{b}^*$.  The two-dimensional domain has a length and height of $75 \, L_\mathrm{c}^*$ and $25 \, L_\mathrm{c}^*$, where the inlet and bottom boundaries are located 10 chord-lengths away from the origin. A dimensionless wave length of $\lambda = \lambda^*/ L^* = 2 \, \pi \, \mathrm{Fn}^2 = 2.0193$ is expected. 

\begin{figure}
\centering
\analytiSolutionPictures
\begin{tikzpicture}
\begin{axis}[
 ylabel style={text width=0.25\textwidth,align=center},
 xlabel={$ x_\mathrm{1}^* / L_\mathrm{c}^* $ [-]},
 ylabel={$ x_\mathrm{2,FS}^* / L_\mathrm{c}^* $ [-]},
 xmin=-0.0,xmax=7.0,
 ymin=-0.06,ymax=0.06,
 legend style={at={(0.02,0.98)},anchor=north west},
 ytick={-0.06,-0.03,0,0.03,0.06},
 yticklabels={-0.06,-0.03,0,0.03,0.06},
 scaled y ticks = false
]

\addplot [line1] table[x expr={\thisrowno{0}},y expr={\thisrowno{1}}]{data/Duncan_CICSAM.dat};
\addplot [line2] table[x expr={\thisrowno{0}},y expr={\thisrowno{1}}]{data/Duncan_HRIC.dat};  
\addplot [mark1,mark repeat=1,only marks] table[x expr={\thisrowno{0}},y expr={\thisrowno{1}}]{data/Duncan_Foil_Experimental_Data.dat}; 

\addlegendentry{HRIC};
\addlegendentry{CICSAM};
\addlegendentry{Duncan (1981)};
 
\end{axis}
\end{tikzpicture}
\begin{tikzpicture}
\begin{axis}[
 ylabel style={text width=0.25\textwidth,align=center},
 xlabel={$ x_\mathrm{1}^* / L_\mathrm{c}^* $ [-]},
 ylabel={$g^* / \mathrm{max}(g_\mathrm{\nu_\mathrm{\hat{c}}=10^{-10}}^*)$ [-]},
 xmin=0.0,xmax=1.0,
 ymin=0.0,ymax=1.0,
 legend style={at={(0.02,0.98)},anchor=north west},
 scaled y ticks = false
]

\addplot [line3] table[x expr={\thisrowno{0}},y expr={\thisrowno{4}}]{data/Duncan_HRIC_Drag.dat};
\addplot [line4] table[x expr={\thisrowno{0}},y expr={\thisrowno{7}}]{data/Duncan_HRIC_Drag.dat};
\addplot [line5] table[x expr={\thisrowno{0}},y expr={\thisrowno{11}}]{data/Duncan_HRIC_Drag.dat};
\addplot [line6] table[x expr={\thisrowno{0}},y expr={\thisrowno{14}}]{data/Duncan_HRIC_Drag.dat};
\addplot [line1] table[x expr={\thisrowno{0}},y expr={\thisrowno{1}}]{data/Duncan_HRIC_Drag.dat};
\addplot [line2] table[x expr={\thisrowno{0}},y expr={\thisrowno{3}}]{data/Duncan_HRIC_Drag.dat};

%\addlegendentry{$\rho_\mathrm{\Delta} = \mu_\mathrm{\Delta} = 0$};
%\addlegendentry{frozen C};
\addlegendentry{$\nu_\mathrm{\hat{c}} = 10^{-10}$};
\addlegendentry{$\nu_\mathrm{\hat{c}} = 10^{-07}$};
\addlegendentry{$\nu_\mathrm{\hat{c}} = 10^{-03}$};
\addlegendentry{$\nu_\mathrm{\hat{c}} = 10^{-00}$};
 
\end{axis}
\end{tikzpicture}
\begin{tikzpicture}
\begin{axis}[
 ylabel style={text width=0.25\textwidth,align=center},
 xlabel={$ x_\mathrm{1}^* / L_\mathrm{c}^* $ [-]},
 ylabel={$g^* / \mathrm{max}(g_\mathrm{\nu_\mathrm{\hat{c}}=10^{-10}}^*)$ [-]},
 xmin=0.0,xmax=0.5,
 ymin=0.0,ymax=0.425,
 legend style={at={(0.98,0.98)},anchor=north east},
 scaled y ticks = false
]

\addplot [line1] table[x expr={\thisrowno{0}},y expr={\thisrowno{1}}]{data/Duncan_HRIC_Drag_Short.dat};
\addplot [line2] table[x expr={\thisrowno{0}},y expr={\thisrowno{3}}]{data/Duncan_HRIC_Drag_Short.dat};
\addplot [line3] table[x expr={\thisrowno{0}},y expr={\thisrowno{4}}]{data/Duncan_HRIC_Drag_Short.dat};
\addplot [line4] table[x expr={\thisrowno{0}},y expr={\thisrowno{7}}]{data/Duncan_HRIC_Drag_Short.dat};
\addplot [line5] table[x expr={\thisrowno{0}},y expr={\thisrowno{11}}]{data/Duncan_HRIC_Drag_Short.dat};
\addplot [line6] table[x expr={\thisrowno{0}},y expr={\thisrowno{14}}]{data/Duncan_HRIC_Drag_Short.dat};
%\addplot [line6] table[x expr={\thisrowno{0}},y expr={\thisrowno{11}}]{data/Duncan_HRIC_Drag_Short.dat};
%\addplot [line6] table[x expr={\thisrowno{0}},y expr={\thisrowno{14}}]{data/Duncan_HRIC_Drag_Short.dat};

\addlegendentry{$\rho_\mathrm{\Delta} = \mu_\mathrm{\Delta} = 0$};
\addlegendentry{frozen C};
%\addlegendentry{$\nu_\mathrm{\hat{c}} = 10^{-10}$};
%\addlegendentry{$\nu_\mathrm{\hat{c}} = 10^{-07}$};
%\addlegendentry{$\nu_\mathrm{\hat{c}} = 10^{-03}$};
%\addlegendentry{$\nu_\mathrm{\hat{c}} = 10^{-00}$};
 
\end{axis}
\end{tikzpicture}
\caption{Submerged hydrofoil case ($\mathrm{Re}_\mathrm{L} = \SI{144 855}{}$, $\mathrm{Fn}=0.567$): (Left) Comparison of predicted wave elevation as well as normalised shape gradients ($\epsilon^* / L_\mathrm{c}^* = 1 / 30$) along the complete (middle) and front half (right) suction side for various adjoint systems ranging from identical fluid properties ($\rho_\mathrm{\Delta} = 0 = \mu_\mathrm{\Delta}$) over a frozen concentration approach ($\alpha = c$) to variable synthetic viscosity's.}
\label{fig:duncan_drag_results}
\end{figure}
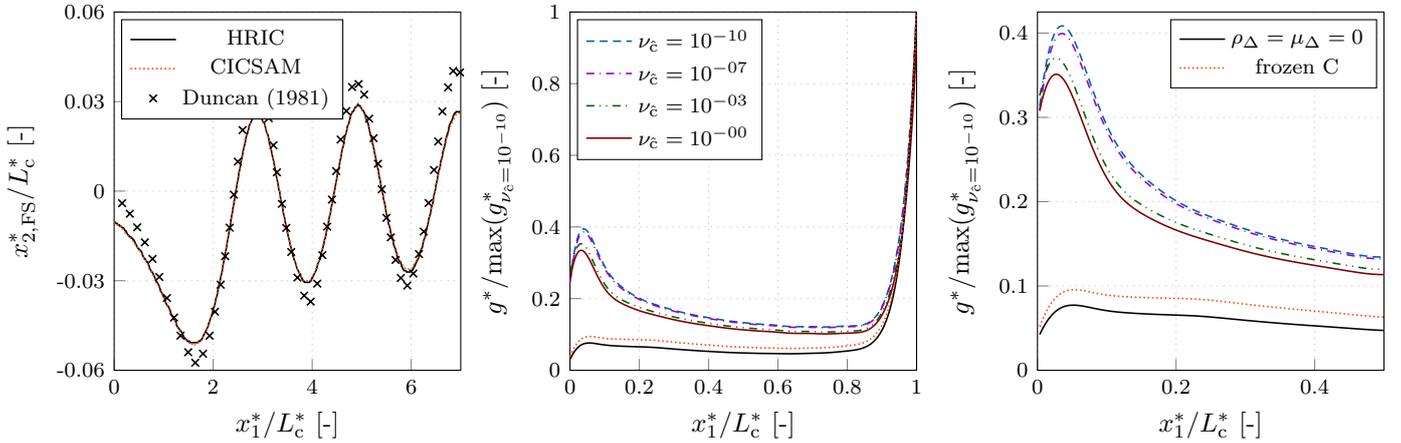

The utilized unstructured numerical grid is displayed in Fig. \ref{fig:dunca_foil} (b) and consists of approximately $\SI{150000}{}$ control volumes. The fully turbulent simulations employ a wall-function based $k-\omega$ SST model \citep{menter2003ten} and convective terms for momentum and turbulence are approximated using the QUICK scheme. At the inlet, velocity and concentration values are prescribed, slip walls are used along the top and bottom boundaries and a hydrostatic pressure boundary is employed along the outlet. The wall normal distance of the first grid layer reads $y^+ \approx \SI{0.1}{}$ and the free surface refinement employs approximately  $\delta x_\mathrm{1}^* / \lambda^* = 1/100$ cells in the longitudinal as well as $\delta x_\mathrm{2}^* / \lambda^* = 1/100$ cells in the normal direction. According to Alg. (\ref{alg:sub_cycling}) the integration in pseudo time applies an adaptive time step size based on $\mathrm{Co}_\mathrm{tar} = 0.4$ embedded in five sub-cycles.
%($N=5$).
%
We aim at a systematic investigation of the influence of the approximation of the adjoint concentration equation on the shape sensitivity (\ref{equ:shape_derivative}).
Fig. \ref{fig:duncan_drag_results} (a) displays the wave elevation predicted by the HRIC and the CICSAM approach in comparison with experimental data. Minor predictive differences are observed, with the CICSAM method being slightly closer to the experimental results for the employed computational set up. An improved predictive accuracy 
can be obtained from local mesh refinement, as for example shown by Manzke \citep{manzke2018development} and Wackers et al. \citep{wackers2011free} for adaptive free surface refinement procedures.

%\nkuehl{Maybe skip right figure and compare integrated sensitivity}

The adjoint system is solved for a force and inverse concentration  
%(\ref{equ:general_objective}) 
criterion considering a steady primal solution. 
%Convergence for adjoint system is a fairly challenging task and might possibly be undoable, which is why the solution is stabilized. 
The adjoint concentration equation is supplemented by the heuristic diffusive term, 
% analogous to the investigation of the planar Couette flow in Sec.  \ref{sec:veri_stud}, 
cf. Eqn. (\ref{equ:fina_adjoi_conce_dile}),  
and the related influence in the predicted sensitivities is investigated for the considered hydrofoil flow.
Note that the test barely converges for $\nu_\mathrm{\hat{c}} \rightarrow 0$, hence the adjoint time step size $\Delta t^\mathrm{adj}$ is significantly reduced by a factor $\Delta t^\mathrm{adj} / \Delta t^\mathrm{pri} = 100$ compared to the primal time step size $\Delta t^\mathrm{pri}$.
%
%In order to achieve convergence of the system even for small synthetic viscosity's,  %For more complex 3D applications, convergence can often no longer be achieved even at very small time increments.
%
During the simulations we noticed that the numerical stability is mainly affected by the Froude term in Eqn. (\ref{equ:fina_adjoi_conce_dile}) which in turn is particularly relevant for a consistent sensitivity as shown by Kroger et al. \citep{kroger2018adjoint} by means of several finite-difference studies.

As indicated in Fig. \ref{fig:dunca_foil}, the foil is investigated for its optimisation potential w.r.t a drift functional, i.e. 50\% drag, 50\% lift or $d_\mathrm{i} = \sqrt{2} / 2$ in  (\ref{equ:general_objective}). According to Tab. (\ref{tab:bound_condi}), the adjoint velocity is prescribed to $\hat{v}_\mathrm{i} = -d_\mathrm{i}$ along the foil. Thus, the adjoint velocity sees the gravitational vector $\hat{v}_\mathrm{i} \, g_\mathrm{i} \neq 0$ in the foil vicinity and we therefore force the Froude term to be active in the sensitive region.
Moreover, the noisy shape derivative (sensitivity) is explicitly transformed into the shape gradient based on the Laplace-Beltrami metric \citep{kroger2015cad, schulz2016computational}. The latter exclusively operates in the tangent plane, viz. $g^* - \epsilon^{*2} \Delta_\mathrm{\Gamma}^* g^* = s^*$ where $g^* = \partial j_\mathrm{\Gamma}^* / \partial u^*$, $s^* = \delta_\mathrm{u} j_\mathrm{\Gamma}^*$ and $\Delta_\mathrm{\Gamma}^* = \partial^2 / \partial x_\mathrm{k}^{*2} - \partial^2 / \partial n^{*2}$ represent the shape gradient, sensitivity derivative as well as the Laplace-Beltrami operator respectively. Dimensional consistency is ensured by $\epsilon^*$,  interpret-able as the filter width of a Gaussian \citep{stuck2011adjoint}. We employ $\epsilon^* / L_\mathrm{c}^* = 1 / 30$.
Various adjoint systems are constructed for the HRIC solution: Starting from a consistent ($\nu_\mathrm{\hat{c}} = 0$) system, the synthetic viscosity $\nu_{\tilde c}$ is carefully increased. Finally, simulations are performed for a frozen concentration approach and an approach based on identical fluid properties ($\Delta \rho = 0 = \Delta \mu$) to better understand and assess the results.
The resulting normalized shape gradients are depicted by Fig. \ref{fig:duncan_drag_results} (center) along the suction side of the foil. Only small qualitative differences are recognized and no changes in sign are observed. The latter would be  disadvantageous in conjunction with a steepest descent approach.
%
%Due to the singularity in the primal shear at the trailing edge, Fig. \ref{fig:duncan_drag_results} (right) depicts the shape gradient along the first half of the suction side only. Even for comparatively large violations of dual consistency, the resulting shape gradient remains qualitatively unchanged.
Only a neglect of the multi phase information within the adjoint momentum equation ($\alpha = c$) results in significant distortions of the shape gradient. Although there is still no sign change, such a frozen C approach now resembles the shape derivative of a single-phase flow or a two-phase flow with the identical material properties. The latter is obtained based on a new primal flow for $\Delta \rho = 0 = \Delta \mu$.
From theses results, we conclude that the impact of the synthetic viscosity is negligible for force objectives as formerly noted in \citep{kroger2016numerical,kroger2018adjoint}.

%An alternative and 
A possibly more crucial objective w.r.t 
consistency 
%a consistent approximation of the adjoint concentration equation 
might be the inverse concentration objective (\ref{equ:special_objective}). In this case, Equ. (\ref{equ:general_objective}) speaks through the adjoint concentration only and enters the adjoint system on the right-hand side of $\hat{C}$ which makes the frozen C approach obsolete. The habitat of $j_\mathrm{\Omega}^*$ reads $\Omega_\mathrm{O}^* = [-L_\mathrm{c}^*,L_\mathrm{c}^*] \times [20L_\mathrm{c}^*,2L_\mathrm{c}^*]$, where the 2D coordinate system originates in the leading edge as depicted in Fig. \ref{fig:dunca_foil}.
\begin{figure}
\centering
\analytiSolutionPictures
\begin{tikzpicture}
\begin{axis}[
 ylabel style={text width=0.25\textwidth,align=center},
 xlabel={$ x_\mathrm{1}^* / L_\mathrm{c}^* $ [-]},
 ylabel={$g^* / \mathrm{max}(g_\mathrm{\nu_\mathrm{\hat{c}}=10^{-10}}^*)$ [-]},
 xmin=0.0,xmax=1.0,
 ymin=-1.0,ymax=0.0,
 legend style={at={(0.02,0.02)},anchor=south west},
 scaled y ticks = false
]

%\addplot [line1] table[x expr={\thisrowno{0}},y expr={\thisrowno{1}}]{data/Duncan_HRIC_Inverse.dat};
%\addplot [line1] table[x expr={\thisrowno{0}},y expr={\thisrowno{2}}]{data/Duncan_HRIC_Inverse.dat};  

\addplot [line1] table[x expr={\thisrowno{0}},y expr={\thisrowno{3}}]{data/Duncan_HRIC_Inverse.dat};
\addplot [line2] table[x expr={\thisrowno{0}},y expr={\thisrowno{4}}]{data/Duncan_HRIC_Inverse.dat};  
\addplot [line3] table[x expr={\thisrowno{0}},y expr={\thisrowno{5}}]{data/Duncan_HRIC_Inverse.dat};
\addplot [line4] table[x expr={\thisrowno{0}},y expr={\thisrowno{6}}]{data/Duncan_HRIC_Inverse.dat};
\addplot [line5] table[x expr={\thisrowno{0}},y expr={\thisrowno{7}}]{data/Duncan_HRIC_Inverse.dat};
\addplot [line6] table[x expr={\thisrowno{0}},y expr={\thisrowno{8}}]{data/Duncan_HRIC_Inverse.dat};

%\addplot [line6] table[x expr={\thisrowno{0}},y expr={\thisrowno{9}}]{data/Duncan_HRIC_Inverse.dat};
%\addplot [line5] table[x expr={\thisrowno{0}},y expr={\thisrowno{10}}]{data/Duncan_HRIC_Inverse.dat};
%\addplot [line6] table[x expr={\thisrowno{0}},y expr={\thisrowno{11}}]{data/Duncan_HRIC_Inverse.dat};

\addlegendentry{$\nu_\mathrm{\hat{c}} = 10^{-8}$};
\addlegendentry{$\nu_\mathrm{\hat{c}} = 10^{-7}$};
\addlegendentry{$\nu_\mathrm{\hat{c}} = 10^{-6}$}; 
 
\end{axis}
\end{tikzpicture}
\begin{tikzpicture}
\begin{axis}[
 ylabel style={text width=0.25\textwidth,align=center},
 xlabel={$ x_\mathrm{1}^* / L_\mathrm{c}^* $ [-]},
 ylabel={$g^* / \mathrm{max}(g_\mathrm{\nu_\mathrm{\hat{c}}=10^{-10}}^*)$ [-]},
 xmin=0.0,xmax=0.5,
 ymin=-0.5,ymax=0.0,
 legend style={at={(0.98,0.02)},anchor=south east},
 scaled y ticks = false
]

%\addplot [line4] table[x expr={\thisrowno{0}},y expr={\thisrowno{5}}]{data/Duncan_HRIC_Inverse_Short.dat};  
%\addplot [line5] table[x expr={\thisrowno{0}},y expr={\thisrowno{1}}]{data/Duncan_HRIC_Inverse_Short.dat};
%\addplot [line6] table[x expr={\thisrowno{0}},y expr={\thisrowno{2}}]{data/Duncan_HRIC_Inverse_Short.dat};
\addplot [line4] table[x expr={\thisrowno{0}},y expr={\thisrowno{6}}]{data/Duncan_HRIC_Inverse_Short.dat};
\addplot [line5] table[x expr={\thisrowno{0}},y expr={\thisrowno{7}}]{data/Duncan_HRIC_Inverse_Short.dat};
\addplot [line6] table[x expr={\thisrowno{0}},y expr={\thisrowno{8}}]{data/Duncan_HRIC_Inverse_Short.dat};
\addplot [line1] table[x expr={\thisrowno{0}},y expr={\thisrowno{3}}]{data/Duncan_HRIC_Inverse_Short.dat};
\addplot [line2] table[x expr={\thisrowno{0}},y expr={\thisrowno{4}}]{data/Duncan_HRIC_Inverse_Short.dat};
\addplot [line3] table[x expr={\thisrowno{0}},y expr={\thisrowno{5}}]{data/Duncan_HRIC_Inverse_Short.dat};
%\addplot [line3] table[x expr={\thisrowno{0}},y expr={\thisrowno{9}}]{data/Duncan_HRIC_Inverse_Short.dat};
%\addplot [line3] table[x expr={\thisrowno{0}},y expr={\thisrowno{10}}]{data/Duncan_HRIC_Inverse_Short.dat};
%\addplot [line3] table[x expr={\thisrowno{0}},y expr={\thisrowno{11}}]{data/Duncan_HRIC_Inverse_Short.dat};

\addlegendentry{$\nu_\mathrm{\hat{c}} = 10^{-5}$};
\addlegendentry{$\nu_\mathrm{\hat{c}} = 10^{-4}$};
\addlegendentry{$\nu_\mathrm{\hat{c}} = 10^{-3}$}; 
 
\end{axis}
\end{tikzpicture}
\begin{tikzpicture}
\begin{axis}[
 ylabel style={text width=0.25\textwidth,align=center},
 xlabel={$\nu_\mathrm{\hat{c}}$ [-]},
 ylabel={$\epsilon_{\mathrm{DD, adHRIC}}$ [-]},
 xmin=1E-10,xmax=1E-00,
 ymin=1E-08,ymax=1E-00,
 legend style={at={(0.02,0.02)},anchor=south west},
 xmode=log,
 ymode=log,
 scaled y ticks = false
]

\addplot [line1, mark1] table[x expr={\thisrowno{0}},y expr={\thisrowno{1}}]{data/Duncan_HRIC_Inverse_DD2UD_Error.dat}; 
\addplot [line1, mark8] table[x expr={\thisrowno{2}},y expr={\thisrowno{3}}]{data/Duncan_HRIC_Inverse_DD2UD_Error.dat};
\addplot [line1, mark9] table[x expr={\thisrowno{2}},y expr={\thisrowno{4}}]{data/Duncan_HRIC_Inverse_DD2UD_Error.dat};
\addplot [line1, mark10] table[x expr={\thisrowno{2}},y expr={\thisrowno{5}}]{data/Duncan_HRIC_Inverse_DD2UD_Error.dat};

\addlegendentry{Comp.};
\addlegendentry{$\mathcal{O}((-\nu_\mathrm{\hat{c}}/1)^{-1})$};
\addlegendentry{$\mathcal{O}((-\nu_\mathrm{\hat{c}}/2)^{-1})$};
\addlegendentry{$\mathcal{O}((-\nu_\mathrm{\hat{c}}/4)^{-1})$};
 
\end{axis}
\end{tikzpicture}
\caption{Submerged hydrofoil case ($\mathrm{Re}_\mathrm{L} = \SI{144 855}{}$, $\mathrm{Fn}=0.567$): Normalised shape gradients ($\epsilon^* / L_\mathrm{c}^* = 1 / 30$) along the complete (left) and front half (middle) suction side as well as the error (right) between a consistent adjoint HRIC and a pure DD approximation of the adjoint convection of $\hat{c}$ based on the maximum norm ($\epsilon_{\mathrm{DD, adHRIC}} = \mathrm{max}(|(g_\mathrm{DD}^* - g_\mathrm{adHRIC}^*)/ g_\mathrm{adHRIC}^*|)$) for various magnitudes of the synthetic viscosity.}
\label{fig:duncan_inverse_results}
\end{figure}
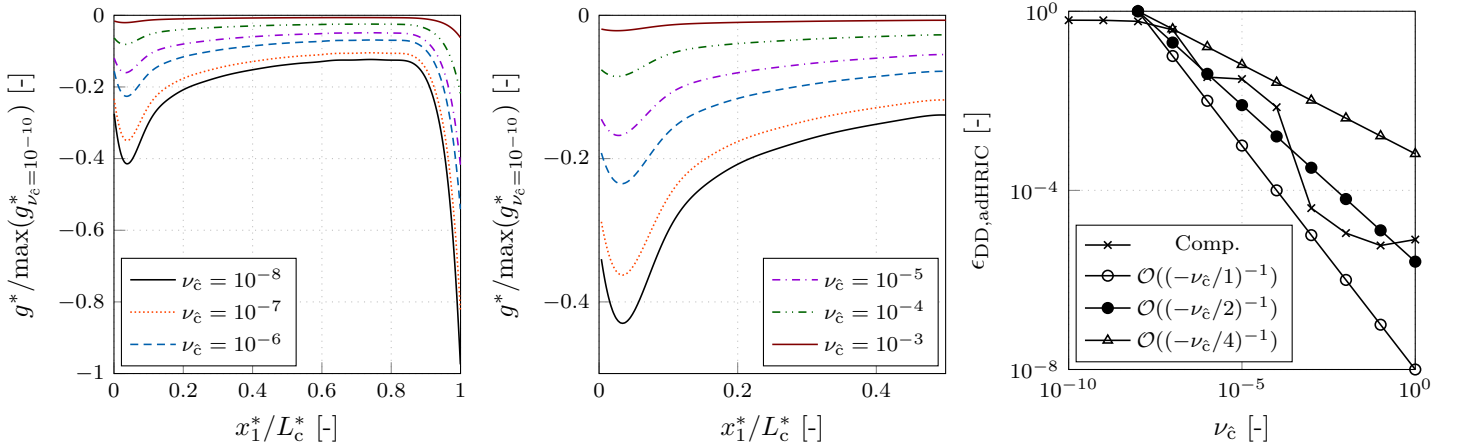
%\nkuehl{Compare Integrated Sensitivity on right side}
Again, various adjoint systems are constructed, which differ only in the amplitude of the synthetic viscosity and result in the normalised shape gradients depicted on the left (center) part in Fig. \ref{fig:duncan_inverse_results} along the complete (front half) suction side. Analogous to the force functional, the increase in synthetic viscosity only leads to a damping of the gradient, which in turn could be treated by an appropriate step-size within a gradient based optimization procedure.

\begin{figure}
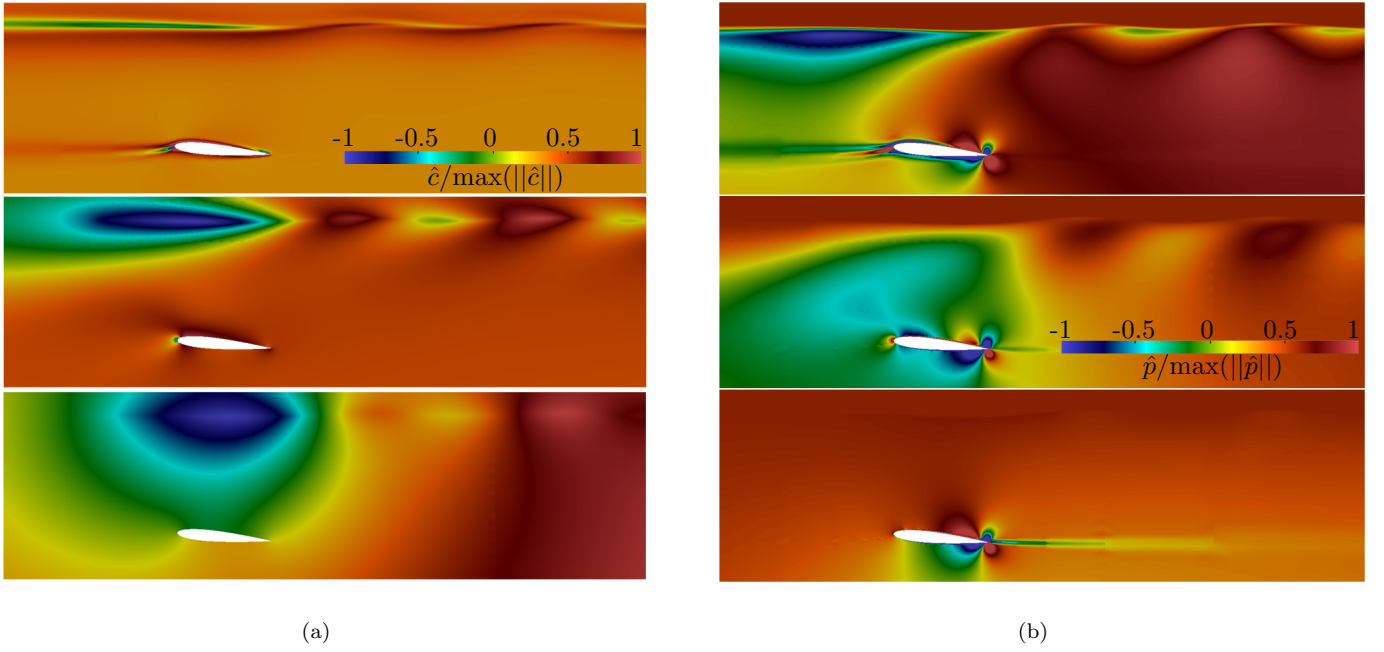

\centering
\subfigure[]{
\input{./tikz/Duncan_adc.tikz}
}
\subfigure[]{
\input{./tikz/Duncan_adp.tikz}
}
\caption{Submerged hydrofoil case ($\mathrm{Re}_\mathrm{L} = \SI{144 855}{}$, $\mathrm{Fn}=0.567$): Normalised left) adjoint concentration and right) adjoint pressure distribution for the inverse concentration objective around the foil for synthetic viscosity increasing from $\nu_\mathrm{\hat{c}} = 10^{-4}$ (top) over $\nu_\mathrm{\hat{c}} = 1$ (middle) to $\nu_\mathrm{\hat{c}} = 10^{4}$ (bottom). The adjoint concentration within the adjoint momentum equation is shifted towards to background fluid ($\alpha = 1$).}
\label{fig:duncan_adc}
\end{figure}

When attention is directed to practical marine engineering applications, the computational effort associated with time stepping compressive primal approximations is substantial. As outlined above, the related efforts increase for an adjoint system. Means to reduce this effort whilst still retaining reliable sensitivity information are appreciated. To address this aspect, simulations were repeated using simple DD approximations of the convective transport in the adjoint concentration equation,
%
%In addition, all simulations are carried out once again: I
instead of swapping from DD to UD along the interface region in line with the adjoint HRIC approach.
%Eqn. (\ref{equ:adjoint_hric_principal}), an additional break of dual consistency is introduced by employing a DD approximation also along the free surface as formerly suggested by Kr\"oger et al. \citep{kroger2018adjoint}. 
This approach was formerly suggested by Kr\"oger et al. \citep{kroger2018adjoint} and circumvents the time-step size dependency of the adjoint system (\ref{equ:adjoint_concentration_matrix}), 
which would basically allow steady adjoint simulations. However, for the current investigation the integration within the pseudo time remains unchanged in order to support the comparison of results. The resulting shape gradients again only differ quantitatively. Hence, their deviation w.r.t a consistent adjoint HRIC approximation is estimated in terms of the maximum norm $\epsilon_{\mathrm{DD, adHRIC}} = \mathrm{max}(||(g_\mathrm{DD}^* - g_\mathrm{adHRIC}^*)/ g_\mathrm{adHRIC}^*||)$ and plotted in Fig. \ref{fig:duncan_inverse_results} (right) over the employed synthetic viscosity. 
The deviation is large for a small synthetic viscosity $\nu_{\hat c}$ and drops significantly for larger values of $\nu_{\hat c}$. Its evolution is located in a corridor limited by $\mathcal{O}(-1/\nu_\mathrm{\hat{c}})$ and $\mathcal{O}(-4/\nu_\mathrm{\hat{c}})$, which results from 
$\nu_{\hat c}$ exceeding the numerical viscosity inherent to any dual (primal) downwind (upwind) biased interpolation method.
%, viz $\nu_\mathrm{ij}^\mathrm{DD,F} = -[\lambda \Delta (-x_\mathrm{j}) v_\mathrm{i}]^\mathrm{F}$ ($\nu_\mathrm{ij}^\mathrm{UD,F} = [\lambda \Delta x_\mathrm{j} v_\mathrm{i}]^\mathrm{F}$) represents an estimation of the tensorial numerical diffusion. Here $v_\mathrm{i}^\mathrm{F}$ denotes the velocity at the face center, $\Delta x_\mathrm{j}$ refers to the connecting vector from the downstream (upstream) to the upstream (downstream) adjacent cell center and $\lambda \Delta x_\mathrm{j}^\mathrm{F}$ approximates the distance between the face and the downstream (upstream) cell. Depending on the time discretization scheme, the related error might be supplemented by additional terms resulting from a modified equation analysis.
%
The diffusive influence of synthetic viscosity on the adjoint concentration field is plotted in Fig. \ref{fig:duncan_adc} (a) for different magnitudes of the synthetic viscosity. Increasing values smear the initially parabolic into an elliptical field.
%, which is underlined by the main diagonal of the characteristic system matrix of $\hat{C}$ (cf. Eqn. (\ref{equ:fina_adjoi_conce_dile})): $A_\mathrm{ii}^\mathrm{\hat{c}} = 0 \rightarrow A_\mathrm{ii}^\mathrm{\hat{c}} = - \nu_\mathrm{\hat{c}}$. 
Furthermore, Fig. \ref{fig:duncan_adc} (b) confirms the observations of the 1D Couette model from Sec. (\ref{sec:couette}), according to which increased synthetic viscosity suppresses the jump of adjoint pressure along the free surface and thus stabilizes the  numerical procedure. Information of the adjoint concentration within the adjoint momentum equation is active in the background fluid ($\alpha = 1$).

%\nkuehl{Adjoint Rhie Chow factor of 5 so far. Maybe increase to 10.}

\newpage
\subsection{Underwater Vehicle}
\begin{figure}
\centering
\subfigure[]{
\centering
\includegraphics[scale=1]{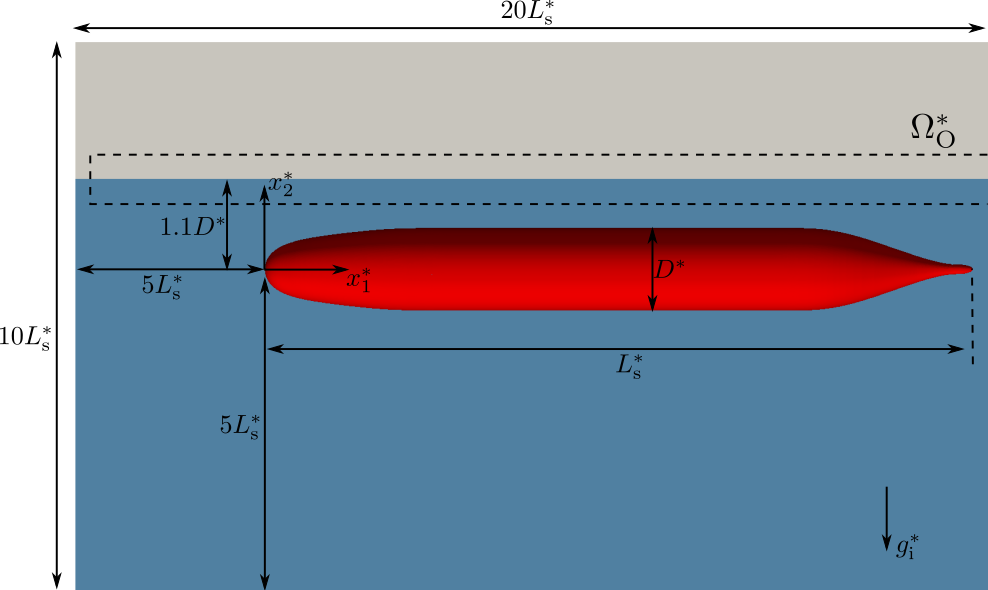}
}
\hspace{1cm}
\subfigure[]{
\centering
\includegraphics[scale=1]{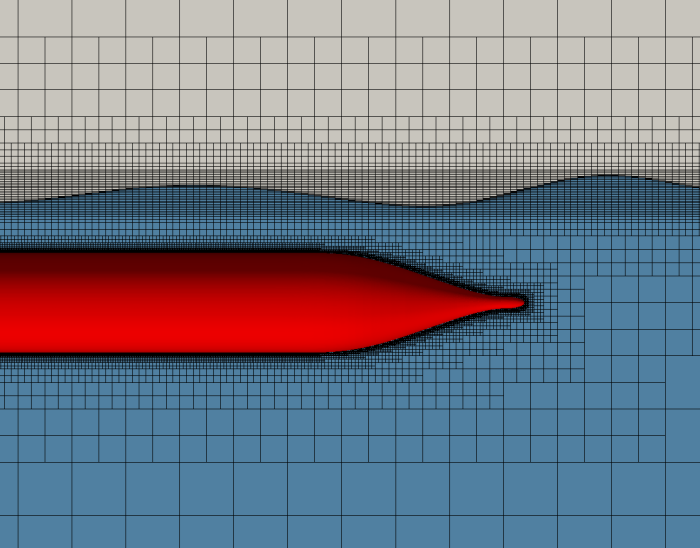}
}
\caption{Submerged DARPA SUBOFF case ($\mathrm{Re}_\mathrm{L} = \SI{8 542 550}{}$, $\mathrm{Fn}=0.3$): (a) Schematic drawing of the initial configuration and (b) unstructured numerical grid around the generic underwater vehicle and the free surface.}
\label{fig:darpa_suboff}
\end{figure}

This final section investigates the influence of the synthetic viscosity on a submerged generic DARPA (Defense Advanced Research Projects Agency) SUBOFF geometry as described in \citep{groves1989geometric} without appendages. The DARPA SUBOFF case is commonly used during studies that focus on e.g. the propulsion and manoeuvring of submarines deeply submerged or in the vicinity of the free surface \citep{wang2020experiments,daum2017effective,chase2013submarine}. When operating close to the free surface, the wave field induced by the interaction of the dynamic pressure field with the free surface is unfavourable in terms of wave resistance as well as the signature of the submarine.
The generic hull basically consists of three components: a bow ($0 \leq x_\mathrm{1}^* \leq 2  D^*$), middle body ($2 D^* \leq x_\mathrm{1}^* \leq 0.745 L_\mathrm{s}^*$) as well as an after body including cap ($0.745 L_\mathrm{s}^* \leq x_\mathrm{1}^* \leq L_\mathrm{s}^*$), measured from the front tip where the coordinate system originates, cf. Fig. (\ref{fig:darpa_suboff}) a).  $D^*$ represents the maximum body diameter along the middle body and its ratio w.r.t. submergence reads $L^*/D^*= 1.1$.
The study is performed for a turbulent flow at $\mathrm{Re}_\mathrm{L} = v_\mathrm{1}^* L_\mathrm{s}^*/\nu_\mathrm{b}^* = \SI{8 542 550}{}$ and $\mathrm{Fn} = v_\mathrm{1}^*/\sqrt{G^* L_\mathrm{s}^*} = \SI{0.3}{}$, based on the gravitational acceleration $G^*$, the inflow velocity $v_\mathrm{1}^*$ and the kinematic viscosity of the water $\nu_\mathrm{b}^*$. The three-dimensional domain has a length, height and width of $20 \, L_\mathrm{s}^*$, $10 \, L_\mathrm{s}^*$ as well as $5 \, L_\mathrm{s}^*$, where the inlet, bottom and outer boundaries are located 5 geometry-lengths away from the origin. A dimensionless wave length of $\lambda = \lambda^*/ L^* = 2 \, \pi \, \mathrm{Fn}^2 = 4.4$ is expected and the wave elevation w.r.t still water should be minimized, viz. $\Omega_\mathrm{O}^* = [-L_\mathrm{s}^*,L_\mathrm{s}^*/10,0] \times [5\, L_\mathrm{s}^*,1.5\, L_\mathrm{s}^*,5\, L_\mathrm{s}^*]$, by modifying only the middle body of the underwater vehicle while conserving its displacement.

\begin{figure}
\centering
\includegraphics[scale=1.0]{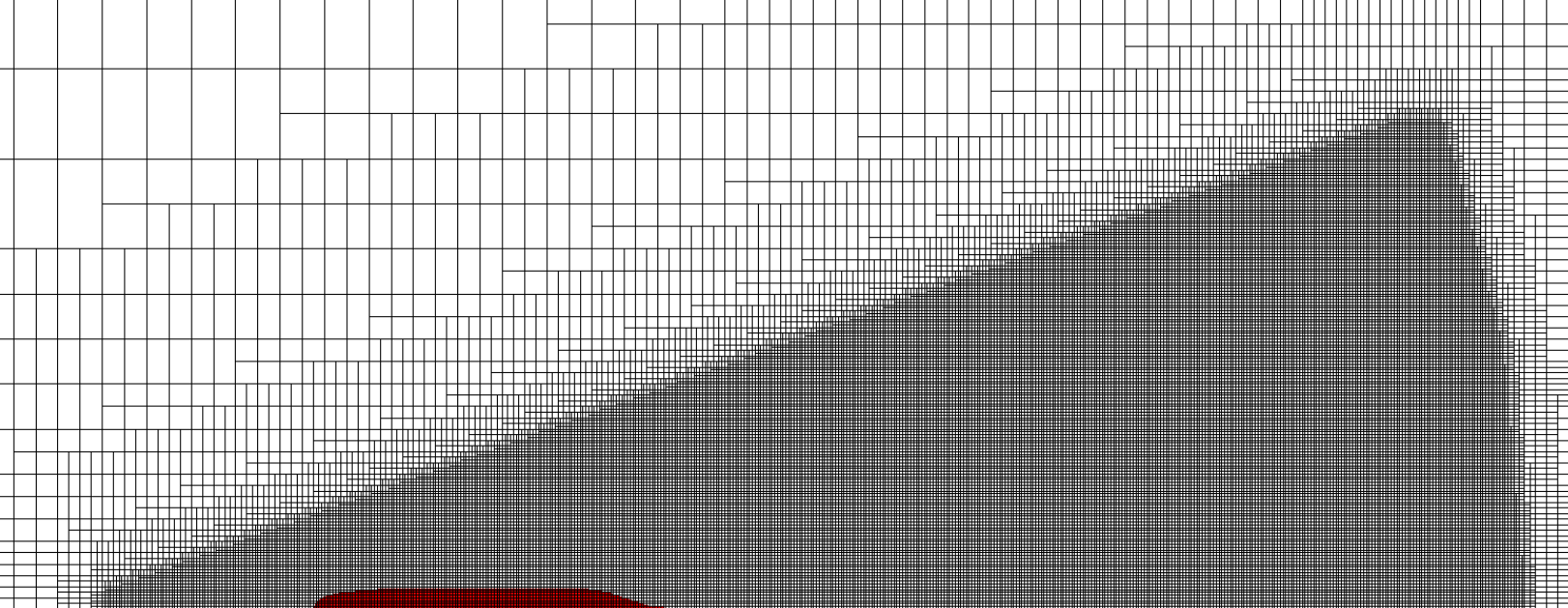}
\caption{Submerged DARPA SUBOFF case ($\mathrm{Re}_\mathrm{L} = \SI{8 542 550}{}$, $\mathrm{Fn}=0.3$): Numerical grid in the still water plane}
\label{fig:darpa_primal}
\end{figure}

The utilized unstructured numerical grid is displayed in Fig. \ref{fig:darpa_suboff} (b) and consists of approximately $\SI{4 000 000}{}$ control volumes. Due to symmetry, only half of the ship
is modelled in the lateral direction. The fully turbulent simulations employ a wall-function based $k-\omega$ SST model \citep{menter2003ten} and convective terms for momentum and turbulence are approximated using the QUICK scheme. The CICSAM scheme is used for the compressive concentration transport. At the inlet, velocity and concentration values are prescribed, slip walls are used along the top, bottom as well as outer boundaries and a hydrostatic pressure boundary is employed along the outlet. Along the midships plane a symmetry condition is declared. The wall normal distance of the first grid layer reads $y^+ \approx \SI{30}{}$ and the free surface refinement employs approximately $\delta x_\mathrm{1}^* / \lambda^* = \delta x_\mathrm{3}^* / \lambda^* = 1/50$ cells in the longitudinal as well as lateral and $\delta x_\mathrm{2}^* / \lambda^* = 1/10$ cells in the normal direction. The tangential resolution of the free-surface region is refined within a Kelvin-wedge to capture the wave field generated by the submerged geometry, cf. Fig. \ref{fig:darpa_primal}. According to Alg. (\ref{alg:sub_cycling}) the integration in pseudo time applies an adaptive time step size based on $\mathrm{Co}_\mathrm{tar} = 0.4$ embedded in five sub-cycles.

Once the shape derivative (\ref{equ:shape_derivative}) is computed, a step into the direction of steepest descent is performed. The necessary shape gradient is evaluated based on the Steklov-Poincare metric \citep{schulz2016computational, haubner2020continuous}
\begin{alignat}{2}
\frac{\partial^2 g_\mathrm{i}^*}{\partial x_\mathrm{k}^{*2}} &= 0 \qquad &&: \Omega^* \label{equ:steklev_poincare} \\
\frac{\partial g_\mathrm{i}^*}{\partial x_\mathrm{k}^*} n_\mathrm{k} &= s^* \, n_\mathrm{i} \qquad &&: \Gamma_\mathrm{D}^* \\
\frac{\partial g_\mathrm{i}^*}{\partial x_\mathrm{k}^*} n_\mathrm{k}  &= 0 \qquad &&: \Gamma \not \subset \Gamma_\mathrm{D}^* \\
g_\mathrm{i}^* &= 0 \qquad &&: \Gamma_\mathrm{in}^*, \Gamma_\mathrm{out}^*, \Gamma_\mathrm{slip}^*, \Gamma_\mathrm{symm}^*,
\end{alignat}
where a deformation along the symmetry plane is suppressed to fix the mid ship plane. Since the purely physical minimum is represented by a disappearing geometry, an additional geometric constraint is introduced to preserve the displacement: $\int_\mathrm{\Gamma^*} g_\mathrm{i}^* n_\mathrm{i} \mathrm{d} \Gamma^* = 0$. The latter enters the gradient-based approach based on a sub-optimization problem, that finally leads to a second set of equations that need to be solved, viz.
\begin{alignat}{2}
\frac{\partial^2 \hat{g}_\mathrm{i}^*}{\partial x_\mathrm{k}^{*2}} &= 0 \qquad &&: \Omega^* \label{equ:adjoint_steklev_poincare}  \\
\frac{\partial \hat{g}_\mathrm{i}^*}{\partial x_\mathrm{k}^*} n_\mathrm{k}  &= n_\mathrm{i} \qquad &&: \Gamma_\mathrm{D}^* \\
\frac{\partial \hat{g}_\mathrm{i}^*}{\partial x_\mathrm{k}^*} n_\mathrm{k}  &= 0 \qquad &&: \Gamma \not \subset \Gamma_\mathrm{D}^* \\
\hat{g}_\mathrm{i}^* &= 0 \qquad &&: \Gamma_\mathrm{in}^*, \Gamma_\mathrm{out}^*, \Gamma_\mathrm{slip}^*, \Gamma_\mathrm{symm}^*.
\end{alignat}
Combining (\ref{equ:steklev_poincare}) and (\ref{equ:adjoint_steklev_poincare}) yields the displacement conserving field deformation $f_\mathrm{i}^* = g_\mathrm{i}^* - \alpha^* \hat{g}_\mathrm{i}^*$ where the constant $\alpha^* = \int_\mathrm{\Gamma^*} g_\mathrm{i}^* n_\mathrm{i} \mathrm{d} \Gamma^* / \int_\mathrm{\Gamma^*} \hat{g}_\mathrm{i}^* n_\mathrm{i} \mathrm{d} \Gamma^*$ results from a conceptually Newton-step when minimizing $ [\int_\mathrm{\Gamma^*} g_\mathrm{i}^* n_\mathrm{i} \mathrm{d} \Gamma^* ]^2$. Equ. (\ref{equ:adjoint_steklev_poincare}) represent the adjoint equations of (\ref{equ:steklev_poincare}) for this minimization problem. Only the middle body of the geometry is considered as design surface.
The employed steepest descent approach uses a step size based on a prescribed maximum deformation for the initial design, viz. $f_\mathrm{i}^* \rightarrow (f_\mathrm{i}^* / f_\mathrm{i}^\mathrm{*,max} ) L_\mathrm{s}^* / 1000$, where we take one per mil of the reference length as maximum deformation. The step size is kept constant over the optimization, leading to a smooth convergence of the objective functional.

Five optimisations are carried out: Four of them carefully increase the adjoint apparent viscosity from $\nu_\mathrm{\hat{c}} \rho_\mathrm{b}^* / \mu_\mathrm{b}^* = 1$ to $\nu_\mathrm{\hat{c}} \rho_\mathrm{b}^* / \mu_\mathrm{b}^* = 1000$. The fifth optimization employs the smallest synthetic viscsosity but neglects all three source terms scaling with the density and viscosity difference within the dual concentration equation and thus resembles a frozen material property approach. These source/sink terms are one of the main reasons for the introduction of apparent viscosity, as they drastically increase the coupling of the adjoint system of equations and by that decrease the numerical stability.

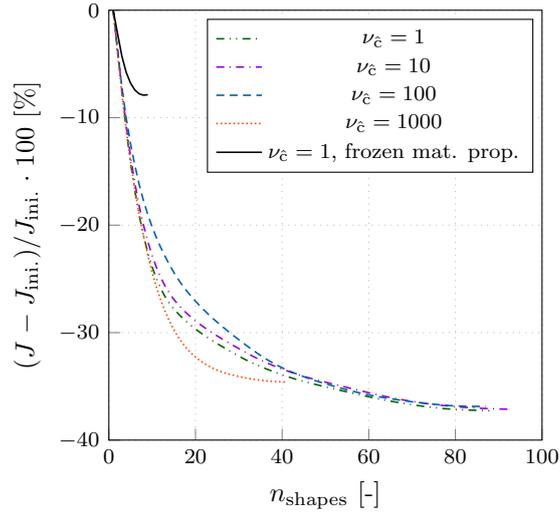
\begin{figure}
\centering
\tinyPicture
\begin{tikzpicture}
\begin{axis}[
 xlabel style={text width=0.25\textwidth,align=center},
 ylabel style={text width=0.25\textwidth,align=center},
 xlabel={$n_\mathrm{shapes}$ [-]},
 ylabel={$(J-J_\mathrm{ini.}) / J_\mathrm{ini.} \cdot 100 $ [\%]},
 xmin=0,xmax=100,
 legend style={at={(0.98,0.98)},anchor=north east},
 %ymode=log
 ymin=-40.0,ymax=0.0
]
\addplot [line5] table[x expr={\thisrowno{0}},y expr={\thisrowno{1}}]{data/DARPA_Wave_Objective_o1.dat};
\addplot [line4] table[x expr={\thisrowno{0}},y expr={\thisrowno{1}}]{data/DARPA_Wave_Objective_o2.dat};
\addplot [line3] table[x expr={\thisrowno{0}},y expr={\thisrowno{1}}]{data/DARPA_Wave_Objective_o3.dat};
\addplot [line2] table[x expr={\thisrowno{0}},y expr={\thisrowno{1}}]{data/DARPA_Wave_Objective_o4.dat};
\addplot [line1] table[x expr={\thisrowno{0}},y expr={\thisrowno{1}}]{data/DARPA_Wave_Objective_o5.dat};
 
\addlegendentry{$\nu_\mathrm{\hat{c}} = 1$};
\addlegendentry{$\nu_\mathrm{\hat{c}} = 10$};
\addlegendentry{$\nu_\mathrm{\hat{c}} = 100$};
\addlegendentry{$\nu_\mathrm{\hat{c}} = 1000$};
\addlegendentry{$\nu_\mathrm{\hat{c}} = 1$, frozen mat. prop.};
 
\end{axis}
\end{tikzpicture}
\caption{Submerged DARPA SUBOFF case ($\mathrm{Re}_\mathrm{L} = \SI{8 542 550}{}$, $\mathrm{Fn}=0.3$): Inverse concentration objective decrease over the number of performed shape updates during a steepest descent procedure. Four optimizations differ in the amplitude of the apparent viscosity and one optimization neglects all adjoint coupling terms that originate from the derivative of material properties.}
\label{fig:darpa_objectives}
\end{figure}

The relative decrease of the cost functional is depicted over the number of the gradient steps (number of the computed geometries) in Fig. (\ref{fig:darpa_objectives}). The optimizations were terminated, as soon as the new value of the objective does not fall below the old one by less than 0.1\%.
Except for the optimization with frozen material property treatment, all optimizations achieve a similar order of magnitude in the minimization of the cost function. The adjoint coupling terms resulting from the derivation of the fluid properties seem to have a much stronger influence on the shape derivation than the adjoint apparent viscosity proposed for stabilization. An increase of the latter results only in a deviating convergence behavior and a somewhat smaller complete reduction of the cost functional. Interestingly, the optimization with the highest synthetic viscosity converges the fastest.

\begin{figure}
\centering
\analytiSolutionPictures
\begin{tikzpicture}
\begin{axis}[
 xlabel style={text width=0.25\textwidth,align=center},
 ylabel style={text width=0.25\textwidth,align=center},
 xlabel={$x_\mathrm{1}^* / L_\mathrm{s}^*$ [-]},
 ylabel={$x_\mathrm{FS,2}^* / L_\mathrm{s}^* \cdot 100$ [-]},
 xmin=-1,xmax=3.5,
 ymin=-0.017,ymax=0.021,
 xtick={-1,0,1,2,3},
 ytick={-0.01, 0.0, 0.01, 0.02},
 yticklabels={-1, 0, 1, 2},
 scaled y ticks = false
]
\addplot [line1] table[x expr={\thisrowno{0}},y expr={\thisrowno{1}}]{data/DARPA_Wave_Cut_0D_o0.dat};
\addplot [line2] table[x expr={\thisrowno{0}},y expr={\thisrowno{1}}]{data/DARPA_Wave_Cut_0D_o1.dat};
\addplot [line3] table[x expr={\thisrowno{0}},y expr={\thisrowno{1}}]{data/DARPA_Wave_Cut_0D_o4.dat};
\addplot [line4] table[x expr={\thisrowno{0}},y expr={\thisrowno{1}}]{data/DARPA_Wave_Cut_0D_o5.dat};

\addlegendentry{initial};

\end{axis}
\end{tikzpicture}
\begin{tikzpicture}
\begin{axis}[
 xlabel style={text width=0.25\textwidth,align=center},
 ylabel style={text width=0.25\textwidth,align=center},
 xlabel={$x_\mathrm{1}^* / L_\mathrm{s}^*$ [-]},
 ylabel={$x_\mathrm{FS,2}^* / L_\mathrm{s}^* \cdot 100$ [-]},
 xmin=-1,xmax=3.5,
 legend style={at={(0.02,0.98)},anchor=north west},
 ymin=-0.012,ymax=0.013,
 xtick={-1,0,1,2,3},
 ytick={-0.01, 0.0, 0.01, 0.02},
 yticklabels={-1, 0, 1, 2},
 scaled y ticks = false
]
\addplot [line2] table[x expr={\thisrowno{0}},y expr={\thisrowno{1}}]{data/DARPA_Wave_Cut_2D_o1.dat};
\addplot [line3] table[x expr={\thisrowno{0}},y expr={\thisrowno{1}}]{data/DARPA_Wave_Cut_2D_o4.dat};
\addplot [line1] table[x expr={\thisrowno{0}},y expr={\thisrowno{1}}]{data/DARPA_Wave_Cut_2D_o0.dat};
\addplot [line4] table[x expr={\thisrowno{0}},y expr={\thisrowno{1}}]{data/DARPA_Wave_Cut_2D_o5.dat};

\addlegendentry{$\nu_\mathrm{\hat{c}} = 1$};
\addlegendentry{$\nu_\mathrm{\hat{c}} = 1000$};

\end{axis}
\end{tikzpicture}
\begin{tikzpicture}
\begin{axis}[
 xlabel style={text width=0.25\textwidth,align=center},
 ylabel style={text width=0.25\textwidth,align=center},
 xlabel={$x_\mathrm{1}^* / L_\mathrm{s}^*$ [-]},
 ylabel={$x_\mathrm{FS,2}^* / L_\mathrm{s}^* \cdot 100$ [-]},
 xmin=-1,xmax=3.5,
 legend style={at={(0.02,0.98)},anchor=north west},
 ymin=-0.01,ymax=0.011,
 xtick={-1,0,1,2,3},
 ytick={-0.01, 0.0, 0.01, 0.02},
 yticklabels={-1, 0, 1, 2},
 scaled y ticks = false
]

\addplot [line4] table[x expr={\thisrowno{0}},y expr={\thisrowno{1}}]{data/DARPA_Wave_Cut_4D_o5.dat};
\addplot [line1] table[x expr={\thisrowno{0}},y expr={\thisrowno{1}}]{data/DARPA_Wave_Cut_4D_o0.dat};
\addplot [line2] table[x expr={\thisrowno{0}},y expr={\thisrowno{1}}]{data/DARPA_Wave_Cut_4D_o1.dat};
\addplot [line3] table[x expr={\thisrowno{0}},y expr={\thisrowno{1}}]{data/DARPA_Wave_Cut_4D_o4.dat};

\addlegendentry{$\nu_\mathrm{\hat{c}} = 1$, frozen mat. prop.};

\end{axis}
\end{tikzpicture}
\caption{Submerged DARPA SUBOFF case ($\mathrm{Re}_\mathrm{L} = \SI{8 542 550}{}$, $\mathrm{Fn}=0.3$): Wave elevation for the initial es well as for three optimized shapes with $\nu_\mathrm{\hat{c}} = 1$ (with and without adjoint two phase coupling terms) as well as $\nu_\mathrm{\hat{c}} = 1000$ along the main flow direction ($x_\mathrm{1}^*$) at three different lateral positions, viz. left) $x_\mathrm{3}^* / D^* = 0$, middle) $x_\mathrm{3}^* / D^* = 2$ and right) $x_\mathrm{3}^* / D^* = 4$.}
\label{fig:darpa_wave_cuts}
\end{figure}
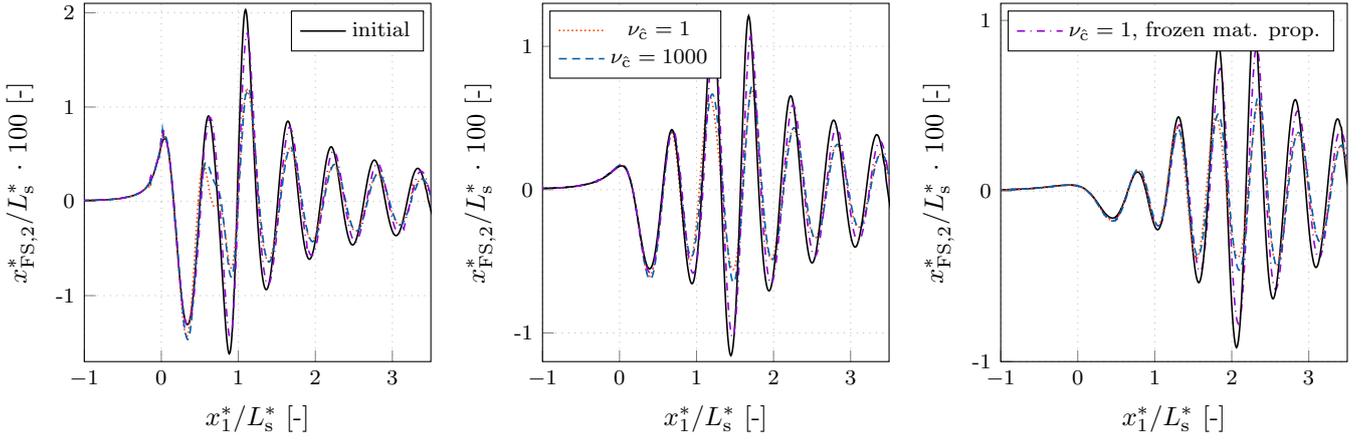

\begin{figure}
\centering
\input{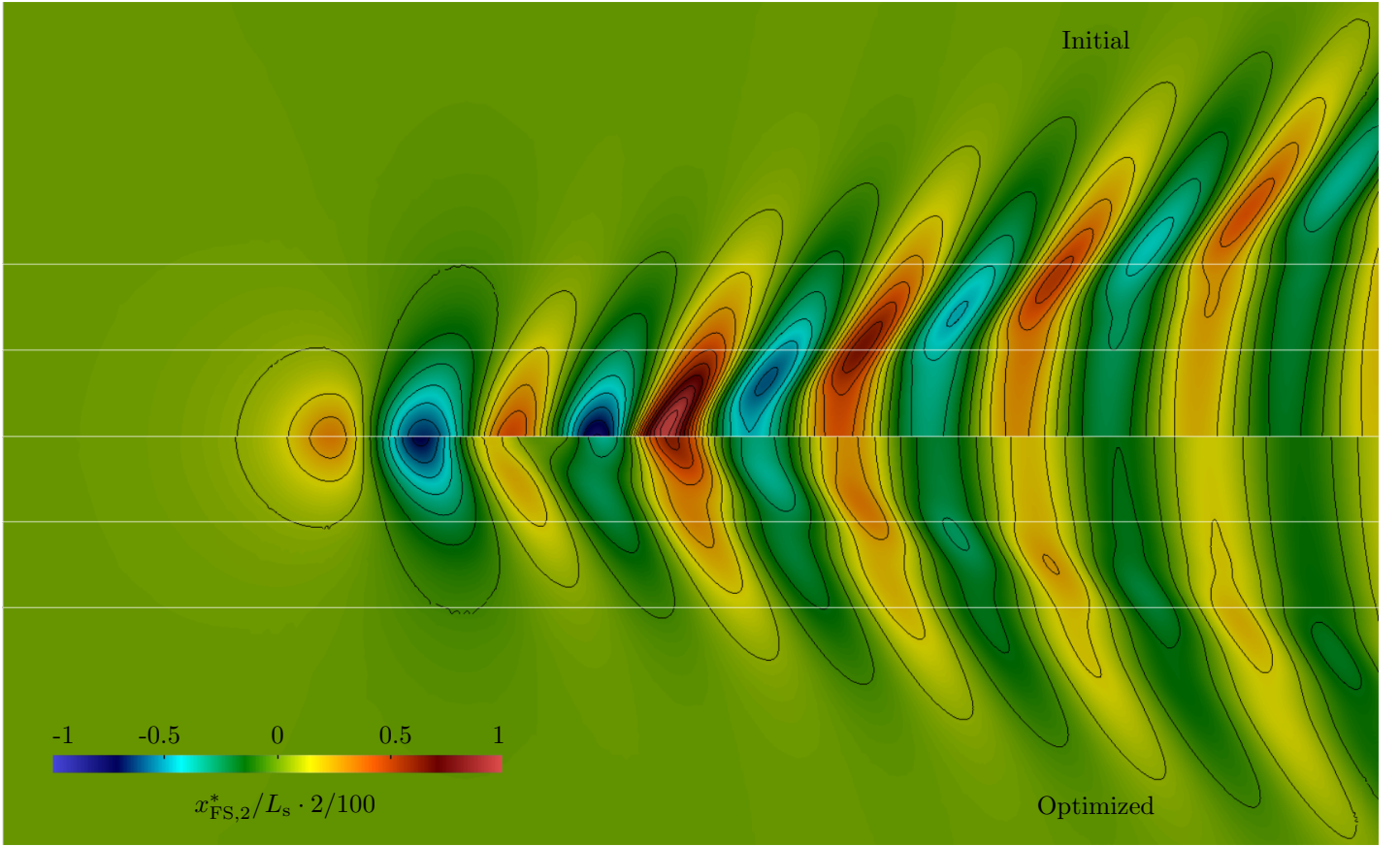}
\caption{Submerged DARPA SUBOFF case ($\mathrm{Re}_\mathrm{L} = \SI{8 542 550}{}$, $\mathrm{Fn}=0.3$): Normalized wave elevation for top) the initial geometry and bottom) the optimized hull resulting from an optimization with $\nu_\mathrm{\hat{c}} = 1$}
\label{fig:darpa_wave_field}
\end{figure}

The apparently lower influence of the adjoint diffusion compared to the coupling terms resulting from differentiation of the material properties is also shown in the wave patterns in Fig. (\ref{fig:darpa_wave_cuts}) at three three different lateral positions, viz. $x_\mathrm{3}^* / D^* = 0$ (left), $x_\mathrm{3}^* / D^* = 2$ (middle) and $x_\mathrm{3}^* / D^* = 4$ (right) as indicated in Fig. (\ref{fig:darpa_wave_field}).

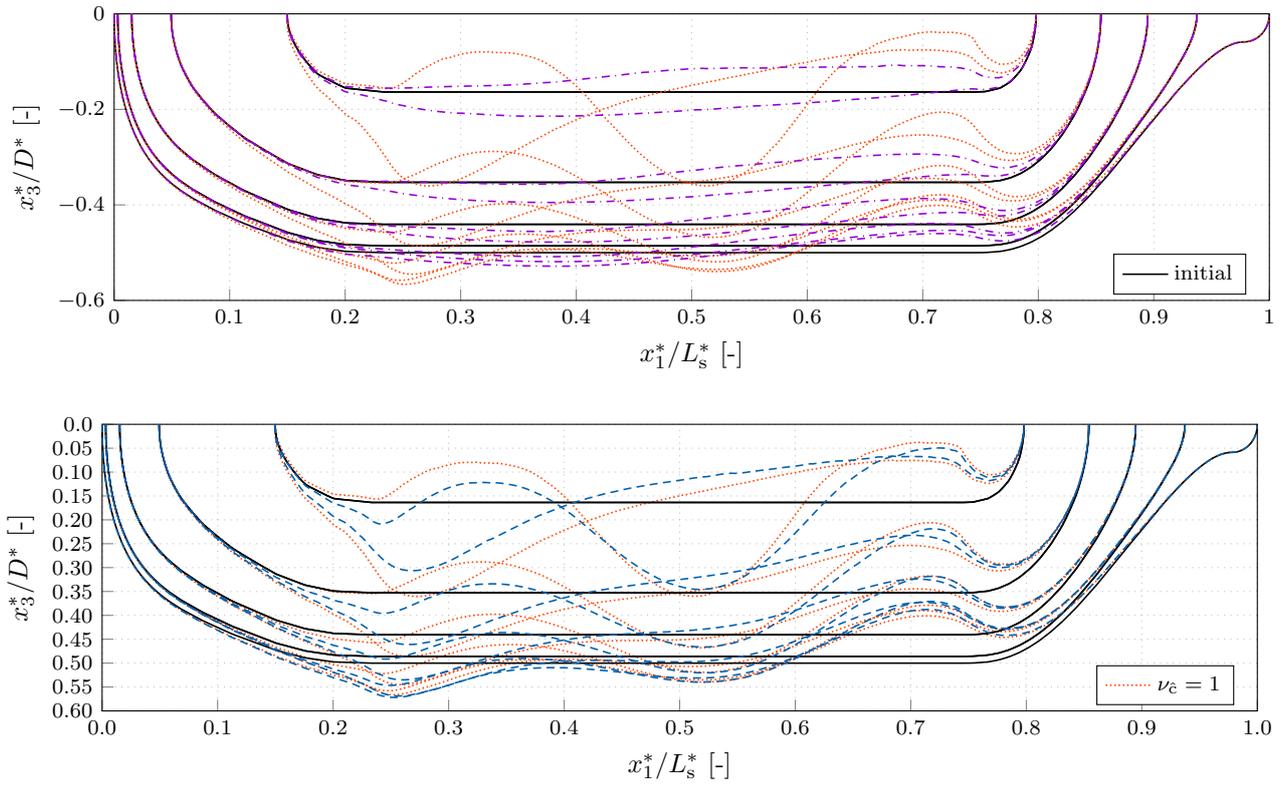
\begin{figure}
\centering
\longPicture
\begin{tikzpicture}
\begin{axis}[
 xlabel style={text width=0.25\textwidth,align=center},
 ylabel style={text width=0.25\textwidth,align=center},
 xlabel={$x_\mathrm{1}^* / L_\mathrm{s}^*$ [-]},
 ylabel={$x_\mathrm{3}^* / D^* $ [-]},
 xmin=0.0,xmax=1.0,
 ymin=-0.6,ymax=0.0,
 legend style={at={(0.98,0.02)},anchor=south east},
 %xmode=log
 %ymode=log
 %xtick={0.0,0.1,0.2,0.3,0.4,0.5,0.6,0.7,0.8,0.9,1.0},
 %xticklabels={0.0,0.0,0.1,0.2,0.3,0.4,0.5,0.6,0.7,0.8,0.9,1.0},
 %ytick={-0.60,-0.55,-0.50,-0.45,-0.40,-0.35,-0.30,-0.25,-0.20,-0.15,-0.10,-0.05,0.0},
 %yticklabels={0.60,0.55,0.50,0.45,0.40,0.35,0.30,0.25,0.20,0.15,0.10,0.05,0.0},
 scaled y ticks = false
]

%\addplot [line1] table[x expr={\thisrowno{0}},y expr={\thisrowno{1}}]{data/DARPA_waterline_o0_01.dat};
\addplot [line1] table[x expr={\thisrowno{0}},y expr={\thisrowno{1}}]{data/DARPA_waterline_o0_02.dat};
\addplot [line1] table[x expr={\thisrowno{0}},y expr={\thisrowno{1}}]{data/DARPA_waterline_o0_03.dat};
\addplot [line1] table[x expr={\thisrowno{0}},y expr={\thisrowno{1}}]{data/DARPA_waterline_o0_04.dat};
\addplot [line1] table[x expr={\thisrowno{0}},y expr={\thisrowno{1}}]{data/DARPA_waterline_o0_05.dat};
\addplot [line1] table[x expr={\thisrowno{0}},y expr={\thisrowno{1}}]{data/DARPA_waterline_o0_06.dat};
\addplot [line1] table[x expr={\thisrowno{0}},y expr={\thisrowno{1}}]{data/DARPA_waterline_o0_07.dat};
\addplot [line1] table[x expr={\thisrowno{0}},y expr={\thisrowno{1}}]{data/DARPA_waterline_o0_08.dat};
\addplot [line1] table[x expr={\thisrowno{0}},y expr={\thisrowno{1}}]{data/DARPA_waterline_o0_09.dat};
\addplot [line1] table[x expr={\thisrowno{0}},y expr={\thisrowno{1}}]{data/DARPA_waterline_o0_10.dat};
%\addplot [line1] table[x expr={\thisrowno{0}},y expr={\thisrowno{1}}]{data/DARPA_waterline_o0_11.dat};

%\addplot [line2] table[x expr={\thisrowno{0}},y expr={\thisrowno{1}}]{data/DARPA_waterline_o1_01.dat};
\addplot [line2] table[x expr={\thisrowno{0}},y expr={\thisrowno{1}}]{data/DARPA_waterline_o1_02.dat};
\addplot [line2] table[x expr={\thisrowno{0}},y expr={\thisrowno{1}}]{data/DARPA_waterline_o1_03.dat};
\addplot [line2] table[x expr={\thisrowno{0}},y expr={\thisrowno{1}}]{data/DARPA_waterline_o1_04.dat};
\addplot [line2] table[x expr={\thisrowno{0}},y expr={\thisrowno{1}}]{data/DARPA_waterline_o1_05.dat};
\addplot [line2] table[x expr={\thisrowno{0}},y expr={\thisrowno{1}}]{data/DARPA_waterline_o1_06.dat};
\addplot [line2] table[x expr={\thisrowno{0}},y expr={\thisrowno{1}}]{data/DARPA_waterline_o1_07.dat};
\addplot [line2] table[x expr={\thisrowno{0}},y expr={\thisrowno{1}}]{data/DARPA_waterline_o1_08.dat};
\addplot [line2] table[x expr={\thisrowno{0}},y expr={\thisrowno{1}}]{data/DARPA_waterline_o1_09.dat};
\addplot [line2] table[x expr={\thisrowno{0}},y expr={\thisrowno{1}}]{data/DARPA_waterline_o1_10.dat};
%\addplot [line2] table[x expr={\thisrowno{0}},y expr={\thisrowno{1}}]{data/DARPA_waterline_o1_11.dat};

%\addplot [line4] table[x expr={\thisrowno{0}},y expr={\thisrowno{1}}]{data/DARPA_waterline_o5_01.dat};
\addplot [line4] table[x expr={\thisrowno{0}},y expr={\thisrowno{1}}]{data/DARPA_waterline_o5_02.dat};
\addplot [line4] table[x expr={\thisrowno{0}},y expr={\thisrowno{1}}]{data/DARPA_waterline_o5_03.dat};
\addplot [line4] table[x expr={\thisrowno{0}},y expr={\thisrowno{1}}]{data/DARPA_waterline_o5_04.dat};
\addplot [line4] table[x expr={\thisrowno{0}},y expr={\thisrowno{1}}]{data/DARPA_waterline_o5_05.dat};
\addplot [line4] table[x expr={\thisrowno{0}},y expr={\thisrowno{1}}]{data/DARPA_waterline_o5_06.dat};
\addplot [line4] table[x expr={\thisrowno{0}},y expr={\thisrowno{1}}]{data/DARPA_waterline_o5_07.dat};
\addplot [line4] table[x expr={\thisrowno{0}},y expr={\thisrowno{1}}]{data/DARPA_waterline_o5_08.dat};
\addplot [line4] table[x expr={\thisrowno{0}},y expr={\thisrowno{1}}]{data/DARPA_waterline_o5_09.dat};
\addplot [line4] table[x expr={\thisrowno{0}},y expr={\thisrowno{1}}]{data/DARPA_waterline_o5_10.dat};
%\addplot [line4] table[x expr={\thisrowno{0}},y expr={\thisrowno{1}}]{data/DARPA_waterline_o5_11.dat};
 
\addlegendentry{initial};

\end{axis}
\end{tikzpicture}
\longPicture
\begin{tikzpicture}
\begin{axis}[
 xlabel style={text width=0.25\textwidth,align=center},
 ylabel style={text width=0.25\textwidth,align=center},
 xlabel={$x_\mathrm{1}^* / L_\mathrm{s}^*$ [-]},
 ylabel={$x_\mathrm{3}^* / D^* $ [-]},
 xmin=0,xmax=1,
 ymin=-0.6,ymax=0.0,
 legend style={at={(0.98,0.02)},anchor=south east},
 %xmode=log
 %ymode=log
 xtick={0.0,0.1,0.2,0.3,0.4,0.5,0.6,0.7,0.8,0.9,1.0},
 xticklabels={0.0,0.1,0.2,0.3,0.4,0.5,0.6,0.7,0.8,0.9,1.0},
 ytick={-0.60,-0.55,-0.50,-0.45,-0.40,-0.35,-0.30,-0.25,-0.20,-0.15,-0.10,-0.05,0.0},
 yticklabels={0.60,0.55,0.50,0.45,0.40,0.35,0.30,0.25,0.20,0.15,0.10,0.05,0.0},
 scaled y ticks = false
]

%\addplot [line2] table[x expr={\thisrowno{0}},y expr={\thisrowno{1}}]{data/DARPA_waterline_o1_01.dat};
\addplot [line2] table[x expr={\thisrowno{0}},y expr={\thisrowno{1}}]{data/DARPA_waterline_o1_02.dat};
\addplot [line2] table[x expr={\thisrowno{0}},y expr={\thisrowno{1}}]{data/DARPA_waterline_o1_03.dat};
\addplot [line2] table[x expr={\thisrowno{0}},y expr={\thisrowno{1}}]{data/DARPA_waterline_o1_04.dat};
\addplot [line2] table[x expr={\thisrowno{0}},y expr={\thisrowno{1}}]{data/DARPA_waterline_o1_05.dat};
\addplot [line2] table[x expr={\thisrowno{0}},y expr={\thisrowno{1}}]{data/DARPA_waterline_o1_06.dat};
\addplot [line2] table[x expr={\thisrowno{0}},y expr={\thisrowno{1}}]{data/DARPA_waterline_o1_07.dat};
\addplot [line2] table[x expr={\thisrowno{0}},y expr={\thisrowno{1}}]{data/DARPA_waterline_o1_08.dat};
\addplot [line2] table[x expr={\thisrowno{0}},y expr={\thisrowno{1}}]{data/DARPA_waterline_o1_09.dat};
\addplot [line2] table[x expr={\thisrowno{0}},y expr={\thisrowno{1}}]{data/DARPA_waterline_o1_10.dat};
%\addplot [line2] table[x expr={\thisrowno{0}},y expr={\thisrowno{1}}]{data/DARPA_waterline_o1_11.dat};

%\addplot [line1] table[x expr={\thisrowno{0}},y expr={\thisrowno{1}}]{data/DARPA_waterline_o0_01.dat};
\addplot [line1] table[x expr={\thisrowno{0}},y expr={\thisrowno{1}}]{data/DARPA_waterline_o0_02.dat};
\addplot [line1] table[x expr={\thisrowno{0}},y expr={\thisrowno{1}}]{data/DARPA_waterline_o0_03.dat};
\addplot [line1] table[x expr={\thisrowno{0}},y expr={\thisrowno{1}}]{data/DARPA_waterline_o0_04.dat};
\addplot [line1] table[x expr={\thisrowno{0}},y expr={\thisrowno{1}}]{data/DARPA_waterline_o0_05.dat};
\addplot [line1] table[x expr={\thisrowno{0}},y expr={\thisrowno{1}}]{data/DARPA_waterline_o0_06.dat};
\addplot [line1] table[x expr={\thisrowno{0}},y expr={\thisrowno{1}}]{data/DARPA_waterline_o0_07.dat};
\addplot [line1] table[x expr={\thisrowno{0}},y expr={\thisrowno{1}}]{data/DARPA_waterline_o0_08.dat};
\addplot [line1] table[x expr={\thisrowno{0}},y expr={\thisrowno{1}}]{data/DARPA_waterline_o0_09.dat};
\addplot [line1] table[x expr={\thisrowno{0}},y expr={\thisrowno{1}}]{data/DARPA_waterline_o0_10.dat};
%\addplot [line1] table[x expr={\thisrowno{0}},y expr={\thisrowno{1}}]{data/DARPA_waterline_o0_11.dat};

%\addplot [line3] table[x expr={\thisrowno{0}},y expr={\thisrowno{1}}]{data/DARPA_waterline_o4_01.dat};
\addplot [line3] table[x expr={\thisrowno{0}},y expr={\thisrowno{1}}]{data/DARPA_waterline_o4_02.dat};
\addplot [line3] table[x expr={\thisrowno{0}},y expr={\thisrowno{1}}]{data/DARPA_waterline_o4_03.dat};
\addplot [line3] table[x expr={\thisrowno{0}},y expr={\thisrowno{1}}]{data/DARPA_waterline_o4_04.dat};
\addplot [line3] table[x expr={\thisrowno{0}},y expr={\thisrowno{1}}]{data/DARPA_waterline_o4_05.dat};
\addplot [line3] table[x expr={\thisrowno{0}},y expr={\thisrowno{1}}]{data/DARPA_waterline_o4_06.dat};
\addplot [line3] table[x expr={\thisrowno{0}},y expr={\thisrowno{1}}]{data/DARPA_waterline_o4_07.dat};
\addplot [line3] table[x expr={\thisrowno{0}},y expr={\thisrowno{1}}]{data/DARPA_waterline_o4_08.dat};
\addplot [line3] table[x expr={\thisrowno{0}},y expr={\thisrowno{1}}]{data/DARPA_waterline_o4_09.dat};
\addplot [line3] table[x expr={\thisrowno{0}},y expr={\thisrowno{1}}]{data/DARPA_waterline_o4_10.dat};
%\addplot [line3] table[x expr={\thisrowno{0}},y expr={\thisrowno{1}}]{data/DARPA_waterline_o4_11.dat};

\addlegendentry{$\nu_\mathrm{\hat{c}} = 1$};
 
\end{axis}
\end{tikzpicture}
\caption{Submerged DARPA SUBOFF case ($\mathrm{Re}_\mathrm{L} = \SI{8 542 550}{}$, $\mathrm{Fn}=0.3$): Water lines for the initial (black) and with $\nu_\mathrm{\hat{c}} = 1$ optimised geometry (orange) as well as the resulting slices for an optimization that employs $\nu_\mathrm{\hat{c}} = 1$ with a frozen material property approach (purple) and $\nu_\mathrm{\hat{c}} = 1000$ (blue).}
\label{fig:darpa_waterline}
\end{figure}

\begin{figure}
\centering
\longPicture
\begin{tikzpicture}
\begin{axis}[
 xlabel style={text width=0.25\textwidth,align=center},
 ylabel style={text width=0.25\textwidth,align=center},
 xlabel={$x_\mathrm{1}^* / L_\mathrm{s}^*$ [-]},
 ylabel={$x_\mathrm{2}^* / D^* $ [-]},
 xmin=0,xmax=1,
 ymin=-0.6,ymax=0.6,
 legend style={at={(0.98,0.02)},anchor=south east},
 %xmode=log
 %ymode=log
 xtick={0.0,0.1,0.2,0.3,0.4,0.5,0.6,0.7,0.8,0.9,1.0},
 xticklabels={0.0,0.0,0.1,0.2,0.3,0.4,0.5,0.6,0.7,0.8,0.9,1.0},
 ytick={-0.60,-0.55,-0.50,-0.45,-0.40,-0.35,-0.30,-0.25,-0.20,-0.15,-0.10,-0.05,0.0,0.05,0.10,0.15,0.20,0.25,0.30,0.35,0.40,0.45,0.50,0.55,0.60},
 yticklabels={-0.60,,-0.50,,-0.40,,-0.30,,-0.20,,-0.10,,0.00,,0.10,,0.20,,0.30,,0.40,,0.50,,0.60},
 scaled y ticks = false
]

%\addplot [line4] table[x expr={\thisrowno{0}},y expr={\thisrowno{1}}]{data/DARPA_buttock_o5_01.dat};
%\addplot [line4] table[x expr={\thisrowno{0}},y expr={\thisrowno{1}}]{data/DARPA_buttock_o5_02.dat};
%\addplot [line4] table[x expr={\thisrowno{0}},y expr={\thisrowno{1}}]{data/DARPA_buttock_o5_03.dat};
\addplot [line4] table[x expr={\thisrowno{0}},y expr={\thisrowno{1}}]{data/DARPA_buttock_o5_03_1.dat};
\addplot [line4] table[x expr={\thisrowno{0}},y expr={\thisrowno{1}}]{data/DARPA_buttock_o5_03_2.dat};
\addplot [line4] table[x expr={\thisrowno{0}},y expr={\thisrowno{1}}]{data/DARPA_buttock_o5_04.dat};
\addplot [line4] table[x expr={\thisrowno{0}},y expr={\thisrowno{1}}]{data/DARPA_buttock_o5_05.dat};
\addplot [line4] table[x expr={\thisrowno{0}},y expr={\thisrowno{1}}]{data/DARPA_buttock_o5_06.dat};
\addplot [line4] table[x expr={\thisrowno{0}},y expr={\thisrowno{1}}]{data/DARPA_buttock_o5_07.dat};
\addplot [line4] table[x expr={\thisrowno{0}},y expr={\thisrowno{1}}]{data/DARPA_buttock_o5_08.dat};
\addplot [line4] table[x expr={\thisrowno{0}},y expr={\thisrowno{1}}]{data/DARPA_buttock_o5_09.dat};
\addplot [line4] table[x expr={\thisrowno{0}},y expr={\thisrowno{1}}]{data/DARPA_buttock_o5_10.dat};
\addplot [line4] table[x expr={\thisrowno{0}},y expr={\thisrowno{1}}]{data/DARPA_buttock_o5_11.dat};

%\addplot [line1] table[x expr={\thisrowno{0}},y expr={\thisrowno{1}}]{data/DARPA_buttock_o0_01.dat};
%\addplot [line1] table[x expr={\thisrowno{0}},y expr={\thisrowno{1}}]{data/DARPA_buttock_o0_02.dat};
\addplot [line1] table[x expr={\thisrowno{0}},y expr={\thisrowno{1}}]{data/DARPA_buttock_o0_03.dat};
\addplot [line1] table[x expr={\thisrowno{0}},y expr={\thisrowno{1}}]{data/DARPA_buttock_o0_04.dat};
\addplot [line1] table[x expr={\thisrowno{0}},y expr={\thisrowno{1}}]{data/DARPA_buttock_o0_05.dat};
\addplot [line1] table[x expr={\thisrowno{0}},y expr={\thisrowno{1}}]{data/DARPA_buttock_o0_06.dat};
\addplot [line1] table[x expr={\thisrowno{0}},y expr={\thisrowno{1}}]{data/DARPA_buttock_o0_07.dat};
\addplot [line1] table[x expr={\thisrowno{0}},y expr={\thisrowno{1}}]{data/DARPA_buttock_o0_08.dat};
\addplot [line1] table[x expr={\thisrowno{0}},y expr={\thisrowno{1}}]{data/DARPA_buttock_o0_09.dat};
\addplot [line1] table[x expr={\thisrowno{0}},y expr={\thisrowno{1}}]{data/DARPA_buttock_o0_10.dat};
\addplot [line1] table[x expr={\thisrowno{0}},y expr={\thisrowno{1}}]{data/DARPA_buttock_o0_11.dat};

%\addplot [line2] table[x expr={\thisrowno{0}},y expr={\thisrowno{1}}]{data/DARPA_buttock_o1_01.dat};
%\addplot [line2] table[x expr={\thisrowno{0}},y expr={\thisrowno{1}}]{data/DARPA_buttock_o1_02.dat};
\addplot [line2] table[x expr={\thisrowno{0}},y expr={\thisrowno{1}}]{data/DARPA_buttock_o1_02_1.dat};
\addplot [line2] table[x expr={\thisrowno{0}},y expr={\thisrowno{1}}]{data/DARPA_buttock_o1_02_2.dat};
\addplot [line2] table[x expr={\thisrowno{0}},y expr={\thisrowno{1}}]{data/DARPA_buttock_o1_03.dat};
%\addplot [line2] table[x expr={\thisrowno{0}},y expr={\thisrowno{1}}]{data/DARPA_buttock_o1_04.dat};
\addplot [line2] table[x expr={\thisrowno{0}},y expr={\thisrowno{1}}]{data/DARPA_buttock_o1_04_1.dat};
\addplot [line2] table[x expr={\thisrowno{0}},y expr={\thisrowno{1}}]{data/DARPA_buttock_o1_04_2.dat};
\addplot [line2] table[x expr={\thisrowno{0}},y expr={\thisrowno{1}}]{data/DARPA_buttock_o1_05.dat};
\addplot [line2] table[x expr={\thisrowno{0}},y expr={\thisrowno{1}}]{data/DARPA_buttock_o1_06.dat};
\addplot [line2] table[x expr={\thisrowno{0}},y expr={\thisrowno{1}}]{data/DARPA_buttock_o1_07.dat};
\addplot [line2] table[x expr={\thisrowno{0}},y expr={\thisrowno{1}}]{data/DARPA_buttock_o1_08.dat};
\addplot [line2] table[x expr={\thisrowno{0}},y expr={\thisrowno{1}}]{data/DARPA_buttock_o1_09.dat};
\addplot [line2] table[x expr={\thisrowno{0}},y expr={\thisrowno{1}}]{data/DARPA_buttock_o1_10.dat};
\addplot [line2] table[x expr={\thisrowno{0}},y expr={\thisrowno{1}}]{data/DARPA_buttock_o1_11.dat};
 
\addlegendentry{$\nu_\mathrm{\hat{c}} = 1$, frozen mat.};

\end{axis}
\end{tikzpicture}
\longPicture
\begin{tikzpicture}
\begin{axis}[
 xlabel style={text width=0.25\textwidth,align=center},
 ylabel style={text width=0.25\textwidth,align=center},
 xlabel={$x_\mathrm{1}^* / L_\mathrm{s}^*$ [-]},
 ylabel={$x_\mathrm{2}^* / D^* $ [-]},
 xmin=0,xmax=1,
 ymin=-0.6,ymax=0.6,
 legend style={at={(0.98,0.02)},anchor=south east},
 %xmode=log
 %ymode=log
 xtick={0.0,0.1,0.2,0.3,0.4,0.5,0.6,0.7,0.8,0.9,1.0},
 xticklabels={0.0,0.1,0.2,0.3,0.4,0.5,0.6,0.7,0.8,0.9,1.0},
 ytick={-0.60,-0.55,-0.50,-0.45,-0.40,-0.35,-0.30,-0.25,-0.20,-0.15,-0.10,-0.05,0.0,0.05,0.10,0.15,0.20,0.25,0.30,0.35,0.40,0.45,0.50,0.55,0.60},
 yticklabels={-0.60,,-0.50,,-0.40,,-0.30,,-0.20,,-0.10,,0.00,,0.10,,0.20,,0.30,,0.40,,0.50,,0.60},
 scaled y ticks = false
]

%\addplot [line3] table[x expr={\thisrowno{0}},y expr={\thisrowno{1}}]{data/DARPA_buttock_o4_01.dat};
%\addplot [line3] table[x expr={\thisrowno{0}},y expr={\thisrowno{1}}]{data/DARPA_buttock_o4_02.dat};
\addplot [line3] table[x expr={\thisrowno{0}},y expr={\thisrowno{1}}]{data/DARPA_buttock_o4_02_1.dat};
\addplot [line3] table[x expr={\thisrowno{0}},y expr={\thisrowno{1}}]{data/DARPA_buttock_o4_02_2.dat};
\addplot [line3] table[x expr={\thisrowno{0}},y expr={\thisrowno{1}}]{data/DARPA_buttock_o4_03.dat};
%\addplot [line3] table[x expr={\thisrowno{0}},y expr={\thisrowno{1}}]{data/DARPA_buttock_o4_04.dat};
\addplot [line3] table[x expr={\thisrowno{0}},y expr={\thisrowno{1}}]{data/DARPA_buttock_o4_04_1.dat};
\addplot [line3] table[x expr={\thisrowno{0}},y expr={\thisrowno{1}}]{data/DARPA_buttock_o4_04_2.dat};
\addplot [line3] table[x expr={\thisrowno{0}},y expr={\thisrowno{1}}]{data/DARPA_buttock_o4_05.dat};
\addplot [line3] table[x expr={\thisrowno{0}},y expr={\thisrowno{1}}]{data/DARPA_buttock_o4_06.dat};
\addplot [line3] table[x expr={\thisrowno{0}},y expr={\thisrowno{1}}]{data/DARPA_buttock_o4_07.dat};
\addplot [line3] table[x expr={\thisrowno{0}},y expr={\thisrowno{1}}]{data/DARPA_buttock_o4_08.dat};
\addplot [line3] table[x expr={\thisrowno{0}},y expr={\thisrowno{1}}]{data/DARPA_buttock_o4_09.dat};
\addplot [line3] table[x expr={\thisrowno{0}},y expr={\thisrowno{1}}]{data/DARPA_buttock_o4_10.dat};
\addplot [line3] table[x expr={\thisrowno{0}},y expr={\thisrowno{1}}]{data/DARPA_buttock_o4_11.dat};

%\addplot [line1] table[x expr={\thisrowno{0}},y expr={\thisrowno{1}}]{data/DARPA_buttock_o0_01.dat};
%\addplot [line1] table[x expr={\thisrowno{0}},y expr={\thisrowno{1}}]{data/DARPA_buttock_o0_02.dat};
\addplot [line1] table[x expr={\thisrowno{0}},y expr={\thisrowno{1}}]{data/DARPA_buttock_o0_03.dat};
\addplot [line1] table[x expr={\thisrowno{0}},y expr={\thisrowno{1}}]{data/DARPA_buttock_o0_04.dat};
\addplot [line1] table[x expr={\thisrowno{0}},y expr={\thisrowno{1}}]{data/DARPA_buttock_o0_05.dat};
\addplot [line1] table[x expr={\thisrowno{0}},y expr={\thisrowno{1}}]{data/DARPA_buttock_o0_06.dat};
\addplot [line1] table[x expr={\thisrowno{0}},y expr={\thisrowno{1}}]{data/DARPA_buttock_o0_07.dat};
\addplot [line1] table[x expr={\thisrowno{0}},y expr={\thisrowno{1}}]{data/DARPA_buttock_o0_08.dat};
\addplot [line1] table[x expr={\thisrowno{0}},y expr={\thisrowno{1}}]{data/DARPA_buttock_o0_09.dat};
\addplot [line1] table[x expr={\thisrowno{0}},y expr={\thisrowno{1}}]{data/DARPA_buttock_o0_10.dat};
\addplot [line1] table[x expr={\thisrowno{0}},y expr={\thisrowno{1}}]{data/DARPA_buttock_o0_11.dat};

%\addplot [line2] table[x expr={\thisrowno{0}},y expr={\thisrowno{1}}]{data/DARPA_buttock_o1_01.dat};
%\addplot [line2] table[x expr={\thisrowno{0}},y expr={\thisrowno{1}}]{data/DARPA_buttock_o1_02.dat};
\addplot [line2] table[x expr={\thisrowno{0}},y expr={\thisrowno{1}}]{data/DARPA_buttock_o1_02_1.dat};
\addplot [line2] table[x expr={\thisrowno{0}},y expr={\thisrowno{1}}]{data/DARPA_buttock_o1_02_2.dat};
\addplot [line2] table[x expr={\thisrowno{0}},y expr={\thisrowno{1}}]{data/DARPA_buttock_o1_03.dat};
%\addplot [line2] table[x expr={\thisrowno{0}},y expr={\thisrowno{1}}]{data/DARPA_buttock_o1_04.dat};
\addplot [line2] table[x expr={\thisrowno{0}},y expr={\thisrowno{1}}]{data/DARPA_buttock_o1_04_1.dat};
\addplot [line2] table[x expr={\thisrowno{0}},y expr={\thisrowno{1}}]{data/DARPA_buttock_o1_04_2.dat};
\addplot [line2] table[x expr={\thisrowno{0}},y expr={\thisrowno{1}}]{data/DARPA_buttock_o1_05.dat};
\addplot [line2] table[x expr={\thisrowno{0}},y expr={\thisrowno{1}}]{data/DARPA_buttock_o1_06.dat};
\addplot [line2] table[x expr={\thisrowno{0}},y expr={\thisrowno{1}}]{data/DARPA_buttock_o1_07.dat};
\addplot [line2] table[x expr={\thisrowno{0}},y expr={\thisrowno{1}}]{data/DARPA_buttock_o1_08.dat};
\addplot [line2] table[x expr={\thisrowno{0}},y expr={\thisrowno{1}}]{data/DARPA_buttock_o1_09.dat};
\addplot [line2] table[x expr={\thisrowno{0}},y expr={\thisrowno{1}}]{data/DARPA_buttock_o1_10.dat};
\addplot [line2] table[x expr={\thisrowno{0}},y expr={\thisrowno{1}}]{data/DARPA_buttock_o1_11.dat};

\addlegendentry{$\nu_\mathrm{\hat{c}} = 1000$};
 
\end{axis}
\end{tikzpicture}
\caption{Submerged DARPA SUBOFF case ($\mathrm{Re}_\mathrm{L} = \SI{8 542 550}{}$, $\mathrm{Fn}=0.3$): Buttocks for the initial (black) and with $\nu_\mathrm{\hat{c}} = 1$ optimised geometry (orange) as well as the resulting slices for an optimization that employs $\nu_\mathrm{\hat{c}} = 1$ with a frozen material property approach (purple) and $\nu_\mathrm{\hat{c}} = 1000$ (blue).}
\label{fig:darpa_buttock}
\end{figure}
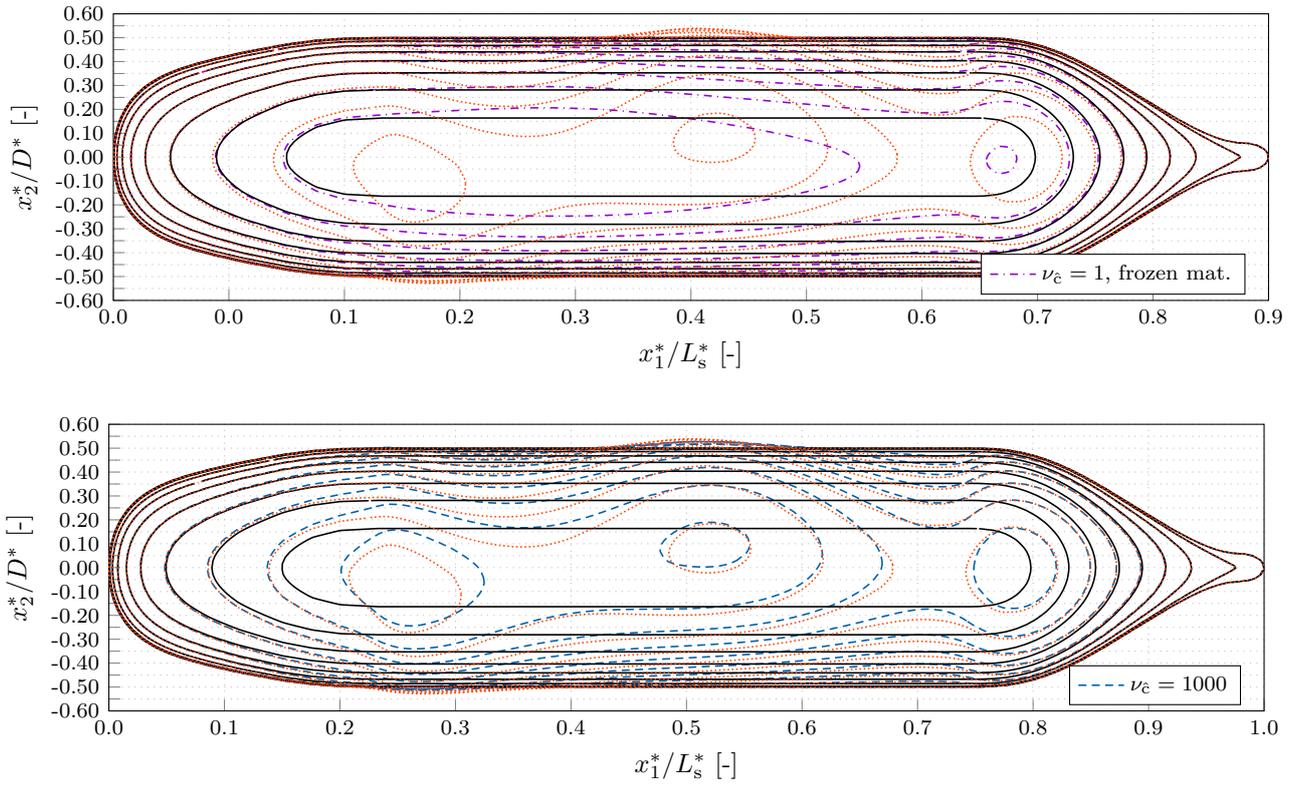

This impression is underlined by the resulting hull geometries. Fig. (\ref{fig:darpa_waterline}) [\ref{fig:darpa_buttock}] shows the water lines [buttocks] of the initial and the optimized geometry with $\nu_\mathrm{\hat{c}} = 1$ against the optimization with frozen material properties (top) and the final shape resulting from $\nu_\mathrm{\hat{c}} = 1000$ (bottom).

\newpage
\section{Discussion and Outlook}
\label{sec:discussion}
The paper has derived and analysed the adjoint complement to an incompressible Volume-of-Fluid (VoF) solver based on the hybrid adjoint method. The latter 
combines the \emph{first-derive-then-discretize} as well as \emph{first-discretize-then-derive} approaches and
derives the continuous adjoint equations (\emph{integration by parts}) as well as the corresponding discretization schemes (\emph{summation by parts}) resulting in a fully consistent numerical Finite-Volume framework.
The deep insight into adjoint relationships associated with this approach allows the use of a pseudo-transient adjoint process for the strongly coupled adjoint system as well as the definition of a model problem for which an analytic primal solution is available. It turns out that the adjoint system cannot be uniquely solved, which is why an additional diffusive term is introduced to the adjoint VoF equation. This term breaks the dual consistency but provides a strongly regularized adjoint system. The lack of adjoint diffusion follows from the underrepresented interface physics inherent to the primal VoF system.

The discrete (compressive) approximation of the discontinuous primal concentration transport based on a generalized Normalised-Variable-Diagram (NVD) was successfully inverted for two prominent convection schemes, namely the High Resolution Interface Capturing Scheme (HRIC) and the Compressive Interface Capturing Scheme for Arbitrary Meshes (CICSAM). Their dual scheme rigorously mirrors the primal NVD stencils. Since attention is restricted to steady state applications, both the primal as well as the dual procedures are performed in pseudo-time and the backward integration of the dual approach is performed around the (pseudo-temporal) converged primal field.  Therefore, the adjoint system experiences the same time step size restrictions as the primal system, is independent of the primal time horizon and forms a robust as well as an a priori stable adjoint solution process.

Numerical results obtained from the modified approach are verified against the analytical solution for the model problem. Subsequently, the consistency of the adjoint system was successfully validated against finite differences for a 2D shape optimization problem. The influence of the modification on the shape sensitivities obtained from simulations for the two-dimensional flow around a submerged hydrofoil at Froude and Reynolds numbers of practical interest are discussed for a range of mobility-parameters. We noted that the influence of the proposed synthetic viscosity varies depending on the cost function, but generally has a more quantitative influence while maintaining the quality, e.g. with respect to zero crossings.

The final 3D application evaluates the proposed approach on integral level via a minimization of the wave elevation above a submerged generic underwater vehicle. Hence, several complete shape optimizations are conducted that vary the apparent viscosity as well as the level of consistency. We found that a comparative small break of the dual consistency introduced through the proposed adjoint diffusion is more acceptable compared to a full neglect of various adjoint coupling terms. The latter provide a strong coupling of the adjoint equations and are therefore the major reason for the introduction of an apparent viscosity.

The advantages of the proposed dual concentration diffusion indicate the general use of diffuse interface models, even on primal side. Therefore, future work might already address the problem of a non-differentiable free surface by e.g. a Cahn-Hilliard Navier-Stokes approach. Furthermore, the integral influence of adjoint coupling terms should also be investigated term-wise in more detail for other geometries of practical relevance.

\section{Acknowledgments}
The current work is a part of the research projects "Drag Optimisation of Ship Shapes’" funded by the German Research Foundation (DFG, Grant No. RU 1575/3-1)
as well as "Dynamic Adaptation of Modular Shape Optimization Processes" funded by the German Federal Ministry for Economic Affairs and Energy (BMWi, Grant No.  03SX453B).
This support is gratefully acknowledged by the authors. 
Selected computations were performed with resources provided by the North-German Supercomputing Alliance (HLRN).
In addition, we would like to mention the Matlab Symbolic Toolbox \citep{matlabsymbolic}, which was a great help when calculating the analytical solutions.

\section{Authorship Contribution Statement}
\textbf{Niklas K{\"u}hl}: Conceptualization, Methodology, Software, Validation, Formal analysis, Investigation, Writing - original draft, Visualization.
\textbf{J{\"o}rn Kr{\"o}ger} : Methodology, Software, Validation, Formal analysis, Writing - review \& editing.
\textbf{Martin Siebenborn}: Methodology, Conceptualization, Writing - review \& editing.
\textbf{Michael Hinze}: Funding acquisition, Methodology, Writing - review \& editing.
\textbf{Thomas Rung}: Project administration, Funding acquisition, Supervision, Conceptualization, Methodology, Resources, Writing - original draft, Writing - review \& editing.

\bibliographystyle{plainnat}
%\bibliography{library.bib}
%\bibliography{/home/beke/ownCloud/Publications/Literature/library.bib}
\bibliography{./library.bib}
%\bibliography{/home/fds211/kuehl/ownCloud/Publications/Literature/library.bib}

%\input{./tex/derivation_area} 

\end{document}